\documentclass[12pt]{article}
\usepackage{amsthm, bm, graphicx, hyperref, mathrsfs, bbm}
\usepackage{amsfonts}
\usepackage{amsmath}
\usepackage{amssymb}
\usepackage{dsfont}
\usepackage{graphicx}
\usepackage{color}
\usepackage[all, knot]{xy}
\usepackage{tikz}
\usepackage{braket}
\usepackage{epstopdf}
\usepackage[footnotesize]{caption}
\usepackage{amsthm}
\usepackage{enumitem}
\usepackage{mathrsfs}
\usepackage{mathtools}     
\usepackage{subeqnarray}         
\usepackage{cases}               
\usepackage{color}
\usepackage{subfigure}
\usepackage{cite}               
\usepackage{hyperref}            
\usepackage{multirow,makecell}   
\usepackage{textcomp}
\usepackage{wasysym}
\usepackage{url}
\usepackage[margin=3cm]{geometry}
\usepackage{geometry}
\usepackage{amsfonts}
\usepackage{graphicx}
\usepackage{float}
\usepackage{amsmath}
\usepackage{hyperref}
\usepackage{tensor}
\usepackage{subfigure}
\usepackage{cases}
\usepackage{appendix}
\usepackage{multirow}
\usepackage{color}
\allowdisplaybreaks[3]

\begin{document}
\begin{titlepage}
\vspace{0.5cm}
\begin{center}
{\Large\bf{Genus Two Correlation Functions in CFTs with $T\bar{T}$ Deformation}}
\lineskip .75em
\vskip 2.5cm
{\large{Song He$^{\clubsuit,\maltese,}$\footnote{hesong@jlu.edu.cn}, Yun-Ze 
Li$^{\clubsuit,}$\footnote{lyz21@mails.jlu.edu.cn}}}
\vskip 2.5em
{\normalsize\it $^\clubsuit$Center for Theoretical Physics and College of Physics, Jilin University,\\
 Changchun 130012, People's Republic of China
\\$^{\maltese}$Max Planck Institute for Gravitational Physics (Albert Einstein Institute),\\ 
Am M\"uhlenberg 1, 14476 Golm, Germany}
\end{center}
\begin{abstract}
Since the definition of $T\bar{T}$ deformation in the curved Riemann surface is obstructive in the literature, we propose a way to do the deformation in the genus two Riemann surfaces by sewing prescription. We construct the correlation functions of conformal field theories (CFTs) on genus two Riemann surfaces with the $T\bar{T}$ deformation in terms of the perturbative CFT approach. Thanks to sewing construction to form higher genus Riemann surfaces from lower genus ones and conformal symmetry, we systematically obtain the first order $T\bar{T}$ correction to the genus two correlation functions in the $T\bar{T}$ deformed CFTs, e.g., partition function and one/higher-point correlation functions.
\end{abstract}
\end{titlepage}
\newpage
\tableofcontents
\section{Introduction}\label{1}
The $T\bar T$ operator was first introduced by Zamolodchikov in \cite{Zamolodchikov:2004ce}. The $T\bar T$ deformation \cite{Smirnov:2016lqw} of 2d quantum field theories (QFTs) has attracted much research interest recently. This deformation has many remarkable properties such as integrability\cite{Cavaglia:2016oda,Rosenhaus:2019utc,LeFloch:2019wlf,Jorjadze:2020ili,Guica:2017lia}. It means that for an undeformed theory with infinite commuting conserved charges, the charges keep conservation and still commute with each other along with the $T \bar T$ deformation.  In a sense, the integrability of the $T \bar T$ deformation makes it solvable. 

Quantum chaotic theories are quite different from integrable theories. It is a natural question to ask how $T\bar{T}$ deformation changes the characteristic behaviors of chaotic/integrable theories. To extract the quantum chaos signals\cite{Guhr:1997ve}, there were several objects capturing quantum chaos, for example, spectrum form factor (SFF)~\cite{Cotler:2016fpe}, out of time-ordered correlation function (OTOC)~\cite{Shenker:2014cwa,Maldacena:2015waa,Roberts:2014ifa}, operator growth \cite{Roberts:2014isa}, eigenstate thermalization hypersis (ETH)\cite{Deutsch,Srednicki,Lashkari:2016vgj}, pole-skipping phenomenon\cite{Blake:2017ris}, Loschmidt echo\cite{Loschmidt}, and quantum entanglement entropies \cite{Calabrese:2004eu}. In deformed theories, there were some preliminary studies \cite{He:2019vzf,He:2020qcs} to extract quantum chaos signals in terms of quantum entanglement and OTOC. To read off quantum chaos signals from these quantities in the $T \bar T$ deformed theories, one has to construct the correlation functions. In this paper, motivated by these reasons, we mainly focus on constructing the correlation functions on a torus and will not calculate the signals of quantum chaos which will be interesting future problems.

There were extensive attempts to construct the correlation functions in deformed theories. The $T\bar{T}$ deformed partition function on a torus, namely the zero-point correlation function, could be computed, and the associated modular properties have been investigated in \cite{Datta:2018thy,Aharony:2018bad,Cardy:2022mhn}. Furthermore the partition function with chemical potentials for KdV charges turning on was also obtained by \cite{Asrat:2020jsh,He:2020cxp}. The correlation functions for stress tensor operators have been obtained in terms of random geometry \cite{Hirano:2020nwq,Hirano:2020ppu} and  holography\cite{Kraus:2018xrn,Li:2020pwa,Li:2020zjb,Ebert:2022cle}. The deformed correlation functions for generic operators have been obtained in perturbative CFT approach \cite{He:2019ahx,He:2019vzf,He:2020qcs,Ebert:2020tuy}. In~\cite{Cardy:2019qao}, they obtained the deformed correlation functions of the deep UV theories in a non-perturbative way. In the context of a massive scalar \cite{Rosenhaus:2019utc} and Dirac fermion \cite{Dey:2021jyl}, integrability was used to fix renormalization ambiguities presented in correlation functions. More recently, authors of~\cite{Guica:2020uhm,Kraus:2021cwf,He:2021bhj} try to construct surface charges of $T\bar{T}$ deformation to constrain the correlation functions.

In the current work, one specific motivation to investigate the theories on higher genus manifold is to explore field theory data for holography; e.g., the field theory defined on torus will be important to understand the boundary theory, which is the holographic dual to BTZ black hole \cite{Keski-Vakkuri:1998gmz}. The other motivation to study the correlation functions in deformed theory on the torus is associated with reading the information about multiple-interval R\'enyi entropy~\cite{Rajabpour:2015nja,Rajabpour:2015uqa,Rajabpour:2015xkj,Headrick:2015gba}. Motivated by these, we focus on the correlation functions in $T\bar{T}-$deformed CFTs live on a higher genus Riemann surface by sewing flat tori\cite{1980Precise, Mason:2006dk,Gilroy:2015gbj}\footnote{In such a situation, the deformation is still available, which is slightly different from the cases discussed in \cite{Jiang:2019tcq}.}. Since the $T\bar{T}$ operator via point splitting in curved backgrounds is ambiguous \cite{Jiang:2019tcq}, we propose a possible sewing prescription in a local patch of the genus 2 Riemann surface with a local flat metric. {There are two main ways called by sewing prescriptions to construct high genus Riemann surface, the one is Schottky Uniformization \cite{Yin:2007gv,Gaberdiel:2010jf, Headrick:2015gba} and the other is offered by \cite{1980Precise, Mason:2006dk,Gilroy:2015gbj}.} In the current work, we focus on the second sewing prescription. The global metric is hyperbolic and not flat on a compact Riemann surface of genus $g \geq 2$. It brings us to the definition of $T \bar{T}$ operator on curved manifolds. It has been proved in \cite{Jiang:2019tcq} that the definition of $T \bar{T}$ operator via point splitting in curved backgrounds is ambiguous. Alternatively, we follow the original $T \bar{T}$ definition \cite{Smirnov:2016lqw} in the flat space-time, e.g., determinant of energy momentun tensor on torus, and we apply the sewing prescription to glue the two tori to form a genus the genus 2 Riemann surface with a local flat metric. We propose a way to locally define a $T\bar T$ deformed action as following
\begin{align}\label{TTbar}
S^{\lambda}=S_{CFT}-\lambda\Big(\int_{S_1\cup A}d^2z_1 O_{T \bar{T}} (z_1,\bar z_1)+\int_{S_2}d^2z_2 O_{T \bar{T}} (z_2,\bar z_2)\Big), 
\end{align}
where $O_{T \bar{T}}$ is the determinant of the stress tensor of the deformed theory. With following \cite{1980Precise, Mason:2006dk,Gilroy:2015gbj}, $\Sigma^{(2)}$ is the genus two Riemann surface formed by sewing two tori $S_1$ and $S_2$, $z_1$ and $z_2$ are the local coordinates on them. On each torus $S_i$, the  deformed operator $O_{T \bar{T}}(z_i,\bar z_i)$ will be reduced to $T\bar T(z_i,\bar z_i)$. We focus on the deformation near the CFTs, which means the $T\bar T$ coupling $\lambda$ is sufficiently small, and the conformal Ward identity still holds when we calculate the first-order deformation. In this sense, eq.(\ref{TTbar}) is not exact ${T \bar{T}}$ deformation proposed by \cite{Zamolodchikov:2004ce}\cite{Smirnov:2016lqw}, since the $O_{T \bar{T}}$ is not global defined on genus two surface instead of on each torus.\par

Since the definition of $T\bar{T}$ deformation in the curved Riemann surface is obstructive in the literature, we propose a way to generalize the deformation in the genus two Riemann surfaces by sewing prescription. We construct the correlation functions of conformal field theories (CFTs) on genus two Riemann surfaces with the $T\bar{T}$ deformation in terms of the perturbative CFT approach. Thanks to sewing construction to form higher genus Riemann surfaces from lower genus ones and conformal symmetry, we obtain the first order $T\bar{T}$ correction to the genus two correlation functions in the $T\bar{T}$ deformed CFTs, e.g., partition function and one/higher-point correlation functions. Our results offer $T\bar{T}$ deformed field theories data to allow us to extract quantum chaos signals and multiple-interval R\'{e}nyi entropies in deformed field theories.

The structure of this paper is as follows. In Section \ref{2}, we review general approaches to form a genus two Riemann surface by sewing two tori and the associated Ward identity used. In Section \ref{3}, we apply the conformal Ward identity on genus two Riemann surfaces to calculate the partition function's first order $T\bar T$ deformation. In Section \ref{4}, we calculate the first-order deformation of correlation functions. Conclusions and discussions are presented in the final section. In the appendices, we would like to list some relevant techniques and notations which are helpful in our analysis.

\section{Mathematical preparation}\label{2}
In this section, we review some mathematical facts about the Riemann surface relevant to the main content. Firstly, we present a typical approach of sewing together two tori to form a genus two Riemann surface and review a method of constructing a high genus partition function. Then we review the genus two Ward identity, which plays a crucial role in calculating the first-order $T\bar{T}$ deformation.
\subsection{Sewing construction}\label{2.1}
\begin{figure}[H]
\centering
\includegraphics[scale=0.8]{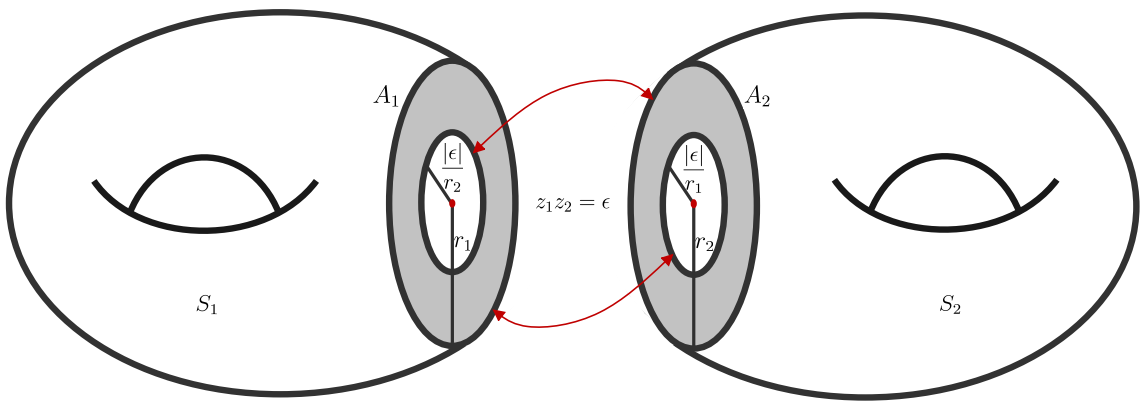}
\caption{The method for constructing a genus two Riemann surface. The annuli $A_1$ and $A_2$ (two gray areas) are introduced on the two tori respectively, and identify them as a single region by sewing relation $z_1z_2=\epsilon$. The two red lines refer to the identification of the boundaries of the two annuli.}
\label{Fig 2}
\end{figure}

We review a general approach to form a genus two Riemann surface by sewing together two tori \cite{1980Precise, Mason:2006dk,Gilroy:2015gbj}
. Two complete tori $T^2_a$ for $a=1,2$ are introduced with modular parameters $\tau_a$ and local coordinates $z_a$, respectively. We construct the torus $T^2_a$ with periods $2\pi i$ and $2\pi i\tau_a$. A closed disk $\lbrace z_a,|z_a|\leq r_a\rbrace$ is introduced on $T^2_a$, which has radius $r_a$ and is centered at $z_a=0$. For the disk, in order not to overlap with itself, its radius $r_a$ must be less than half the minimum period $D_a=\text{min}\lbrace{2\pi,2\pi|\tau_a|}\rbrace$ of the torus. A complex parameter $\epsilon$ is introduced to sew the two tori together, and an annulus $A_a=\lbrace z_a,|\epsilon|/r_{\bar a}\leq|z_a|\leq r_a\rbrace$ is introduced on each torus $T^2_a$. The modulus of $\epsilon$ is upper bounded:
\begin{equation}\label{condition1}
|\epsilon|\leq r_ar_{\bar a}<\frac{1}{4}D_aD_{\bar a},
\end{equation}
where we use the convention $\bar{1}=2,\ \bar{2}=1$. The genus two Riemann surface, which is denoted as $\Sigma^{(2)}$ in the following context, can be formed by identifying two annuli $A_1$ and $A_2$ as a single region $A$ via the sewing relation $z_1z_2=\epsilon$. After removing the small disk $\lbrace z_a, |z_a|\leq|\epsilon|/r_{\bar a}\rbrace$ and annulus $A$, the remainder of the torus $T^2_a$ is denoted as $S_a$, thus $\Sigma^{(2)}$ can be divided into three parts, as shown in Fig.\ref{Fig 2}:
\begin{align}
\Sigma^{(2)}=S_1\cup A\cup S_2. \label{sewing prescription}
\end{align}
\noindent By this construction, a sewn genus two Riemann surface can be completely described by five parameters $(\tau_1,\tau_2,\epsilon,r_1,r_2)$. 
The period matrix $\Omega$ of $\Sigma^{(2)}$, which is used to parameterize the moduli space of Riemann surfaces, can be completely determined by the parameters $\tau_1$, $\tau_2$ and $\epsilon$, by a holomorphic map $F^{\Omega}:(\tau_1,\tau_2,\epsilon)\mapsto\Omega(\tau_1,\tau_2,\epsilon)$ in \cite{Mason:2006dk}.
Here the parameters $\tau_1,\tau_2\in\mathbb{H}_1$ and $\epsilon\in\lbrace{\epsilon,|\epsilon|<\frac{1}{4}D_1D_2}\rbrace\subset\mathbb{C}$, and the period matrix $\Omega\in\mathbb{H}_2$, where $\mathbb{H}_n$ is the Siegel upper half-space. The explicit form of $\Omega(\tau_1,\tau_2,\epsilon)$ is given by
\begin{align}
    2\pi i\Omega_{11}&=2\pi i\tau_1+\epsilon[A_2(\mathbbm{1}-A_1A_2)^{-1}](1,1),\\
    2\pi i\Omega_{22}&=2\pi i\tau_2+\epsilon[A_1(\mathbbm{1}-A_2A_1)^{-1}](1,1),\\
    2\pi i\Omega_{12}&=2\pi i\Omega_{21}=-\epsilon[(\mathbbm{1}-A_1A_2)^{-1}](1,1),
\end{align}
where $(1,1)$ refers to the component of a matrix. $A_a$ is the infinite matrix with components
\begin{equation}
    A_a(m,n)=\epsilon^{\frac{m+n}{2}}(-1)^{m+1}\frac{1}{\sqrt{mn}}\frac{(m+n-1)!}{(m-1)!(n-1)!}E_{m+n}(\tau_a).
\end{equation}
It has been proved in \cite{Mason:2006dk} that the holomorphic map
 $F^{\Omega}$ is injective but not surjective.\par
The general method of calculating the high genus partition function is decomposing it into correlation functions on the lower genus Riemann surface.
A Riemann surface $\Sigma^{(g)}$ can be divided into two lower genus ones by cutting it along a particular closed curve. For instance, the sewn Riemann surface with genus-$2$ in Fig.\ref{Fig 2} can be divided into two tori by cutting it along a non-contractible closed curve in the annulus $A$. By inserting the complete set of boundary states, the partition function on $\Sigma^{(2)}$ can be decomposed into a set of one-point functions on two tori. The generic construction of correlation function of CFTs on higher genus is offered by \cite{Friedan:1986ua,Sonoda:1988mf, Vafa:1987ea,Mason:2009ely}. The correlation function on the higher genus can be expressed by the correlation function on the lower genus, which can be shown in eq.(\ref{sewing correlation function}). It is nontrivial relation itself. It dates from the Conway-Norton conjectures \cite{Conway:1979qga} proved by Borcherds \cite{Borcherds:1983sq}, and Zhu \cite{1996Modular}. The generic correlation function should depend on the CFT data indeed. Here the sewing construction can apply to some CFTs owning the vertex operator algebra structure. In particular, one can put rational CFTs on the generic higher genus surface. To define the CFTs on the higher genus, there is a prior assumption that the path integral exists on the higher genus surface. The path integral can be separated into several pieces as shown in \cite{Friedan:1986ua, Sonoda:1988mf, Vafa:1987ea}. The genus two correlation function with $T\bar{T}$ deformation means that we start from a CFT and do perturbation $T\bar{T}$ theory around the original CFTs up to the first order. Here the conformal symmetry on the higher genus is still held, and we can follow the sewing construction to define a correlation function in the first order $T\bar{T}$ deformed CFTs. In general, the sewing construction can not apply to higher-order deformed cases.\par
We will follow the convention in \cite{Gilroy:2015gbj} throughout the rest of the discussion. We will use the square bracket Vertex Operator Algebra, which was first introduced by Zhu in \cite{1996Modular}. A new vertex operator $V[v,z]$ is defined as 
\begin{equation}
    V[v,z]=V(e^{zL(0)}v,e^z-1).\label{VO}
\end{equation}
In Vertex Operator Algebra, Virasoro generators are the Laurent modes of a particular vertex operator:
\begin{equation}
    V[\tilde\omega,z]=\sum_{n\in\mathbb{Z}}L[n]z^{-n-2},
\end{equation}
where $\tilde\omega$ is called conformal vector with conformal weight $L[0]\tilde\omega=wt[\tilde\omega]\tilde\omega=2\tilde\omega$. For $v$ is a primary state, $wt[v]=wt(v)$. $\mathcal{H}$ is used to denote the complex vector space
in which quantum states live, which can be written as a sum of subspaces spanned by states with the same conformal weight:
\begin{equation}
\mathcal{H}=\bigoplus_{n\in{\mathbb{Z}}}\mathcal{H}_{[n]}, 
\end{equation}
where $\mathcal{H}_{[n]}=\lbrace{v\in \mathcal{H}|L[0]v=wt[v]v=nv}\rbrace$. $\lbrace{u^{[n]}_i\in{\mathcal{H}_{[n]}},i=1,...,\text{dim}\mathcal{H}_{[n]}}\rbrace$ is a basis for $\mathcal{H}_{[n]}$, and its dual basis $\lbrace{\bar u^{[n]}_i}\rbrace$ is defined by $\langle{u^{[n]}_i,\bar u^{[n]}_j}\rangle_{\text{sq}}=\delta_{i,j}$, where $\langle{,}\rangle_{\text{sq}}$ is the square bracket Li-Zamolodchikov metric in \cite{Mason:2009ely}. The partition function $Z^{(2)}(\tau_1,\tau_2,\epsilon)$ on the sewn Riemann surface $\Sigma^{(2)}$ can be decomposed into a combination of product of one-point functions on two tori. $Z^{(1)}(v,x;\tau_a)$ is used to denote a unnormalized one-point function on torus with a general state $v$ inserted at $x\in T^2_a$. The tours one-point function is independent of coordinate $x$ because of translation invariance. The genus two partition function of the sewing construction is given by \cite{Mason:2009ely}
\begin{equation}
Z^{(2)}(\tau_1,\tau_2,\epsilon)=\sum_{n\geq 0}\epsilon^n\sum_{i=1}^{\text{dim}\mathcal{H}_{[n]}}Z^{(1)}(u^{[n]}_i;\tau_1)Z^{(1)}(\bar u^{[n]}_i;\tau_2). \label{genus two partition function}
\end{equation}
In particular, the subspace $\mathcal{H}_{[0]}$ is spanned by the vacuum state $\mathbf{1}$ in CFT, and for vacuum state the one-point function $Z^{(1)}(\mathbf{1};\tau_a)$ is the partition function on torus $T^2_a$. {More generally, we insert states $v_1,...,v_L$ at $x_1,...,x_L\in S_1\cup A$ and $w_1,...,w_R$ at $y_1,...,y_R\in S_2\cup A$, and the genus two $(L+R)$-point function of the sewing construction takes the form \cite{Mason:2009ely}:
\begin{align}
    &Z^{(2)}(v_1,x_1;...;v_L,x_L;w_1,y_1;...;w_R,y_R;\tau_1,\tau_2,\epsilon)\notag\\
    =&\sum_{n\geq 0}\epsilon^n\sum_{i=1}^{\text{dim}\mathcal{H}_{[n]}}Z^{(1)}(V[v_1,x_1]...V[v_L,x_L]u^{[n]}_i;\tau_1)Z^{(1)}(V[w_1,y_1]...V[w_R,y_R]\bar u^{[n]}_i;\tau_2), \label{sewing correlation function}
\end{align}
where $V[v,x]$ is the vertex operator in eq.(\ref{VO}). The correlation function of sewing construction depends on three parameters $\tau_1$, $\tau_2$ and $\epsilon$. The leading term for $n=0$ in eq.(\ref{sewing correlation function}) is the product of two torus correlation functions.}

\par
There are other sewing constructions to form a genus two Riemann surface. In \cite{Headrick:2015gba, Gaberdiel:2010jf}, four disks are removed from the Riemann sphere, and the boundaries of each pair of disks are identified to obtain two handles. Besides, the annuli $A_1$ and $A_2$ mentioned above can be introduced on the same torus and are centered at two different points. One can identify the two annuli by sewing relation to form a self-sewn torus \cite{Mason:2006dk}. One can also obtain the Riemann surface of an arbitrary genus by sewing some spheres with three punctures on each \footnote{One can refer to \cite{Sen:1990bt} and section 9.3 of \cite{Polchinski:1998rq} for construction details.}. In this paper, our calculation is based on the sewing construction in Fig.\ref{Fig 2}, and we follow the conventions in \cite{Gilroy:2015gbj}. 
\subsection{Genus two Ward identity}\label{2.2}
We review the genus two Ward identity derived by Gilroy and Tuite in \cite{Gilroy:2015gbj}. They generalize the Zhu recursion \cite{1996Modular} to genus two correlation function, which provide an approach to represent $(n+1)$-point function in terms of $n$-point functions. The genus-2 Ward identity is a special case of their result.
On the genus two Riemann surface $\Sigma^{(2)}$, some general states $u_1,...,u_L$ and $v_1,...,v_R$ are inserted at $x_1,...,x_L\in S_1\cup A$ and $y_1,...,y_R\in S_2\cup A$, respectively, which provides an unnormalized $(L+R)$-point correlation function denoted as $Z(u_1,x_1;...;u_L,x_L;v_1,y_1;...;v_R,y_R;\tau_1,\tau_2,\epsilon)$. 
The conformal vector $\tilde{\omega}$ are inserted at $z\in\Sigma^{(2)}$ to obtain an $(L+R+1)$-point function, which satisfies the genus two Ward identity \cite{Gilroy:2015gbj}
\begin{align}
&Z(\tilde\omega,z;{\emph{u}},{\emph{x}};{\emph{v}},{\emph{y}};\tau_1,\tau_2,\epsilon)\notag\\=&D_zZ({\emph{u}},{\emph{x}};{\emph{v}},{\emph{y}};\tau_1,\tau_2,\epsilon)\notag\\
&+\sum_{l=1}^L\mathop{\sum}\limits_{j\geq0}{}^2\mathcal{P}_{1+j}(z,x_l;\tau_1,\tau_2,\epsilon)Z(...;L[j-1]u_l,x_l;...)\notag\\
&+\sum_{r=1}^R\mathop{\sum}\limits_{j\geq0}{}^2\mathcal{P}_{1+j}(z,y_r;\tau_2,\tau_1,\epsilon)Z(...;L[j-1]v_r,y_r;...), \label{WI0}
\end{align}
where $({\emph{u}},{\emph{x}};{\emph{v}},{\emph{y}})$ is used to mark $(u_1,x_1;...;u_L,x_L;v_1,y_1;...;v_R,y_R)$. In what follows, we will explain the notations in eq.(\ref{WI0}) and introduce some of our conventions to facilitate the subsequent calculation. The operator $D_z$ contains the derivative operator of the sewing parameters $(\tau_1,\tau_2,\epsilon)$ which is defined as
\begin{gather}
D_z=\sideset{^2}{_1}{\mathop{\mathcal{F}}}(z;\tau_1,\tau_2,\epsilon)\frac{1}{2\pi i}\partial_{\tau_1}+\sideset{^2}{_2}{\mathop{\mathcal{F}}}(z;\tau_1,\tau_2,\epsilon)\frac{1}{2\pi i}\partial_{\tau_2}+{}^2\mathcal{F}^{\Pi}(z;\tau_1,\tau_2,\epsilon)\epsilon^{\frac{1}{2}}\partial_{\epsilon}. \label{Dz}
\end{gather}
The index $2$ at the top left of the $\mathcal{F}$ function represents the conformal weight of $\tilde{\omega}$ in \cite{Gilroy:2015gbj}. The definitions of $\sideset{^2}{_a}{\mathop{\mathcal{F}}}(z;\tau_1,\tau_2,\epsilon)$ for $a=1,2$ are
\begin{gather}
\sideset{^2}{_a}{\mathop{\mathcal{F}}}(z;\tau_1,\tau_2,\epsilon)=
\begin{cases}
1+\mathop{\sum}\limits_{k\geq1}P_{k+3}(z,\tau_a)\alpha_a(k),&z\in S_a\cup A,\\
\mathop{\sum}\limits_{k\geq1}P_{k+3}(z,\tau_{\bar{a}})\beta_{\bar{a}}(k),&z\in S_{\bar{a}}\cup A,
\end{cases} \label{Fa}
\end{gather}
where $P_{k}(z,\tau_a)$ is the Elliptic function defined in Appendix \ref{A}. When the conformal vector $\tilde\omega$ is inserted on the different torus, the $\sideset{^2}{_a}{\mathop{\mathcal{F}}}(z;\tau_1,\tau_2,\epsilon)$ takes various forms. The column vectors with components $\alpha_a(k)$ and $\beta_a(k)$ are defined as follows. Firstly, the infinite matrix $\tilde\Lambda_a$ is introduced with components 
\begin{gather}
\tilde\Lambda_a(m,n)=\epsilon^{\frac{m+n+2}{2}}(-1)^{n+1}\left(\begin{array}{ccc}m+n+1\\n+2\end{array}\right)E_{m+n+2}(\tau_a),
\end{gather}
where $E_n(\tau_a)$ is the {Eisenstein series} (see Appendix \ref{A}), which is equal to zero for $n$ odd. Then we define the $\Theta_a$ matrix by $\Theta_a=(\mathbbm{1}-\tilde\Lambda_{\bar a}\tilde\Lambda_a)^{-1}$, where $\mathbbm{1}$ is the identity matrix. We expand the components of $\Theta_a$ by the powers of $\epsilon$ and have
\begin{align}
&\Theta_a(m,n)\notag\\
=&\delta_{m,n}+\sum_{k\geq1}\epsilon^{\frac{m+n+2k+4}{2}}A^{m,n+2}_{k,\bar a}E_{k+n+2}(\tau_a)+\Big(\sum_{k\geq1}\epsilon^{\frac{m+n+2k+4}{2}}A^{m,n+2}_{k,\bar a}E_{k+n+2}(\tau_a)\Big)^2+...,
\end{align}
where the coefficient $A^{m,n}_{k,a}$ is
\begin{gather}
A^{m,n}_{k,a}=(-1)^{k+n}\left(\begin{array}{ccc}m+k+1\\k+2\end{array}\right)\left(\begin{array}{ccc}k+n-1\\n\end{array}\right)E_{m+k+2}(\tau_{a}). \label{coefficient A}
\end{gather}
By the properties of the Eisenstein series, only if $(m+n)$ is even, $A^{m,n}_{k,a}$ does not all equal to zero. The components $\alpha_a(k)$ and $\beta_a(k)$ are defined as
\begin{align}
\alpha_a(k)=&\mathop{\sum}\limits_{l\geq1}\epsilon^{\frac{k+l+6}{2}}\Theta_a(k,l)A_{1,\bar a}^{l,1},  \label{alpha}\\
\beta_a(k)=&\epsilon^{\frac{k+3}{2}}\Theta_{a}(k,1).  \label{beta}
\end{align}
The definition of ${}^2\mathcal{F}^{\Pi}(z;\tau_1,\tau_2,\epsilon)$ that appears in eq.(\ref{Dz}) is 
\begin{equation}
{}^2\mathcal{F}^{\Pi}(z;\tau_1,\tau_2,\epsilon)=\epsilon^{-\frac{1}{2}}\sum_{k\geq1}P_{k+1}(z,\tau_{a})\theta_a(k) \label{Fpi}	
\end{equation}
for $z\in S_a\cup A$. $\theta_a$ is a column vector with components
\begin{align}
\theta_a(k)=\epsilon^{\frac{k+1}{2}}\times
\begin{cases}
1,&k=1,\\
0,&k=2,\\
\mathop{\sum}\limits_{l\geq1}\epsilon^{\frac{l+1}{2}}\Theta_a(k-2,l)\Big(lE_{l+1}(\tau_{\bar a})+\mathop{\sum}\limits_{m\geq1}\epsilon^{m+1}A^{l,1}_{m,\bar a}E_{m+1}(\tau_a)\Big),&k\geq3.
\end{cases} \label{theta}
\end{align}\par
The function ${}^2\mathcal{P}_{1+j}(z,x;\tau_a,\tau_{\bar a},\epsilon)$ that appears in Ward identity eq.(\ref{WI0}) is referred to as the Genus Two Generalised Weierstrass Function in \cite{Gilroy:2015gbj}, and the index 2 at the top left of it has the same meaning as $\sideset{^2}{}{\mathop{\mathcal{F}}}(z;\tau_1,\tau_2,\epsilon)$. Firstly, the definition of ${}^2\mathcal{P}_{1}(z,x;\tau_a,\tau_{\bar a},\epsilon)$ is
\begin{align}
&{}^2\mathcal{P}_1(z,x;\tau_a,\tau_{\bar a},\epsilon)\notag\\=&
\begin{cases}
P_1(z-x,\tau_a)-P_1(z,\tau_a)+\mathop{\sum}\limits_{k\geq 1} P_{k+3}(z,\tau_a)[\xi^{(0)}_a(x)](k),&z,x\in{S_a}\cup A, \\
-\epsilon P_3(z,\tau_{\bar a})+\mathop{\sum}\limits_{k\geq 1} P_{k+3}(z,\tau_{\bar a})[\zeta^{(0)}_{\bar a}(x)](k),&z\in{S_{\bar a}}\cup A,x\in{S_{a}\cup A}. \label{P1}
\end{cases} 
\end{align}
We define the two column vectors $\xi^{(0)}_a(x)$ and $\zeta^{(0)}_a(x)$ that depend on the coordinate $x$ with components
\begin{align}
[\xi^{(0)}_a(x)](k)=&\sum_{l\geq 1}\epsilon^{\frac{k+l+4}{2}}\Theta_a(k,l)\Big[\frac{l(l+1)}{2}E_{l+2}(\tau_{\bar{a}})+\sum_{m\geq 1}\epsilon^{m}A_{m,\bar a}^{l,0}P'_m(x,\tau_a)\Big],   \label{xi 0}\\
[\zeta^{(0)}_a(x)](k)=&-\sum_{l\geq 1}\epsilon^{\frac{k+l+2}{2}}\Theta_a(k,l)\Big[P'_{l}(x,\tau_{\bar{a}})+\sum_{m\geq 1}\epsilon^{m+2}A_{m,\bar a}^{l,2}E_{m+2}(\tau_a)\Big],   \label{zeta 0}
\end{align}
where we define $P'_k(x,\tau)=P_k(x,\tau)-E_k(x,\tau)$, which is the elliptic function $P_k$ minus the constant term of its Laurent expansion. The functions ${}^2\mathcal{P}_{1+j}(z,x;\tau_1,\tau_2,\epsilon)$ for $j\geq 1$ are defined by taking the derivative of ${}^2\mathcal{P}_{1}(z,x;\tau_1,\tau_2,\epsilon)$ multiple times ${}^2\mathcal{P}_{1+j}(z,x)=(j!)^{-1}\partial^{j}_{x}[{}^2\mathcal{P}_{1}(z,x)]$ in \cite{Gilroy:2015gbj}. We introduce two column vectors $\xi^{(j)}_a(x)$ and $\zeta^{(j)}_a(x)$ to write ${}^2\mathcal{P}_{1+j}(z,x)$ as
\begin{align}
&{}^2\mathcal{P}_{1+j}(z,x;\tau_a,\tau_{\bar a},\epsilon)\notag\\
=&\begin{cases}
P_{1+j}(z-x,\tau_a)+\mathop{\sum}\limits_{k\geq 1} P_{k+3}(z,\tau_a)[\xi^{(j)}_a(x)](k),&z,x\in{S_a}\cup A, \\
\mathop{\sum}\limits_{k\geq 1} P_{k+3}(z,\tau_{\bar a})[\zeta^{(j)}_{\bar a}(x)](k),&z\in{S_{\bar a}}\cup A,x\in{S_{a}\cup A}, \label{P1+j}
\end{cases}
\end{align}
where 
\begin{align}
[\xi^{(j)}_a(x)](k)=&\sum_{l,m\geq 1}\epsilon^{\frac{k+l+2m+4}{2}}\Theta_a(k,l)A_{m,\bar a}^{l,j}P_{m+j}(x,\tau_a),   \label{xi j}\\
[\zeta^{(j)}_a(x)](k)=&(-1)^{1+j}\sum_{m\geq 1}\epsilon^{\frac{k+m+2}{2}}\Theta_a(k,m)\left(\begin{array}{ccc}m+j-1\\j\end{array}\right)P_{m+j}(x,\tau_{\bar{a}}).   \label{zeta j}
\end{align}
In eq.(\ref{WI0}), we are only considering the Ward identity by inserting the holomorphic stress tensor. The Ward identity for inserting an anti-holomorphic stress tensor is similar to eq.(\ref{WI0}). We conjugate all the functions, coordinates, and parameters and replace $L[j-1]$ by $\bar L[j-1]$.
\section{Partition function}\label{3}
In this section, we apply the genus two Ward identity eq.(\ref{WI0}) to calculate the first-order deformation of partition function under  $T\bar{T}$ on $\Sigma^{(2)}$. The deformed partition function $\delta_{\lambda}Z$ up to $|\epsilon|^1$ is shown as eq.(\ref{1st 1ord PF}), which is the main result of this section. At the end of this section, we comment on some potential applications of the deformed partition function.\par
The deformed action $S^{\lambda}$ with parameter $\lambda$ satisfies the flow equation
\begin{align}
\partial_{\lambda}S^{\lambda}=&-\int_{S_1\cup A}d^2z_1 O_{T\bar T}^{\lambda}(z_1,\bar z_1)-\int_{S_2}d^2z_2 O_{T\bar T}^{\lambda}(z_2,\bar z_2). \label{flow equation}
\end{align}
On each torus, the deformed operator $O_{T\bar T}(z_i,\bar z_i)$ will be reduced to $T\bar T(z_i,\bar z_i)$. We can expand the $S^{\lambda}$ and $O_{T\bar T}^{\lambda}$ to the power of $\lambda$:
\begin{equation}
S^{\lambda}=\sum_{n=0}^{\infty}\lambda^{n}S^{(n)},\	\ \ O_{T\bar T}^{\lambda}=\sum_{n=0}^{\infty}\lambda^{n}O_{T\bar T}^{(n)}.
\end{equation}
The index $(n)$ here and below refers to the expansion coefficient of order $\lambda^n$. From the flow equation eq.(\ref{flow equation}), we can derive a recursion relation
\begin{equation}
S^{(n+1)}=-\frac{1}{n+1}\Big[\int_{S_1\cup A}d^2z_1 O_{T\bar T}^{(n)}(z_1,\bar z_1)+\int_{S_2}d^2z_2 O_{T\bar T}^{(n)}(z_2,\bar z_2)\Big].	
\end{equation}
We write down the deformed action
$S^{\lambda}$ and expand it to the first-order of $\lambda$ 
\begin{equation}
S^{\lambda}=S^{(0)}-\lambda\Big[\int_{S_1\cup A}d^2z_1 O_{T\bar T}^{(0)}(z_1,\bar z_1)+\int_{S_2}d^2z_2 O_{T\bar T}^{(0)}(z_2,\bar z_2)\Big]+O(\lambda^2).
\end{equation}
The deformed partition function $Z^{\lambda}$ can be derived using the path integral formula
\begin{align}
Z^{\lambda}=&\int D\phi\ e^{-S^{\lambda}[\phi]}\notag\\
=&\int D\phi\ \Big[1+\lambda\big(\int_{S_1\cup A}d^2z_1 O_{T\bar T}^{(0)}(z_1,\bar z_1)+\int_{S_2}d^2z_2 O_{T\bar T}^{(0)}(z_2,\bar z_2)\big)\Big]e^{-S^{(0)}[\phi]}+O(\lambda^{2})\notag\\
=&Z^{(0)}+\lambda Z^{(0)}\big(\int_{S_1\cup A}d^2z_1\langle{O_{T\bar T}^{(0)}(z_1,\bar z_1)}\rangle^{(0)}+\int_{S_2}d^2z_2\langle{O_{T\bar T}^{(0)}(z_2,\bar z_2)}\rangle^{(0)}\big)+O(\lambda^2),\label{partition}
\end{align}
Here $Z^{(0)}$ in the second term of eq.(\ref{partition}) comes from the normalization. $Z^{(0)}$ and $T^{(0)}$ are the undeformed partition function and stress tensor, respectively, and we will omit the index $(0)$ below. The radius parameters $r_1,r_2$ affect the first-order deformation of the partition function (the domain of integration depends on $r_1$ and $r_2$, see Fig.\ref{Fig 2}). 
We set up a relationship between radius parameter and $\epsilon$:
\begin{equation}
r_1=r_2=\sqrt{|\epsilon|}. \label{radius parameter}
\end{equation}
In the current work, for simplicity, we choose the above condition, which satisfies the prerequisite conditions given by eq.(\ref{condition1}). Intuitively, the condition causes the width of the annulus $A$ to be 0. Even though the condition eq.(\ref{radius parameter}) is introduced to constrain the shape of the Riemann surface $\Sigma^{(2)}$ largely, we can still read two characteristics from the sewing parameter $\epsilon$. The modulus $|\epsilon|$ determines the coupling degree of the two tori, and the argument $e^{i\varphi}=\epsilon/|\epsilon|$ determines the relative rotation between them. The undeformed theory is assumed to be a CFT, which means that the expectation value $\langle{T\bar T}\rangle$ (at zeroth-order of $\lambda$) can be calculated by CFT Ward identity. From the genus two Ward identity eq.(\ref{WI0}), $\langle{T\bar T(z,\bar z)}\rangle$ is regular on the annulus $A$, and its integral over $A$ vanishes under condition eq.(\ref{radius parameter}). Finally, the first-order deformation of the partition function can be divided into two parts
\begin{align}
\delta_{\lambda}Z=&\int_{S_1}d^2z_1Z\langle{O_{T\bar T}(z_1,\bar z_1)}\rangle+\int_{S_2}d^2z_2Z\langle{O_{T\bar T}(z_2,\bar z_2)}\rangle\notag\\
=&\int_{S_1}d^2z_1D_{z_1}\bar D_{\bar{z}_1}Z+\int_{S_2}d^2z_2D_{z_2}\bar D_{\bar{z}_2}Z. \label{deltalambdaZ}
\end{align}
The integral of $D_z\bar D_{\bar z}$ is calculated in detail in Appendix \ref{C.2}. Here we consider the weak coupling between $S_1$ and $S_2$, which means that $|\epsilon|$ is sufficiently small. We approximate $\delta_{\lambda}Z$ to $|\epsilon|^1$ and obtain:
\begin{align}
\delta_{\lambda}Z=&\sum_{a=1}^2\Big[\frac{1}{2}\big(\text{Im}[\tau_a]-\frac{|\epsilon|}{6\pi}\big)\partial_{\tau_a}\partial_{\bar{\tau}_a}Z-\frac{i}{2}\bar\epsilon\partial_{\tau_a}\partial_{\bar{\epsilon}}Z\Big]+\pi|\epsilon|\partial_\epsilon\partial_{\bar{\epsilon}}Z+c.c.+O(|\epsilon|^2), \label{1st 1ord PF}
\end{align}
where $c.c.$ refers to complex conjugate term.\par
{In particular, we extract the leading term in eq.(\ref{1st 1ord PF}). According to eq.(\ref{genus two partition function}), the leading term of the genus two partition function $Z$ can be written as the product of two torus partition functions\footnote{{One can not directly count the dimension of moduli space for a genus-two surface to make sure the limit $\epsilon\to 0$ is legal. The limit leads the surface to become special. That is to say, some pinch point or degenerate point present during the limit. This is a particular example of a boundary of moduli space. The boundary of moduli space is important since any divergences must from there. During this limit process, it is inevitable, which is given by \cite{Polchinski:1998rq}. It will be a non-trivial mathematical problem. However, $|\epsilon|$ can be kept infinitesimal value by Weyl transformation, and we can do series expansion of correlation function. Here we expand the genus two partition function in terms of $\epsilon$. We find that the leading terms are associated with genus one partition function, and subleading terms will be involved in higher powers of $\epsilon$.}} $Z_1$ and $Z_2$:}
\begin{align}
\delta_{\lambda}Z=&Z_2\text{Im}[\tau_1]\partial_{\tau_1}\partial_{\bar{\tau}_1}Z_1+Z_1\text{Im}[\tau_2]\partial_{\tau_2}\partial_{\bar{\tau}_2}Z_2+O(|\epsilon|)\notag\\
=&Z_2\delta_{\lambda}Z_1+Z_1\delta_{\lambda}Z_2+O(|\epsilon|).
\end{align}
{$Z$ on the left hand side is the genus two partition function, which corresponds to $Z^{(2)}(\tau_1,\tau_2,\epsilon)$  in eq.(\ref{genus two partition function}). $Z_1$ and $Z_2$ on the right hand side are the genus one partition functions, which correspond to $Z^{(1)}(\mathbf{1};\tau_1)$ and $Z^{(1)}(\mathbf{1};\tau_2)$ in eq.(\ref{genus two partition function}) respectively.} One can read off the deformation of the torus partition function $\delta_{\lambda}Z_a$, which coincides with the first order deformation of the torus partition function obtained by \cite{He:2020udl} up to a normalization factor. \par
To close this section, we would like to add some potential applications of the deformed genus two partition function. One can apply the first order deformation of the partition function to investigate the multiple-interval R\'enyi entropy~\cite{Rajabpour:2015nja,Rajabpour:2015uqa,Rajabpour:2015xkj,Headrick:2015gba}. There has been a resurgence of interest in higher-genus partition functions of two-dimensional deformed CFTs, partly motivated by the perturbative study of entanglement entropies. The computation of entanglement entropies via the replica trick involves evaluating R\'enyi entropy, which can be regarded as certain higher-genus partition functions with the deformation perturbatively. Fascinating research applies calculations of R\'enyi entropy to check whether the holographic Ryu-Takayanagi formula \cite{Ryu:2006bv, Hubeny:2007xt} exists or not in the $T\bar{T}$ deformed theories. This is a check of our basic understanding of $AdS_3/CFT_2$ duality with $T\bar{T}$ deformation.

\section{Correlation functions}\label{4}
In this section, we further consider the first-order deformation of correlation functions on genus two Riemann surfaces $\Sigma^{(2)}$. We calculate deformed one-point functions of the primary field $\langle{V}\rangle$ and the stress tensor $\langle{T}\rangle$ , which are shown as eq.(\ref{1st 1ord V}) and eq.(\ref{1st T}), respectively. The deformed two-point function of primary fields $\langle{V_1V_2}\rangle$ is shown as eqs.(\ref{1st V1V2})(\ref{1st VxVy}). For the deformed stress tensor two-point function, we compute two types $\langle{T_1T_2}\rangle$ and $\langle{T_1\bar T_2}\rangle$, which are shown as eqs.(\ref{1st 1ord T1T2})(\ref{1st 1ord TxTy}) and eqs.(\ref{1st 1ord T1T2bar})(\ref{1st 1ord TxTybar}) respectively. At the end of this section, we comment on some potential applications of the deformed correlation function. \par
$X$ denotes the product of a series of fields, which also flow under $T\bar T$. We expand $X^{\lambda}$ to the power of $\lambda$:
\begin{equation}
X^{\lambda}=\sum_{n=0}^{\infty}\lambda^nX^{(n)}=X+\lambda\delta_{\lambda}X+O(\lambda^2).
\end{equation}
The deformed correlation function $\langle{X^{\lambda}}\rangle^{\lambda}$ can be derived by the path integral formula
\begin{align}
\langle{X^{\lambda}}\rangle^{\lambda}=&\frac{1}{Z^{\lambda}}\int D\phi\ X^{\lambda}[\phi]e^{-S^{\lambda}[\phi]}\notag\\
=&\Big(1-\frac{\lambda\delta_{\lambda}Z}{Z}\Big)\frac{1}{Z}\int D\phi\ \Big(X+\lambda\delta_{\lambda}X\Big)\Big(1+\lambda\sum_{a=1,2}\int_{S_a}d^2z_aO_{T\bar T}^{(0)}\Big)e^{-S[\phi]}+O(\lambda^2)\notag\\
=&\langle{X}\rangle+\lambda\Big[-\frac{\delta_{\lambda}Z}{Z}\langle{X}\rangle+\langle{\delta_{\lambda}X}\rangle+\sum_{a=1,2}\int_{S_a}d^2z_a\langle{O_{T\bar T}(z_a,\bar z_a)X}\rangle\Big]+O(\lambda^2). \label{1st of CF}
\end{align}
In eq.(\ref{1st of CF}), the first-order deformation $\delta_{\lambda}\langle{X}\rangle$ can be divided into three parts. The first part $-\frac{\delta_{\lambda}Z}{Z}\langle{X}\rangle$ is contributed by the deformation of the normalization factor(i.e. the partition function), which has been calculated in eq.(\ref{1st 1ord PF}). The second part $\langle{\delta_{\lambda}X}\rangle$ depends on the flow of the field $X$ under $T\bar T$. On the plane, the flow equation of a generic field under $T\bar T$ has been well studied in \cite{Cardy:2019qao, Kruthoff:2020hsi}
\begin{align}
    \partial_{\lambda}X^{\lambda}(x)=&2\pi\epsilon^{ab}\epsilon^{ij}\int_{x}^{X}dx'_jT^{\lambda}_{ai}(x'+\varepsilon)\partial_{x^{b}}X^{\lambda}(x)\notag\\
    =&\sum_{m,n=0}^{+\infty}2\pi\lambda^{m+n}\epsilon^{ab}\epsilon^{ij}\int_{x}^{X}dx'_jT^{(m)}_{ai}(x'+\varepsilon)\partial_{x^{b}}X^{(n)}(x),\label{flow equation plane}
    \end{align}
and $\langle{\delta_{\lambda}X}\rangle$ corresponds to the expectation value of the zero-order of eq.(\ref{flow equation plane})
    \begin{align}
    \langle{\delta_{\lambda}X}\rangle=&2\pi\epsilon^{ab}\epsilon^{ij}\int_{x}^{X}dx'_j\langle{T^{(0)}_{ai}(x'+\varepsilon)\partial_{x^b}X^{(0)}(x)}\rangle.
\end{align}
However, generalizing this flow equation to the torus and higher genus Riemann surface is still unknown, which has been in our consideration recently. Since $\lambda$ is small enough, the conformal symmetry still hold approximately and we do not take the $T\bar{T}$ flow effect of $\delta_\lambda X$ here as shown in \cite{He:2019vzf,He:2019ahx,He:2020udl}.
In this paper we focus on the contribution of the third term $\sum_{a=1,2}\int_{S_a}d^2z_a\langle{O_{T\bar T}(z_a,\bar z_a)X}\rangle$. We normalize the Ward identity eq.(\ref{WI0}) by dividing both sides by $Z$:
\begin{align}
&\langle{T(z)X({\emph{u}},{\emph{x}};{\emph{v}},{\emph{y}})}\rangle\notag\\ =&D_z\langle{X({\emph{u}},{\emph{x}};{\emph{v}},{\emph{y}})}\rangle+\langle{X({\emph{u}},{\emph{x}};{\emph{v}},{\emph{y}})}\rangle D_z\log Z\notag\\
&+\sum_{l=1}^L\mathop{\sum}\limits_{j\geq0}{}^2\mathcal{P}_{1+j}(z,x_l;\tau_1,\tau_2,\epsilon)\langle{X(...;L[j-1]u_l,x_l;...)}\rangle\notag\\
&+\sum_{r=1}^R\mathop{\sum}\limits_{j\geq0}{}^2\mathcal{P}_{1+j}(z,y_r;\tau_2,\tau_1,\epsilon)\langle{X(...;L[j-1]v_r,y_r;...)}\rangle.\label{WI1}
\end{align}
In what follows, we will consider the $T\bar T$ deformed correlation function of primary fields and of stress tensors respectively. We will calculate the first-order deformation of one-point and two-point functions, and generalize to higher-point functions.
\subsection{One-point function of primary field}\label{4.1}
The vertex operator of a primary state $u$ satisfies
\begin{align}
V[L[-1]u,x]=\partial_x V[u,x],\ V[L[0]u,x]=wt[u]V[u,x],\ V[L[j>0]u,x]=0.
\end{align}
Let $X$ denote the product of primary fields, and Ward identity eq.(\ref{WI1}) can be simplified to
\begin{align}
\langle{T(z)X({\emph{u}},{\emph{x}};{\emph{v}},{\emph{y}})}\rangle =&D_z\langle{X({\emph{u}},{\emph{x}};{\emph{v}},{\emph{y}})}\rangle+\langle{X({\emph{u}},{\emph{x}};{\emph{v}},{\emph{y}})}\rangle D_z\log Z\notag\\
&+\Big(\sum_{l=1}^L\mathcal{P}_{z,x_l}+\sum_{r=1}^R\mathcal{P}_{z,y_r}\Big)\langle{X({\emph{u}},{\emph{x}};{\emph{v}},{\emph{y}})}\rangle, \label{WI1.5}
\end{align}
where the operator $\mathcal{P}_{z,x}$ is defined as
\begin{equation}
    \mathcal{P}_{z,x}={}^2\mathcal{P}_1(z,x;\tau_a,\tau_{\bar a},\epsilon)\partial_x+{}^2\mathcal{P}_2(z,x;\tau_a,\tau_{\bar a},\epsilon)wt[u], \label{Pzx}
\end{equation}
for $x\in S_a$, and $wt[u]$ is the conformal weight of $u$.
Following the prescription in \cite{Dijkgraaf:1996iy}, we remove a small open disk $D_{\delta}=\lbrace{z,|z-x|<\delta}\rbrace$ centered on the singularity $z=x$, which introduces an additional boundary $\partial D_{\delta}$ to the domain of integration demonstrated by Fig.\ref{Fig 3}. In the rest of $\Sigma^{(2)}$, the commutativity\footnote{The commutativity of $\mathcal{P}_{z,x}$ and $\bar{\mathcal{P}}_{\bar z,\bar x}$ is broken at singularities, due to
 \begin{align}
 \partial_{x}\bar P_1(\bar z,\bar x)=&\partial_{x}\frac{1}{\bar z-\bar x}=-\pi\delta^{(2)}(z-x),\notag\\
 \partial_{x}\bar P_2(\bar z,\bar x)=&\partial_{x}\Big(\frac{1}{\bar z-\bar x}\Big)^2=-\pi\partial_{\bar x}\delta^{(2)}(z-x).
 \end{align}} of $\mathcal{P}_{z,x}$ and $\bar{\mathcal{P}}_{\bar z,\bar x}$ is preserved. After integration, we regularize it by taking the limit $\delta\rightarrow0$ and simply discarding the divergence. \par
Let us begin with the first-order deformed one-point function of primary field. We insert a pair of primary states $(u,\bar u)$ at $(x,\bar x)\in S_a$ and apply Ward identity eq.(\ref{WI1.5}) to obtain
\begin{align}
&\int_{S_a\backslash{D_{\delta}}}d^2z_a\langle{O_{T\bar T}(z_a,\bar z_a)V}\rangle+\int_{S_{\bar a}}d^2z_{\bar a}\langle{O_{T\bar T}(z_{\bar a},\bar z_{\bar a})V}\rangle\notag\\
=&\frac{1}{Z}\sum_{b=a,\bar a}\Big[\int_{S_b\backslash{D_{\delta}}}d^2z_b\Big(D_{z_b}\bar{D}_{\bar{z}_b}+D_{z_b}\bar{\mathcal{P}}_{\bar z_b,\bar x}+\bar{D}_{\bar{z}_b}\mathcal{P}_{z_b,x}+\mathcal{P}_{z_b,x}\bar{\mathcal{P}}_{\bar z_b,\bar x}\Big)\Big]\big(Z\langle{V}\rangle\big). \label{1pt correction}
\end{align}
The integrals of $D_z\bar{\mathcal{P}}_{\bar z,\bar x}$, $\bar{D}_{\bar{z}}\mathcal{P}_{z,x}$ and $\mathcal{P}_{z,x}\bar{\mathcal{P}}_{\bar z,\bar x}$ is calculated in detail in Appendix \ref{C.3}. Using eqs.(\ref{1pt correction})(\ref{1st PF})(\ref{1st 1p DP})(\ref{1p PP}) we obtain the first-order correction $\delta_{\lambda}\langle{V}\rangle-\langle{\delta_{\lambda}V}\rangle$ up to $|\epsilon|^1$:
\begin{align}
&\delta_{\lambda}\langle{V(x)}\rangle-\langle{\delta_{\lambda}V(x)}\rangle\notag\\
=&\pi wt[u]wt[\bar u]\langle{V}\rangle\frac{1}{\delta}-\frac{\langle{V}\rangle}{Z}\bigg\lbrace\sum_{b=a,\bar a}\Big[\frac{1}{2}\big(\text{Im}[\tau_b]-\frac{|\epsilon|}{6\pi}\big)\partial_{\tau_b}\partial_{\bar{\tau}_b}Z-\frac{i}{2}\bar\epsilon\partial_{\tau_b}\partial_{\bar{\epsilon}}Z\Big]+\pi|\epsilon|\partial_\epsilon\partial_{\bar{\epsilon}}Z\bigg\rbrace\notag\\
&+\frac{1}{Z}\bigg\lbrace\sum_{b=a,\bar a}\Big[\frac{1}{2}\big(\text{Im}[\tau_b]-\frac{|\epsilon|}{6\pi}\big)\partial_{\tau_b}\partial_{\bar{\tau}_b}-\frac{i}{2}\bar\epsilon\partial_{\tau_b}\partial_{\bar{\epsilon}}]\Big]+\pi|\epsilon|\partial_\epsilon\partial_{\bar{\epsilon}}\notag\\
&+i\big(\text{Re}[x]-\frac{|\epsilon|}{3}\bar P_1(\bar x,\bar \tau_a)\big)\partial_{\tau_a}\partial_{\bar x}+i\big(\frac{1}{2}+\frac{|\epsilon|}{3}\bar P_2(\bar x,\bar \tau_a)\big)wt[\bar u]\partial_{\tau_a}+\pi\epsilon P_1(x,\tau_a)\partial_{\epsilon}\partial_{\bar x}\notag\\
&+\pi\big(\log|Q(x,\tau_a)|^2+\frac{1}{4}-\frac{|\epsilon|}{3}|P_1(x,\tau_a)|\big)\partial_x\partial_{\bar x}-\frac{\pi|\epsilon|}{3}|P_{2}(x,\tau_a)|^2wt[u]wt[\bar u]\notag\\
&+\pi\big(\bar P_{1}(\bar x,\bar\tau_a)+\frac{2|\epsilon|}{3}P_1(x,\tau_a)\bar P_2(\bar x,\bar\tau_a)\big)\partial_x wt[\bar u]\bigg\rbrace\big(Z\langle{V}\rangle\big)+c.c.+O(|\epsilon|^2), \label{1st 1ord V}
\end{align}
where $\log Q(x,\tau_a)$ is introduced by eq.(\ref{Q}) as a primitive function of $P_1(x,\tau_a)$. The first term is proportional to $\delta^{-1}$, which is the purely divergent term. In the torus case, the $T\bar T$ deformed one-point function has a similar divergence. We adopt the same regularization as in \cite{He:2020cxp}. The purely divergent term (proportional to $\delta^{-n}$ for $n\geq 1$ in our calculations) is discarded.
\subsection{Higher-point function of primary field}\label{4.2}
We consider the $T\bar T$ deformed correlation function with two pairs of primary states $(u_1,\bar u_1)$ and $(u_2,\bar u_2)$ inserted at $(x_1,\bar x_1)$ and $(x_2,\bar x_2)$, respectively. 
According to eq.(\ref{WI1.5}) the first-order correction depends on
\begin{align}
&\int_{S_a\backslash{D_{\delta}}}d^2z_a\langle{O_{T\bar T}(z_a,\bar z_a)V_1V_2}\rangle+\int_{S_{\bar a}\backslash{D_{\delta}}}d^2z_{\bar a}\langle{O_{T\bar T}(z_{\bar a},\bar z_{\bar a})V_1V_2}\rangle\notag\\
=&\frac{1}{Z}\sum_{b=a,\bar a}\int_{S_b\backslash{D_{\delta}}}d^2z_b\Big[D_{z_b}\bar{D}_{\bar z_b}+D_{z_b}\bar{\mathcal{P}}_{\bar z_b,\bar{x}_1}+D_{z_b}\bar{\mathcal{P}}_{\bar z_b,\bar{x}_2}+\bar{D}_{\bar z_b}\mathcal{P}_{z_b,x_1}+\bar{D}_{\bar z_b}\mathcal{P}_{z_b,x_2}\notag\\
&+\mathcal{P}_{z_b,x_1}\bar{\mathcal{P}}_{\bar z_b,\bar{x}_1}+\mathcal{P}_{z_b,x_2}\bar{\mathcal{P}}_{\bar z_b,\bar{x}_2}+\mathcal{P}_{z_b,x_1}\bar{\mathcal{P}}_{\bar z_b,\bar{x}_2}+\mathcal{P}_{z_b,x_2}\bar{\mathcal{P}}_{\bar z_b,\bar{x}_1}\Big]\big(Z\langle{V_1V_2}\rangle\big). \label{2p WI}
\end{align}
Let us consider two different profiles. The one profile is that two points are inserted in the same torus $x_1,x_2\in S_a$, one can obtain the first-order correction $\delta_{\lambda}\langle{V_1V_2}\rangle-\langle{\delta_{\lambda}(V_1V_2)}\rangle$ up to $|\epsilon|^1$ from eqs.(\ref{2p WI})(\ref{1st PF})(\ref{1st 1p DP})(\ref{1p PP})(\ref{2p PP 1}):
\begin{align}
&\delta_{\lambda}\langle{V_1V_2}\rangle-\langle{\delta_{\lambda}(V_1V_2)}\rangle\notag\\
=&-\frac{\langle{V_1V_2}\rangle}{Z}\bigg\lbrace\sum_{b=a,\bar a}\Big[\frac{1}{2}\big(\text{Im}[\tau_b]-\frac{|\epsilon|}{6\pi}\big)\partial_{\tau_b}\partial_{\bar{\tau}_b}Z-\frac{i}{2}\bar\epsilon\partial_{\tau_b}\partial_{\bar{\epsilon}}Z\Big]+\pi|\epsilon|\partial_\epsilon\partial_{\bar{\epsilon}}Z\bigg\rbrace\notag\\
&+\frac{1}{Z}\bigg\lbrace\sum_{b=a,\bar a}\Big[\frac{1}{2}\big(\text{Im}[\tau_b]-\frac{|\epsilon|}{6\pi}\big)\partial_{\tau_b}\partial_{\bar{\tau}_b}-\frac{i}{2}\bar\epsilon\partial_{\tau_b}\partial_{\bar{\epsilon}}\Big]+\pi|\epsilon|\partial_\epsilon\partial_{\bar{\epsilon}}\notag\\
&+\sum_{i=1}^2\Big[i\big(\text{Re}[x_i]-\frac{|\epsilon|}{3}\bar P_1(\bar x_i,\bar \tau_a)\big)\partial_{\tau_a}\partial_{\bar x_i}+i\big(\frac{1}{2}+\frac{|\epsilon|}{3}\bar P_2(\bar x_i,\bar \tau_a)\big)wt[\bar u_i]\partial_{\tau_a}\notag\\
&+\pi\epsilon P_1(x_i,\tau_a)\partial_{\epsilon}\partial_{\bar x_i}+\pi\big(\log|Q(x_i,\tau_a)|^2+\frac{1}{4}-\frac{|\epsilon|}{3}|P_1(x_i,\tau_a)|\big)\partial_{x_i}\partial_{\bar x_i}\notag\\
&-\frac{\pi|\epsilon|}{3}|P_{2}(x_i,\tau_a)|^2wt[u_i]wt[\bar u_i]+\pi\big(\bar P_{1}(\bar x_i,\bar\tau_a)+\frac{2|\epsilon|}{3}P_1(x_i,\tau_a)\bar P_2(\bar x_i,\bar\tau_a)\big)\partial_{x_i}wt[\bar u_i]\Big]\notag\\
&+\pi\big(\log\frac{|Q(x_1,\tau_a)Q(x_2,\tau_a)|^2}{|Q(x_1-x_2,\tau_a)|^2}+\frac{1}{2}-\frac{2|\epsilon|}{3}P_1(x_1,\tau_a)\bar P_1(\bar x_2,\bar\tau_a)\big)\partial_{x_1}\partial_{\bar x_2}\notag\\
&+\pi\big(\bar P_1(\bar x_1-\bar x_2,\bar\tau_a)+\bar P_1(\bar x_2,\bar\tau_a)+\frac{2|\epsilon|}{3}P_1(x_1,\tau_a)\bar P_2(\bar x_2,\bar\tau_a)\big)\partial_{x_1} wt[\bar u_2]\notag\\
&+\pi\big(P_1(x_2-x_1,\tau_a)+P_1(x_1,\tau_a)+\frac{2|\epsilon|}{3}P_2(x_1,\tau_a)\bar P_1(\bar x_2,\bar\tau_a)\big)wt[u_1]\partial_{\bar x_2}\notag\\
&-\frac{2\pi|\epsilon|}{3}P_2(x_1,\tau_a)\bar P_2(\bar x_2,\bar\tau_a)wt[u_1]wt[\bar u_2])\bigg\rbrace\big(Z\langle{V_1V_2}\rangle\big)+c.c.+O(|\epsilon|^2). \label{1st V1V2}
\end{align}
The other profile is that two points are inserted in the different tori. We set $x_1=x\in S_{a_1}$ and $x_2=y\in S_{a_2}$, and insert primary states $u_1=u$ and $u_2=v$ at $x$ and $y$ respectively. The first-order correction $\delta_{\lambda}\langle{V_xV_y}\rangle-\langle{\delta_{\lambda}(V_xV_y)}\rangle$ is given by eqs.(\ref{1st PF})(\ref{1st 1p DP})(\ref{1p PP})(\ref{2p PP 2}):
\begin{align}
    &\delta_{\lambda}\langle{V_xV_y}\rangle-\langle{\delta_{\lambda}(V_xV_y)}\rangle\notag\\
=&-\frac{\langle{V_xV_y}\rangle}{Z}\bigg\lbrace\sum_{a=1}^2\Big[\frac{1}{2}\big(\text{Im}[\tau_a]-\frac{|\epsilon|}{6\pi}\big)\partial_{\tau_a}\partial_{\bar{\tau}_a}Z-\frac{i}{2}\bar\epsilon\partial_{\tau_a}\partial_{\bar{\epsilon}}Z\Big]+\pi|\epsilon|\partial_\epsilon\partial_{\bar{\epsilon}}Z\bigg\rbrace\notag\\
&+\frac{1}{Z}\bigg\lbrace\sum_{a=1}^2\Big[\frac{1}{2}\big(\text{Im}[\tau_a]-\frac{|\epsilon|}{6\pi}\big)\partial_{\tau_a}\partial_{\bar{\tau}_a}-\frac{i}{2}\bar\epsilon\partial_{\tau_a}\partial_{\bar{\epsilon}}\Big]+\pi|\epsilon|\partial_\epsilon\partial_{\bar{\epsilon}}\notag\\
&+\sum_{i=1}^2\Big[i\big(\text{Re}[x_i]-\frac{|\epsilon|}{3}\bar P_1(\bar x_i,\bar \tau_{a_i})\big)\partial_{\tau_{a_i}}\partial_{\bar x_i}+i\big(\frac{1}{2}+\frac{|\epsilon|}{3}\bar P_2(\bar x_i,\bar\tau_{a_i})\big)wt[\bar u_i]\partial_{\tau_{a_i}}\notag\\
&+\pi\epsilon P_1(x_i,\tau_{a_i})\partial_{\epsilon}\partial_{\bar x_i}+\pi\big(\log|Q(x_i,\tau_{a_i})|^2+\frac{1}{4}-\frac{|\epsilon|}{3}|P_1(x_i,\tau_{a_i})|\big)\partial_{x_i}\partial_{\bar x_i}\notag\\
&-\frac{\pi|\epsilon|}{3}|P_{2}(x_i,\tau_{a_i})|^2wt[u_i]wt[\bar u_i]+\pi\big(\bar P_{1}(\bar x_i,\bar\tau_{a_i})+\frac{2|\epsilon|}{3}P_1(x_i,\tau_{a_i})\bar P_2(\bar x_i,\bar\tau_{a_i})\big)\partial_{x_i}wt[\bar u_i]\Big]\notag\\
&-\frac{\pi}{2}\big(\bar\epsilon\bar P'_{2}(\bar x,\bar\tau_{a_1})+\epsilon P'_{2}(y,\tau_{a_2})\big)\partial_{x}\partial_{\bar y}\bigg\rbrace\big(Z\langle{V_xV_y}\rangle\big)+c.c.+O(|\epsilon|^2). \label{1st VxVy}
\end{align}
\begin{figure}[H]
\centering
\includegraphics[scale=0.5]{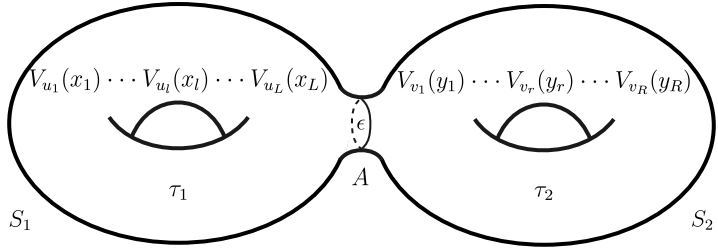}
\caption{A primary $(L+R)$-point function on sewn Riemann surface with genus-2.}
\label{Fig primary}
\end{figure}\par
We generalize our result to a $(L+R)$-point function of primary fields. As shown in Fig.{\ref{Fig primary}}, we insert $L$ pairs of primary states $(u_1,\bar u_1),...,(u_L,\bar u_L)$ and $R$ pairs of primary states $(v_1,\bar v_1),...,(v_R,\bar v_R)$ at $(x_1,\bar x_1),...,(x_L,\bar x_L)\in S_1$ and $(y_1,\bar y_1),...,(y_R,\bar y_R)\in S_2$, respectively. For a primary $(L+R)$-point function\footnote{For simplicity, the vertex operator corresponding to each pair of primary states $(u_l,\bar u_l)$ at $(x_l,\bar x_l)$ is denoted as $V_{u_l}$, and $(v_r,\bar v_r)$ at $(y_r,\bar y_r)$ is denoted as $V_{v_r}$} $\langle{X_LX_R}\rangle=\langle{\prod_{l=1}^LV_{u_l}\prod_{r=1}^RV_{v_r}}\rangle$, the first-order correction is derived by Ward identity eq.(\ref{WI1.5}) as follows:
\begin{align}
&\int_{S_1\backslash{D_{\delta}}}d^2z_1\langle{O_{T\bar T}(z_1,\bar z_1)X_LX_R}\rangle+\int_{S_2\backslash{D_{\delta}}}d^2z_2\langle{O_{T\bar T}(z_2,\bar z_2)X_LX_R}\rangle\notag\\
=&\frac{1}{Z}\sum_{a=1,2}\int_{S_a\backslash{D_{\delta}}}d^2z_a\bigg\lbrace D_{z_a}\bar{D}_{\bar z_a}+\sum_{l=1}^{L}\Big[D_{z_a}\bar{\mathcal{P}}_{\bar{z}_a,\bar{x}_l}+\bar{D}_{\bar{z}_a}\mathcal{P}_{z_a,x_l}+\mathcal{P}_{z_a,x_l}\bar{\mathcal{P}}_{\bar{z}_a,\bar{x}_l}\Big]\notag\\
&+\sum_{r=1}^{R}\Big[D_{z_a}\bar{\mathcal{P}}_{\bar{z}_a,\bar{y}_r}+\bar{D}_{\bar{z}_a}\mathcal{P}_{z_a,y_r}+\mathcal{P}_{z_a,y_r}\bar{\mathcal{P}}_{\bar{z}_a,\bar{y}_r}\Big]+\Big[\sum_{l'\neq l''}\mathcal{P}_{z_a,x_{l'}}\bar{\mathcal{P}}_{\bar{z}_a,\bar{x}_{l''}}+\sum_{r'\neq r''}\mathcal{P}_{z_a,y_{r'}}\bar{\mathcal{P}}_{\bar{z}_a,\bar{y}_{r''}}\Big]\notag\\
&+\sum_{l,r}\Big[\bar{\mathcal{P}}_{\bar{z}_a,\bar{x}_l}\mathcal{P}_{z_a,y_{r}}+\mathcal{P}_{z_a,x_{l}}\bar{\mathcal{P}}_{\bar{z}_a,\bar{y}_r}\Big]\bigg\rbrace\Big[Z\langle{X_LX_R}\rangle\Big]. \label{higher-point primary}
\end{align}
The first term in eq.(\ref{higher-point primary}) is analogous to eq.(\ref{1st PF}) (we call it the zero-point contribution). The second and third terms in eq.(\ref{higher-point primary}) are the sum of all one-point contribution eqs.(\ref{1p secondterm})(\ref{1st fourthterm}). The fourth and fifth terms in eq.(\ref{higher-point primary}) contain all the two-point contribution eqs.(\ref{2p PP 1})(\ref{2p PP 2}). No matter how many points are inserted, the first-order correction of correlation function contains at most two-point contribution. The 
contribution of three or more points can be found in higher-order correction. 
\subsection{One-point function of stress tensor} \label{4.3}
In the next two subsections, we consider the $T\bar T$ deformed correlation functions of stress tensor based on first-order perturbation theory. The holomorphic stress tensor $T(z)$ is the vertex operator of conformal vector $\tilde\omega$, and it is a quasi-primary field with conformal weight $wt[\tilde\omega]=2$, which satisfies
\begin{align}
V[L[-1]\tilde\omega,z]=\partial_zV[\tilde\omega,z],\ V[L[0]\tilde\omega,z]=2V[\tilde\omega,z],\ V[L[j\geq 1]\tilde\omega,z]=\frac{c}{2}\mathbbm{1}\delta_{j,2}
\end{align}
where $\mathbbm{1}$ is the identity operator and $c$ is the central charge.  The one-point function of stress tensor can be obtained in undeformed CFT by Ward identity eq.(\ref{WI1}) and operator $D_x$ in eq.(\ref{Dz}):
\begin{equation}
\langle{T(x)}\rangle=\frac{1}{Z}D_xZ=\frac{1}{2\pi iZ}\sideset{^2}{_a}{\mathop{\mathcal{F}}}(x)\partial_{\tau_a}Z+\frac{1}{2\pi iZ}\sideset{^2}{_{\bar a}}{\mathop{\mathcal{F}}}(x)\partial_{\tau_{\bar a}}Z+\frac{1}{Z}{}^2\mathcal{F}^{\Pi}(x)\epsilon^{\frac{1}{2}}\partial_{\epsilon}Z, \label{T(x)}
\end{equation}
where $x\in S_a$. According to the definition of $\mathcal{F}$ functions eqs.(\ref{Fa})(\ref{Fpi}), the expectation value $\langle{T(x)}\rangle$ is biperiodic on each torus. The first-order $T\bar{T}$ deformation $\delta_{\lambda}\langle{T(x)}\rangle$ is obtained by eq.(\ref{1st of CF}):
\begin{align}
\delta_{\lambda}\langle{T(x)}\rangle=-\frac{1}{Z^2}\delta_{\lambda}ZD_xZ+\langle{\delta_{\lambda}T(x)}\rangle+\sum_{a=1,2}\int_{S_a\backslash D_{\delta}}d^2z_a\langle{O_{T\bar T}(z_a,\bar z_a)T(x)}\rangle. \label{1st of T}
\end{align}
The first term is the flow effect of the normalization factor, which has been calculated in eq.(\ref{1st 1ord PF}). The second term depends on specific Lagrangian. The third term $\int_{S_a\backslash D_{\delta}}d^2z_a\langle{O_{T\bar T}(z_a,\bar z_a)T(x)}\rangle$ can be calculated by eq.(\ref{WI1}):
\begin{align}
&\int_{S_1\backslash D_{\delta}}d^2z_1\langle{O_{T\bar T}(z_1,\bar z_1)T(x)}\rangle+\int_{S_2\backslash D_{\delta}}d^2z_2\langle{O_{T\bar T}(z_2,\bar z_2)T(x)}\rangle\notag\\
=&\frac{1}{Z}\sum_{a=1,2}\int_{S_a\backslash{D_{\delta}}}d^2z_a\bigg\lbrace D_{z_a}\bar D_{\bar z_a}\big(D_xZ\big)+\bar D_{\bar z_a}\mathcal{P}_{z_a,x}\big(D_xZ\big)+\frac{c}{2}{}^2{\mathcal{P}}_4(z_a,x)\bar D_{\bar z_a}Z\bigg\rbrace, \label{1st int T}
\end{align}
which is discussed in detail in Appendix \ref{C.2}. We approximate the first-order correction of $\langle{T(x)}\rangle$ to $|\epsilon|^1$ using eqs.(\ref{1st PF})(\ref{1st 1p DP})(\ref{1st T PD}):
\begin{align}
&\delta_{\lambda}\langle{T(x)}\rangle-\langle{\delta_{\lambda}T(x)}\rangle\notag\\
=&-\frac{1}{2\pi}\big(1+\frac{2|\epsilon|}{3}P_2(x,\tau_a)\big)\frac{\partial_{\tau_a}\partial_{\bar\tau_a}Z}{Z}-\frac{ic}{6}|\epsilon|P_4(x,\tau_a)\frac{\partial_{\bar\tau_a}Z}{Z}\notag\\
&-\frac{i}{2\pi}\sum_{b=a,\bar a}\Big[\Big(\big(\text{Im}[\tau_b]-\frac{|\epsilon|}{6\pi}\big)\partial_{\tau_a}+\epsilon \text{Im}[\tau_b]P_2(x,\tau_a)\partial_{\epsilon}\Big)\big(\frac{\partial_{\tau_b}\partial_{\bar\tau_b}Z}{Z}\big)\Big]\notag\\
&-\frac{1}{4\pi}\sum_{b=a,\bar a}\partial_{\tau_a}\Big[\bar\epsilon\frac{\partial_{\tau_b}\partial_{\bar\epsilon}Z}{Z}-\epsilon\frac{\partial_{\bar\tau_b}\partial_{\epsilon}Z}{Z}\Big]+i\epsilon\Big[\big(\text{Re}[x]\partial_x-i\text{Im}[\tau_a]\partial_{\tau_a}+2\big)P_2(x,\tau_a)\Big]\frac{\partial_{\bar\tau_a}{\partial_{\epsilon}Z}}{Z}\notag\\
&-i|\epsilon|\big(\partial_{\tau_a}+2\pi iP_2(x,\tau_a)\big)\frac{\partial_{\epsilon}\partial_{\bar\epsilon}Z}{Z}+O(|\epsilon|^2). \label{1st T}
\end{align}\par
{The leading term of the genus two partition function $Z$ is $Z_aZ_{\bar a}$, where $Z_a$ denotes torus partition function on $T^2_a$. According to eq.(\ref{T(x)}), the leading term of the one-point function of stress tensor is  $\frac{1}{2\pi iZ_a}\partial_{\tau_a}Z_a=\langle{T(x)}\rangle_{T^2_a}$, where $\langle{T(x)}\rangle_{T^2_a}$ is completely defined on torus $T^2_a$. One can read off the first-order deformation of $\langle{T(x)}\rangle_{T^2_a}$ from the leading order of eq.(\ref{1st T}):}
\begin{align}
&\delta_{\lambda}\langle{T(x)}\rangle-\langle{\delta_{\lambda}T(x)}\rangle\notag\\
=&-\frac{1}{2\pi}\frac{\partial_{\tau_a}\partial_{\bar\tau_a}{Z_aZ_{\bar a}}}{Z_aZ_{\bar a}}-\frac{i}{2\pi}\text{Im}[\tau_a]\partial_{\tau_a}\Big[\frac{\partial_{\tau_a}\partial_{\bar\tau_a}Z_aZ_{\bar a}}{Z_aZ_{\bar a}}\Big]-\frac{i}{2\pi}\text{Im}[\tau_{\bar a}]\partial_{\tau_a}\Big[\frac{\partial_{\tau_{\bar a}}\partial_{\bar\tau_{\bar a}}Z_aZ_{\bar a}}{Z_aZ_{\bar a}}\Big]+O(|\epsilon|)\notag\\
=&-\frac{1}{2\pi}\frac{\partial_{\tau_a}\partial_{\bar\tau_a}{Z_a}}{Z_a}-\frac{i}{2\pi}\text{Im}[\tau_a]\frac{\partial^2_{\tau_a}\partial_{\bar\tau_a}Z_a}{Z_a}+\frac{i}{2\pi}\text{Im}[\tau_a]\frac{\partial_{\tau_a}\partial_{\bar\tau_a}Z_a}{Z_a}\frac{\partial_{\tau_a}Z_a}{Z_a}+O(|\epsilon|)\notag\\
=&\delta_{\lambda}\langle{T(x)}\rangle_{T^2_a}-\langle{\delta_{\lambda}T(x)}\rangle_{T^2_a}+O(|\epsilon|),
\end{align}
This result coincides with the first order deformation of the expectation value of stress tensor on torus obtained by \cite{He:2020udl} up to a normalization factor.
\subsection{Higher-point function of stress tensor} \label{4.4}
We consider two types of the two-point function of stress tensor, $\langle{T(x_1)T(x_2)}\rangle$ and $\langle{T(x_1)\bar T(\bar x_2)}\rangle$ respectively. In undeformed CFT, these two expectation values can be obtained by eq.(\ref{WI1})
\begin{align}
    \langle{T(x_1)T(x_2)}\rangle=&\frac{1}{Z}\Big[D_{x_1}+{}^2{\mathcal{P}}_1(x_1,x_2)\partial_{x_2}+{}^2{\mathcal{P}}_2(x_1,x_2)2\Big]\big(D_{x_2}Z\big)+{}^2{\mathcal{P}}_4(x_1,x_2)\frac{c}{2},\notag\\
    \langle{T(x_1)\bar T(\bar x_2)}\rangle=&\frac{1}{Z}D_{x_1}\bar D_{\bar x_2}Z.\label{TT}
\end{align}
As in the previous sections, we concentrate on using Ward identity eq.(\ref{WI1}) to compute integrals $\int_{S_a\backslash D_\delta}d^2z_a\langle{O_{T\bar T}(z_a,\bar z_a)T_1T_2}\rangle$ and $\int_{S_a\backslash D_\delta}d^2z_a\langle{O_{T\bar T}(z_a,\bar z_a)T_1\bar T_2}\rangle$, which contribute to first-order deformation. For the first type $\langle{T_1T_2}\rangle$ we have
\begin{align}
&\int_{S_1\backslash D_{\delta}}d^2z_1\langle{O_{T\bar T}(z_1,\bar z_1)T_1T_2}\rangle+\int_{S_2\backslash D_{\delta}}d^2z_2\langle{O_{T\bar T}(z_2,\bar z_2)T_1T_2}\rangle\notag\\
=&\frac{1}{Z}\sum_{a=1,2}\int_{S_a\backslash{D_{\delta}}}d^2z_a\bigg\lbrace\Big[D_{z_a}\bar D_{\bar z_a}+\bar D_{\bar z_a}\mathcal{P}_{z_a,x_1}+\bar D_{\bar z_a}\mathcal{P}_{z_a,x_2}\Big]\big(Z\langle{T_1T_2}\rangle\big)\notag\\
&+\frac{c}{2}{}^2{\mathcal{P}}_4(z_a,x_1)\bar D_{\bar z_a}\big(Z\langle{T_2}\rangle\big)+\frac{c}{2}{}^2{\mathcal{P}}_4(z_a,x_2)\bar D_{\bar z_a}\big(Z\langle{T_1}\rangle\big)\bigg\rbrace. \label{1st T1T2}
\end{align}
The one profile is that two insertion points live on the same torus $x_1,x_2\in S_a$. the first-order correction is obtained by eqs.(\ref{1st T1T2})(\ref{1st PF})(\ref{1st 1p DP})(\ref{1st T PD}), up to $|\epsilon|^1$:
\begin{align}
&\delta_\lambda\langle{T_1T_2}\rangle-\langle{\delta_\lambda(T_1T_2)}\rangle\notag\\
=&-\frac{\langle{T_1T_2}\rangle}{Z}\bigg\lbrace\sum_{b=a,\bar a}\Big[\frac{1}{2}\big(\text{Im}[\tau_b]-\frac{|\epsilon|}{6\pi}\big)\partial_{\tau_b}\partial_{\bar{\tau}_b}Z-\frac{i}{2}\bar\epsilon\partial_{\tau_b}\partial_{\bar{\epsilon}}Z\Big]+\pi|\epsilon|\partial_\epsilon\partial_{\bar{\epsilon}}Z\bigg\rbrace\notag\\
&+\frac{1}{Z}\bigg\lbrace\Big[\sum_{b=a,\bar a}\big[\big(\text{Im}[\tau_b]-\frac{|\epsilon|}{6\pi}\big)\partial_{\tau_b}\partial_{\bar\tau_b}+\frac{i}{2}\bar\epsilon\partial_{\tau_b}\partial_{\bar\epsilon}-\frac{i}{2}\epsilon\partial_{\bar\tau_b}\partial_{\epsilon}\big]+2\pi|\epsilon|\partial_{\epsilon}\partial_{\bar\epsilon}-2i\partial_{\bar\tau_a}\notag\\
&-\frac{2i|\epsilon|}{3}\sum_{i=1,2}P_2(x_i,\tau_a)\partial_{\bar\tau_a}\Big]\big(-\frac{1}{4\pi^2}\partial^2_{\tau_a}Z+\frac{1}{\pi i}P_2(x_1-x_2,\tau_a)\partial_{\tau_a}Z+\frac{c}{2}P_4(x_1-x_2,\tau_a)Z\big)\notag\\
&+\Big[\big(2\text{Re}[x_1-x_2]-\frac{2|\epsilon|}{3}[P_1(x_1,\tau_a)-P_1(x_2,\tau_a)]\big)\partial_{\bar\tau_a}-2\pi i\bar\epsilon[\bar P_1(\bar x_1,\bar\tau_a)-\bar P_1(\bar x_2,\bar\tau_a)]\partial_{\bar\epsilon}\Big]\notag\\
&\times\big(\frac{1}{\pi}P_3(x_1-x_2,\tau_a)\partial_{\tau_a}Z+icP_5(x_1-x_2,\tau_a)Z\big)\notag\\
&+\epsilon\partial_{\epsilon}\big(\sum_{b=a,\bar a}\text{Im}[\tau_b]\partial_{\tau_b}\partial_{\bar\tau_b}-i\sum_{i=1,2}\text{Re}[x_i]\partial_{x_i}\partial_{\bar\tau_a}-2i\partial_{\bar\tau_a}\big)\Big[\frac{1}{2\pi i}\sum_{i=1,2}P_2(x_i,\tau_a)\partial_{\tau_a}Z\notag\\
&+\big(\frac{1}{2\pi i}\partial_{\tau_a}+[P_1(x_1-x_2,\tau_a)-P_1(x_1,\tau_a)]\partial_{x_2}+2P_2(x_1-x_2,\tau_a)\big)P_2(x_2,\tau_a)Z\Big]\notag\\
&-\frac{c|\epsilon|}{12\pi}\sum_{i=1,2}P_4(x_i,\tau_a)\partial_{\tau_a}\partial_{\bar\tau_a}Z\bigg\rbrace+O(|\epsilon|^2). \label{1st 1ord T1T2}
\end{align}
The other profile is that two points are inserted in different tori $x\in S_{a}$ and $y\in S_{\bar a}$. We combine eqs.(\ref{TT})(\ref{1st PF})(\ref{1st 1p DP})(\ref{1st T PD}) and have
\begin{align}
&\delta_\lambda\langle{T_xT_y}\rangle-\langle{\delta_\lambda(T_xT_y)}\rangle\notag\\
=&-\frac{\langle{T_xT_y}\rangle}{Z}\bigg\lbrace\sum_{b=a,\bar a}\Big[\frac{1}{2}\big(\text{Im}[\tau_b]-\frac{|\epsilon|}{6\pi}\big)\partial_{\tau_b}\partial_{\bar{\tau}_b}Z-\frac{i}{2}\bar\epsilon\partial_{\tau_b}\partial_{\bar{\epsilon}}Z\Big]+\pi|\epsilon|\partial_\epsilon\partial_{\bar{\epsilon}}Z\bigg\rbrace\notag\\
&+\frac{1}{Z}\bigg\lbrace\Big[\sum_{b=a,\bar a}\big[\big(\text{Im}[\tau_b]-\frac{|\epsilon|}{6\pi}\big)\partial_{\tau_b}\partial_{\bar\tau_b}+\frac{i}{2}\bar\epsilon\partial_{\tau_b}\partial_{\bar\epsilon}-\frac{i}{2}\epsilon\partial_{\bar\tau_b}\partial_{\epsilon}\big]+2\pi|\epsilon|\partial_{\epsilon}\partial_{\bar\epsilon}-i\partial_{\bar\tau_a}-i\partial_{\bar\tau_{\bar a}}\notag\\
&-\frac{i|\epsilon|}{3}P_2(x,\tau_a)\partial_{\bar\tau_a}-\frac{i|\epsilon|}{3}P_2(y,\tau_{\bar a})\partial_{\bar\tau_{\bar a}}\Big]\big(-\frac{1}{4\pi^2}\partial_{\tau_a}\partial_{\tau_{\bar a}}Z\big)\notag\\
&+\epsilon\partial_{\epsilon}\big(\sum_{b=a,\bar a}\text{Im}[\tau_b]\partial_{\tau_b}\partial_{\bar\tau_b}-i\partial_{\bar\tau_a}-i\partial_{\bar\tau_{\bar a}}\big)\Big[\frac{1}{2\pi i}\big(P_2(x,\tau_a)\partial_{\tau_{\bar a}}+P_2(y,\tau_{\bar a})\partial_{\tau_a}\big)Z\Big]\notag\\
&+\epsilon\partial_{\epsilon}\frac{1}{\pi}\big(\text{Re}[x]P_3(x,\tau_a)\partial_{\tau_{\bar a}}\partial_{\bar\tau_a}Z+\text{Re}[y]P_3(y,\tau_{\bar a})\partial_{\tau_a}\partial_{\bar\tau_{\bar a}}Z\big)\notag\\
&-\frac{c|\epsilon|}{12\pi}\big(P_4(x,\tau_a)\partial_{\tau_{\bar a}}\partial_{\bar\tau_a}Z+P_4(y,\tau_{\bar a})\partial_{\tau_a}\partial_{\bar\tau_{\bar a}}Z\big)\bigg\rbrace+O(|\epsilon|^2). \label{1st 1ord TxTy}
\end{align}
For the second type $\langle{T_1\bar T_2}\rangle$ we have
\begin{align}
&\int_{S_1\backslash{D_{\delta}}}d^2z_1\langle{O_{T\bar T}(z_1,\bar z_1)T_1\bar T_2}\rangle+\int_{S_2\backslash{D_{\delta}}}d^2z_2\langle{O_{T\bar T}(z_2,\bar z_2)T_1\bar T_2}\rangle\notag\\
=&\frac{1}{Z}\sum_{a=1,2}\int_{S_a\backslash{D_{\delta}}}d^2z_a\bigg\lbrace\Big[D_{z_a}\bar D_{\bar z_a}+D_{z_a}\bar{\mathcal{P}}_{\bar z_a,\bar x_2}+\bar D_{\bar z_a}{\mathcal{P}}_{z_a,x_1}+{\mathcal{P}}_{z_a,x_1}\bar{\mathcal{P}}_{\bar z_a,\bar x_2}\Big]\big(Z\langle{T_1\bar T_2}\rangle\big)\notag\\
&+\frac{c}{2}{}^2\bar{\mathcal{P}}_4(\bar z_a,\bar x_2)\Big[D_{z_a}+{\mathcal{P}}_{z_a,x_1}\Big]\big(Z\langle{T_1}\rangle\big)+\frac{c}{2}{}^2{\mathcal{P}}_4(z_a,x_1)\Big[\bar D_{\bar z_a}+\bar{\mathcal{P}}_{\bar z_a,\bar x_2}\Big]\big(Z\langle{\bar T_2}\rangle\big)\notag\\
&+\frac{c^2}{4}{}^2{\mathcal{P}}_4(z_a,x_1){}^2\bar{\mathcal{P}}_4(\bar z_a,\bar x_2)Z\bigg\rbrace, \label{1st T1T2bar}
\end{align}
In the case of two points inserted in the same torus $x_1,x_2\in S_a$, one can obtain first-order correction using eqs.(\ref{1st T1T2bar})(\ref{1st PF})(\ref{1st 1p DP})(\ref{1st T PD})(\ref{2p PP 1})(\ref{P4P 12})(\ref{P4P4 12}):
\begin{align}
&\delta_\lambda\langle{T_1\bar T_2}\rangle-\langle{\delta_\lambda(T_1\bar T_2)}\rangle\notag\\
=&-\frac{\langle{T_1\bar T_2}\rangle}{Z}\bigg\lbrace\sum_{b=a,\bar a}\Big[\frac{1}{2}\big(\text{Im}[\tau_b]-\frac{|\epsilon|}{6\pi}\big)\partial_{\tau_b}\partial_{\bar{\tau}_b}Z-\frac{i}{2}\bar\epsilon\partial_{\tau_b}\partial_{\bar{\epsilon}}Z\Big]+\pi|\epsilon|\partial_\epsilon\partial_{\bar{\epsilon}}Z\bigg\rbrace\notag\\
&+\frac{1}{Z}\bigg\lbrace\Big[\sum_{b=a,\bar a}\big[\big(\text{Im}[\tau_b]-\frac{|\epsilon|}{6\pi}\big)\partial_{\tau_b}\partial_{\bar\tau_b}+\frac{i}{2}\bar\epsilon\partial_{\tau_b}\partial_{\bar\epsilon}-\frac{i}{2}\epsilon\partial_{\bar\tau_b}\partial_{\epsilon}\big]+2\pi|\epsilon|\partial_{\epsilon}\partial_{\bar\epsilon}+i\partial_{\tau_a}-i\partial_{\bar\tau_a}\notag\\
&-\frac{2i|\epsilon|}{3}P_2(x_1,\tau_a)\partial_{\bar\tau_a}+\frac{2i|\epsilon|}{3}\bar P_2(\bar x_2,\bar\tau_a)\partial_{\tau_a}\Big]\big(\frac{1}{4\pi^2}\partial_{\tau_a}\partial_{\bar\tau_a}Z\big)\notag\\
&+\big(\sum_{b=a,\bar a}\text{Im}[\tau_b]\partial_{\tau_b}\partial_{\bar\tau_b}+i\partial_{\tau_a}-i\partial_{\bar\tau_a}\big)\Big[\frac{1}{2\pi i}\big(\bar\epsilon\bar P_2(\bar x_2,\bar\tau_a)\partial_{\bar\epsilon}\partial_{\tau_a}-\epsilon P_2(x_1,\tau_a)\partial_{\epsilon}\partial_{\bar\tau_a}\big)Z\Big]\notag\\
&-\frac{c|\epsilon|}{12\pi}\Big[P_4(x_1,\tau_a)\big(\partial_{\bar\tau_a}-4\pi i\bar P_2(\bar x_2,\bar\tau_a)\big)\partial_{\bar\tau_a}+\bar P_4(\bar x_2,\bar\tau_a)\big(\partial_{\tau_a}+4\pi iP_2(x_1,\tau_a)\big)\partial_{\tau_a}\Big]Z\notag\\
&-\epsilon\partial_{\epsilon}P_3(x_1,\tau_a)\Big[\frac{1}{\pi}\text{Re}[x_1]\partial^2_{\bar\tau_a}+2i\big(\bar P_1(\bar x_1-\bar x_2,\bar\tau_a)+\bar P_1(\bar x_2,\bar\tau_a)\big)\partial_{\bar\tau_a}\notag\\
&-\frac{\pi c}{3}\big(\bar P_3(\bar x_1-\bar x_2,\bar\tau_a)+\bar P_3(\bar x_2,\bar\tau_a)\big)\Big]Z\notag\\
&-\bar\epsilon\partial_{\bar\epsilon}\bar P_3(\bar x_2,\bar\tau_a)\Big[\frac{1}{\pi}\text{Re}[x_2]\partial^2_{\tau_a}-2i\big(P_1(x_2-x_1,\tau_a)+P_1(x_1,\tau_a)\big)\partial_{\tau_a}\notag\\
&-\frac{\pi c}{3}\big(P_3(x_2-x_1,\tau_a)+P_3(x_1,\tau_a)\big)\Big]Z\bigg\rbrace+O(|\epsilon|^2). \label{1st 1ord T1T2bar}
\end{align}
In the case of two points inserted in different tori $x\in S_{a}$ and $y\in S_{\bar a}$, one can obtain first-order correction from eqs.(\ref{1st T1T2bar})(\ref{1st PF})(\ref{1st 1p DP})(\ref{1st T PD})(\ref{2p PP 2})(\ref{P4P xy})(\ref{P4P4 xy}):
\begin{align}
&\delta_\lambda\langle{T_x\bar T_y}\rangle-\langle{\delta_\lambda(T_x\bar T_y)}\rangle\notag\\
=&-\frac{\langle{T_x\bar T_y}\rangle}{Z}\bigg\lbrace\sum_{b=a,\bar a}\Big[\frac{1}{2}\big(\text{Im}[\tau_b]-\frac{|\epsilon|}{6\pi}\big)\partial_{\tau_b}\partial_{\bar{\tau}_b}Z-\frac{i}{2}\bar\epsilon\partial_{\tau_b}\partial_{\bar{\epsilon}}Z\Big]+\pi|\epsilon|\partial_\epsilon\partial_{\bar{\epsilon}}Z\bigg\rbrace\notag\\
&+\frac{1}{Z}\bigg\lbrace\Big[\sum_{b=a,\bar a}\big[\big(\text{Im}[\tau_b]-\frac{|\epsilon|}{6\pi}\big)\partial_{\tau_b}\partial_{\bar\tau_b}+\frac{i}{2}\bar\epsilon\partial_{\tau_b}\partial_{\bar\epsilon}-\frac{i}{2}\epsilon\partial_{\bar\tau_b}\partial_{\epsilon}\big]+2\pi|\epsilon|\partial_{\epsilon}\partial_{\bar\epsilon}+i\partial_{\tau_{\bar a}}-i\partial_{\bar\tau_a}\notag\\
&-\frac{2i|\epsilon|}{3}P_2(x,\tau_a)\partial_{\bar\tau_a}+\frac{2i|\epsilon|}{3}\bar P_2(\bar y,\bar\tau_{\bar a})\partial_{\tau_{\bar a}}\Big]\big(\frac{1}{4\pi^2}\partial_{\tau_a}\partial_{\bar\tau_{\bar a}}Z\big)\notag\\
&+\big(\sum_{b=a,\bar a}\text{Im}[\tau_b]\partial_{\tau_b}\partial_{\bar\tau_b}+i\partial_{\tau_{\bar a}}-i\partial_{\bar\tau_a}\big)\Big[\frac{1}{2\pi i}\big(\bar\epsilon\bar P_2(\bar y,\bar\tau_{\bar a})\partial_{\bar\epsilon}\partial_{\tau_a}-\epsilon P_2(x,\tau_a)\partial_{\epsilon}\partial_{\bar\tau_{\bar a}}\big)Z\Big]\notag\\
&-\frac{c|\epsilon|}{12\pi}\Big[P_4(x,\tau_a)\partial_{\bar\tau_a}\partial_{\bar\tau_{\bar a}}+\bar P_4(\bar y,\bar\tau_{\bar a})\partial_{\tau_a}\partial_{\tau_{\bar a}}\Big]Z\notag\\
&-\frac{\epsilon}{\pi}\text{Re}[x]P_3(x,\tau_a)\partial_{\epsilon}\partial_{\bar\tau_a}\partial_{\bar\tau_{\bar a}}Z-\frac{\bar\epsilon}{\pi}\text{Re}[y]\bar P_3(\bar x,\bar\tau_{\bar a})\partial_{\bar\epsilon}\partial_{\tau_a}\partial_{\tau_{\bar a}}Z\bigg\rbrace+O(|\epsilon|^2). \label{1st 1ord TxTybar}
\end{align}
In particular, we consider the leading term of the first-order correction of $\langle{T_1\bar T_2}\rangle$ in eq.(\ref{1st 1ord T1T2bar}). Accroding to eq.(\ref{TT}), $\langle{T_1\bar T_2}\rangle=\frac{1}{4\pi^2Z_a}\partial_{\tau_a}\partial_{\bar\tau_a}Z_a+O(|\epsilon|)=\langle{T_1\bar T_2}\rangle_{T^2_a}+O(\epsilon)$, which is completely defined on torus $T^2_a$. Thus we can read off the leading term of the first-order correction of $\langle{T_1\bar T_2}\rangle_{T^2_a}$ from eq.(\ref{1st 1ord T1T2bar}):
\begin{align}
&\delta_{\lambda}\langle{T_1\bar T_2}\rangle-\langle{\delta_{\lambda}T_1\bar T_2}\rangle\notag\\
=&\frac{1}{4\pi^2Z_a}\Big[\text{Im}[\tau_a]\partial^2_{\tau_a}\partial^2_{\bar\tau_a}Z_a+i\big(\partial^2_{\tau_a}\partial_{\bar\tau_a}-\partial_{\tau_a}\partial^2_{\bar\tau_a}\big)Z_a\Big]-\frac{\delta_{\lambda}Z_a}{Z_a}\langle{T_1\bar T_2}\rangle_{T^2_a}+O(|\epsilon|)\notag\\
=&\delta_{\lambda}\langle{T_1\bar T_2}\rangle_{T^2_a}-\langle{\delta_{\lambda}T_1\bar T_2}\rangle_{T^2_a}+O(|\epsilon|).
\end{align}
\begin{figure}[H]
\centering
\includegraphics[scale=0.5]{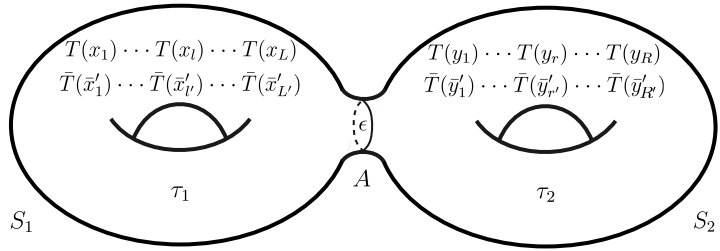}
\caption{An $(L+R+L'+R')$-point function of stress tensors on sewn Riemann surface with genus-2. }
\label{Fig stress}
\end{figure}
Now we consider the first-order deformation of a general $(L+R+L'+R')$-point function of stress tensors, as shown in Fig.\ref{Fig stress}, $(L+R)$ holomorphic stress tensor are inserted at $x_1,...,x_L\in S_1,\ y_1,...,y_L\in S_2$ and $(L'+R')$ anti-holomorphic stress tensor are inserted at $\bar x'_1,...,\bar x'_{L'}\in S_1,\ \bar y'_1,...,\bar y'_{R'}\in S_2$, respectively. For simplicity, we define\footnote{For simplicity, $ \backslash_lX_{L}$ stands for delelating the $l-$th factor in $X_{L}$.}
\begin{align}
    X_L\equiv&T(x_1)\cdot\cdot\cdot T(x_L),\ X_R\equiv T(y_1)\cdot\cdot\cdot T(y_R),\\
    \bar X_{L'}\equiv&\bar T(\bar x'_1)\cdot\cdot\cdot\bar T(\bar x'_{L'}),\ \bar X_{R'}\equiv\bar T(\bar y'_1)\cdot\cdot\cdot\bar T(\bar y'_{R'}),\\
    \backslash_lX_{L}\equiv&T(x_1)\cdot\cdot\cdot T(x_{l-1})T(x_{l+1})\cdot\cdot\cdot T(x_L),\\
    \backslash_{l'}\bar X_{L'}\equiv&\bar T(\bar x'_1)\cdot\cdot\cdot \bar T(\bar x'_{l'-1})\bar T(\bar x'_{l'+1})\cdot\cdot\cdot \bar T(\bar x'_{L'}).
\end{align}
Then we use eq.(\ref{WI1}) to calculate the first-order deformation of $\langle{X_L\bar X_{L'}X_R\bar X_{R'}}\rangle$:
\begin{align}
&\int_{S_1\backslash{D_{\delta}}}d^2z_1\langle{O_{T\bar{T}}(z_1,\bar z_1)X_L\bar X_{L'}X_R\bar X_{R'}}\rangle+\int_{S_2\backslash{D_{\delta}}}d^2z_2\langle{O_{T\bar{T}}(z_2,\bar z_2)X_L\bar X_{L'}X_R\bar X_{R'}}\rangle\notag\\
=&\frac{1}{Z}\sum_{a=1,2}\int_{S_a\backslash{D_{\delta}}}d^2z\bigg\lbrace D_{z_a}\bar{D}_{\bar{z}_a}+\sum_{l=1}^L\Big[\bar D_{\bar z_a}\mathcal{P}_{z_a,x_l}+\frac{c}{2}\bar D_{\bar z}{}^2\mathcal{P}_4^{(z_a,x_l)}\backslash_l\Big]\notag\\
&+\sum_{r=1}^R\Big[\bar D_{\bar z_a}\mathcal{P}_{z_a,y_r}+\frac{c}{2}\bar D_{\bar z}{}^2\mathcal{P}_4^{(z_a,y_r)}\backslash_r\Big]+\sum_{l'=1}^{L'}\Big[D_{z_a}\bar{\mathcal{P}}_{\bar z_a,\bar x'_{l'}}+\frac{c}{2}D_{z_a}{}^2\bar{\mathcal{P}}_4^{(\bar z,\bar x'_{l'})}\backslash_{l'}\Big]\notag\\
&+\sum_{r'=1}^{R'}\Big[D_{z}\bar{\mathcal{P}}_{\bar z_a,\bar y'_{r'}}+\frac{c}{2}D_{z_a}{}^2\bar{\mathcal{P}}_4^{(\bar z_a,\bar y'_{r'})}\backslash_{r'}\Big]+\sum_{l,l'}\Big[\mathcal{P}_{z_a,x_l}\bar{\mathcal{P}}_{\bar z_a,\bar x'_{l'}}+\frac{c^2}{4}{}^2\mathcal{P}_4^{(z_a,x_l)}{}^2\bar{\mathcal{P}}_4^{(\bar z_a,\bar x'_{l'})}\backslash_{l,l'}\Big]\notag\\
&+\sum_{r,l'}\Big[\mathcal{P}_{z_a,y_r}\bar{\mathcal{P}}_{\bar z_a,\bar x'_{l'}}+\frac{c^2}{4}{}^2\mathcal{P}_4^{(z_a,y_r)}{}^2\bar{\mathcal{P}}_4^{(\bar z_a,\bar x'_{l'})}\backslash_{r,l'}\Big]+\sum_{l,r'}\Big[\mathcal{P}_{z_a,x_l}\bar{\mathcal{P}}_{\bar z_a,\bar y'_{r'}}+\frac{c^2}{4}{}^2\mathcal{P}_4^{(z_a,x_l)}{}^2\bar{\mathcal{P}}_4^{(\bar z_a,\bar y'_{r'})}\backslash_{l,r'}\Big]\notag\\
&+\sum_{r,r'}\Big[\mathcal{P}_{z_a,y_r}\bar{\mathcal{P}}_{\bar z_a,\bar y'_{r'}}+\frac{c^2}{4}{}^2\mathcal{P}_4^{(z_a,y_r)}{}^2\bar{\mathcal{P}}_4^{(\bar z_a,\bar y'_{r'})}\backslash_{r,r'}\Big]\bigg\rbrace\Big[Z\langle{X_L\bar X_{L'}X_R\bar X_{R'}}\rangle\Big].
 \label{higher-point stress tensor}
\end{align}
The first term in eq.(\ref{higher-point stress tensor}) is calculated in eq.(\ref{1st PF}). The terms from second to fifth contains all the one-point contribution of stress tensors eqs.(\ref{1st 1p DP})(\ref{1st T PD}). The remaining terms contains all the two-point contribution eqs.(\ref{P4P 12})(\ref{P4P xy})(\ref{P4P4 12})(\ref{P4P4 xy}). \par
To close this section, we would like to add a potential application of these deformed correlation functions. To check the $AdS_3/CFT_2$ with $T\bar{T}$ deformation\cite{McGough:2016lol,Kraus:2018xrn,Guica:2019nzm}, one has to match the correlation functions in both field theory side and gravity side. As reviewed in the introduction, it is a nontrivial attempt to construct the non-perturbative deformed correlation functions. Here we apply the perturbative field theory approach to construct the generic deformed correlation functions in the boundary field theories with nontrivial topology. Since the holographic CFTs show that maximally quantum chaotic behavior \cite{Shenker:2014cwa, Maldacena:2015waa, Roberts:2014ifa}, extracting the chaos signals of the deformed holographic CFTs is an important step in checking the holographic dictionary. In particular, one can directly apply the higher point correlation functions in deformed theory on higher genus Riemann surface to calculate OTOC and multiple-interval R\'eny entropies as following previous works \cite{He:2019vzf, He:2020qcs}. Further, one can also do Fourier transformation of deformed two-point functions of stress tensor to look at the pole structure to read off the chaos signals; namely, pole-skipping phenomenon proposed by \cite{Blake:2017ris}.
\section{Conclusions and perspectives}
To understand the quantum chaos of $T\bar{T}$ deformed conformal field theories, one has to calculate OTOC, spectrum form factor, and pole skipping phenomenon, which is associated with the correlation function in deformed field theory. Furthermore, the definition of $T\bar{T}$ deformation in the curved Riemann surface is ambiguous in the literature. We propose a way to generalize the deformation in the higher genus Riemann surface. Furthermore, it is highly nontrivial to construct the correlation function in a non-perturbative approach. Alternatively, one can follow a perturbative approach~\cite{He:2019vzf, He:2020qcs} to learn some lessons about the quantum chaos of deformed CFTs. In this work, we have applied the perturbative conformal field theory approach to construct higher genus correlation functions of $T\bar{T}$-deformed theories to offer field theories data to achieve our final goal of understanding the quantum chaos of deformed CFTs. The most important ingredients are sewing construction and the conformal ward identity of CFTs on higher genus two-dimensional Riemann surface. Thanks to sewing construction, one can construct the higher genus correlation functions in terms of the correlation functions on the low genus Riemann surface. In the current work, we apply a particular sewing construction and perturbative conformal field theory approach to obtain the first order $T\bar{T}$ deformation of the partition function and correlation functions on a genus two Riemann surface. {As a consistency check, we extract the leading term and find that the leading term of genus two correlation functions can be expressed by genus one partition function and correlation functions presented in the literature.} To obtain the final results, we apply a systematic renormalization \cite{Dijkgraaf:1996iy} by following the calculation given in the $T\bar{T} $ deformation in genus one CFTs \cite{He:2020udl, He:2020cxp}.

It is a preliminary attempt to calculate the correlation functions in the higher genus $T\bar{T}$ deformed CFTs in the perturbative approach. To go beyond the first-order calculation will be a highly nontrivial project in this direction, even in the free field theory \cite{He:2020cxp}. In higher-order deformation, one must take the flow effects of $T\bar{T}$ operator into account. Further, one can follow up on the resulting correlation functions to investigate quantum chaos signals or quantum integrability structure of $T\bar{T}$ deformed theories, as we mentioned at the beginning. We would like to report further progress in future works.

\subsection*{Acknowledgements} We thank Miao He, Hao Ouyang, Hongfei Shu, Yuan Sun, and Yu-Xuan Zhang for valuable discussions. S.H. would like to appreciate the financial support from Jilin University, Max Planck Partner group, and Natural Science Foundation of China Grants (No.12075101, No.12047569).
\appendix
\section{Elliptic functions}\label{A}
In this appendix, we list the elliptic functions that appear in the context and their properties. We follow the conventions in \cite{Mason:2006dk,Mason:2009ely}. The torus $T^2$ is defined by the identification on complex plane $z\sim z+2\pi i$ and $z\sim z+2\pi i\tau$. $P_k(z,\tau)$ is used to denote an elliptic function with subscript $k$, and its Laurent expansion in the neighborhood of $z=0$ is
\begin{equation}
P_{k\geq 1}(z,\tau)=\frac{1}{z^k}+(-1)^{k}\sum_{n\geq k}E_n(\tau)\left(\begin{array}{ccc}n-1\\k-1\end{array}\right)z^{n-k}, \tag{S1}\label{Laurent expansion}
\end{equation}
where $\tau$ is modular parameter of the torus. $E_n(\tau)$ is the \emph{Eisenstein series} for $n\geq 2$ which equals to zero for $n$ odd, and for $n$ even $E_n(\tau)$ can be defined as
\begin{equation}
    E_n(\tau)=-\frac{B_n}{n!}+\frac{2}{(n-1)!}\sum_{m\geq 1}\sigma_{n-1}(m)q^m,\tag{S2}
\end{equation}
where $B_n$ is the $n^{\text{th}}$ Bernoulli number, $\sigma_{n-1}(m)=\sum_{d|m}d^{n-1}$ and $q=e^{2\pi i\tau}$. In this paper, we also use the convention $P_0=1$. There are simple relations among $P_1(z,\tau), P_2(z,\tau)$ and classical Weierstrass functions:
\begin{align}
P_2(z,\tau)=&\wp(z,\tau)+E_2(\tau), \tag{S3}\\
P_1(z,\tau)=&\zeta(z,\tau)-E_2(\tau)z, \tag{S4}
\end{align}
where $\wp(z,\tau)$ is \emph{Weierstrass P-function} and $\zeta(z,\tau)$ is \emph{Weierstrass $\zeta$-function}. These two functions are defined as
\begin{align}
\wp(z,\tau)=&\frac{1}{z^2}+\sum_{(m,n)\neq(0,0)}\Big[\frac{1}{(z-w_{m,n})^2}-\frac{1}{w_{m,n}^2}\Big],\tag{S5}\\
\zeta(z,\tau)=&\frac{1}{z}+\sum_{(m,n)\neq(0,0)}\Big[\frac{1}{z-w_{m,n}}+\frac{1}{w_{m,n}}+\frac{z}{w_{m,n}^2}\Big],\tag{S6}
\end{align}
where $w_{m,n}=2\pi im+2\pi i\tau n$ is the coordinate of the lattice. The $k$-th derivative of $P_1(z,\tau)$ with respect to $z$ gives $P_{k+1}(z,\tau)$:
\begin{equation}
P_{k+1}(z,\tau)=\frac{(-1)^{k}}{k!}\partial_z^{k}P_1(z,\tau)=-\frac{1}{k}\partial_zP_k(z,\tau). \tag{S7} \label{recursion}
\end{equation}
$P_1(z,\tau)$ is quasi-periodic, and $P_k(z,\tau)$ is periodic for $k\geq 2$, and they satisfy
\begin{align}
P_1(z+2\pi i,\tau)=&P_1(z,\tau)+2P_1(\pi i,\tau)=P_1(z,\tau),\tag{S8} \\
P_1(z+2\pi i\tau,\tau)=&P_1(z,\tau)+2P_1(\pi i\tau,\tau)=P_1(z,\tau)-1,\tag{S9} \\
P_k(z+2\pi i,\tau)=&P_k(z+2\pi i\tau,\tau)=P_k(z,\tau).\tag{S11}
\end{align}
We define $Q(z,\tau)$ function by
\begin{align}
\int_{0}^{z}dz'\Big(P_1(z',\tau)-\frac{1}{z'}\Big)=&\log\frac{Q(z,\tau)}{z},\notag\\
P_1(z,\tau)=&\frac{\partial_z Q(z,\tau)}{Q(z,\tau)}=\partial_z \log Q(z,\tau),\notag\\
\log Q(z,\tau)=&\log z-\sum_{n\geq 1}\frac{1}{n}E_n(\tau)z^n.\tag{S11} \label{Q}
\end{align}
The relationship between $Q(z,\tau)$ and \emph{Weierstrass $\sigma$-function} is
\begin{align}
Q(z,\tau)=e^{-\frac{1}{2}E_2(\tau)z^2}\sigma(z,\tau), \tag{S12}
\end{align}
and \emph{Weierstrass $\sigma$-function} is defined as
\begin{equation}
\sigma(z,\tau)=z\prod_{(m,n)\neq(0,0)}\Big[\Big(1-\frac{z}{w_{m,n}}\Big)\text{exp}\Big(\frac{z}{w_{m,n}}+\frac{z^2}{2w_{m,n}^2}\Big)\Big].\tag{S13}
\end{equation}
$Q(z,\tau)$ is an odd function just like $\sigma(z,\tau)$, and it is quasi-periodic with
\begin{align}
Q(z+2\pi i,\tau)=&-Q(z,\tau),\tag{S14}\\
Q(z+2\pi i\tau,\tau)=&-e^{-(z+i\pi\tau)}Q(z,\tau).\tag{S15}	\label{properties Q}
\end{align}
\section{Some useful integrals}\label{B}
\begin{figure}[H]
\centering
\includegraphics[scale=0.75]{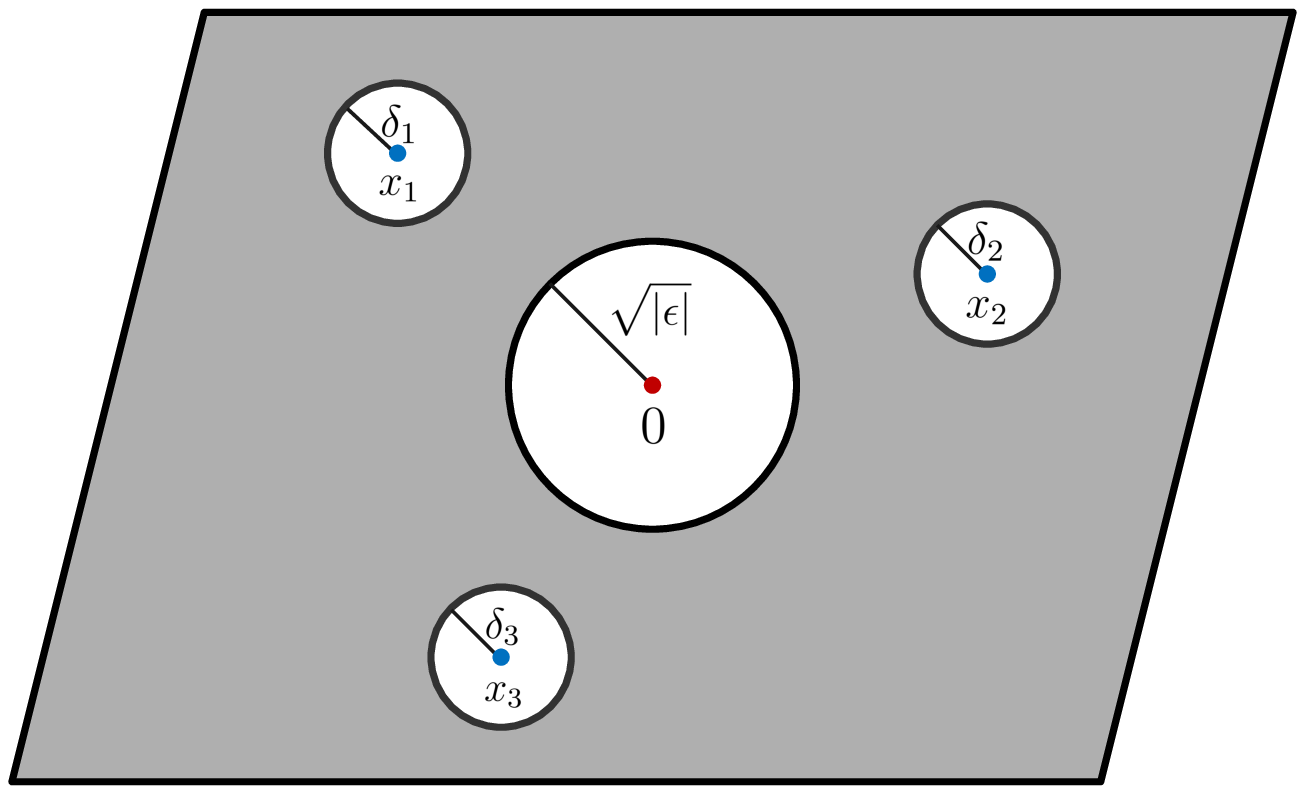}
\caption*{Figure S1: An example for the regularized domain of integration. The red point is the origin of the torus coordinates, and the circle of radius $\sqrt{|\epsilon|}$ surrounding it is the sewing region. The blue points are the singularities of the integrand, and the small disks surrounding them have been removed. The gray area is the domain of integration.}
\label{Fig 3}
\end{figure}

In this appendix, we discuss the details of integrals $I_a$, $J_a(x,\bar x)$ and $K_a(x_1,\bar x_1;x_2,\bar x_2)$ that appear in the context. The integrands over a torus may contain singularities. Following the prescription in \cite{Dijkgraaf:1996iy}, we regularize the domain of integration by removing some small disks $(D_{\delta_1},...,D_{\delta_N})$ of radiuses $(\delta_1,...,\delta_N)$ centered on the singularities $(x_1,...,x_N)$ (see Fig.\ref{Fig 3}). $f(z,\bar z)$ is used to denote a general function on torus $T^2$ with these singularities, and we assume that it can be written as the divergence of some vector field $F^{\mu}$:
\begin{equation}
f(z,\bar z)=\partial_{\mu}F^{\mu}(z,\bar z)=\partial_{z}F^{z}(z,\bar z)+\partial_{\bar z}F^{\bar z}(z,\bar z).\tag{S16}
\end{equation}
The integral of $f(z,\bar z)$ over the regularized domain can be calculated using the Stoke's theorem
\begin{equation}
\int_{T^2\backslash D_{\delta}}d^2z\partial_{\mu}F^{\mu}(z,\bar z)=\frac{i}{2}\Big(\oint_{\partial T^2}-\sum_{i=1}^N\oint_{\partial D_{\delta_i}}\Big)(F^zd\bar z-F^{\bar z}dz). \tag{S17}\label{stoke's theorem}
\end{equation}
After integrating eq.(\ref{stoke's theorem}), we take the limit $(\delta_1,...,\delta_N)\rightarrow(0,...,0)$ and discard the divergent term (if it exists) to regularize the integral. In this paper, we introduce constraints on the integrand $f(z,\bar z)$. The integrand may diverge at the boundary of the torus $\partial T^2$, for instance, the elliptic function $P_k(z,\tau)$ diverges at points on the lattice $z=2\pi im+2\pi i\tau n$ with $m,n \in\mathds{Z}$. To avoid contact between the boundary of
\noindent disk and the boundary of the torus, we require that the integrand $f(z,\bar z)$ be biperiodic on the torus concerning both the variables $z$ and $\bar z$. Under this condition, we can translate the parallelogram of the torus on the complex plane to ensure that the singularities only appear inside. Next, we require that the integrand can be factorized into two parts:
\begin{equation}
f(z,\bar z)=g(z)h(\bar z),\tag{S18}
\end{equation}
and we assume that $g(z)=\partial_zG(z)$ and $h(\bar z)=\partial_{\bar z}H(\bar z)$. Since we have removed the disks around the singularities, $g(z)$ and $h(\bar z)$ are holomorphic and antiholomorphic respectively over the domain of integration. 
Thus we can construct the vector fields $F^{\mu}$ and $F'^{\mu}$ as
\begin{align}
&F^z(z,\bar z)=G(z)h(\bar z),\ F^{\bar z}(z,\bar z)=0,\tag{S19}\\
&F'^z(z,\bar z)=0,\ F'^{\bar z}(z,\bar z)=g(z)H(\bar z).\tag{S20}
\end{align}
Take the first construction for example, eq.(\ref{stoke's theorem}) can be further written as
\begin{align}
\int_{T^2\backslash D_{\delta}}d^2zf(z,\bar z)=&\frac{i}{2}\int_{\bar z_0}^{\bar z_0-2\pi i}d\bar z[G(z)-G(z+2\pi i\tau)]h(\bar z)\notag\\
&-\int_{\bar z_0}^{\bar z_0-2\pi i\bar\tau}d\bar z[G(z)-G(z+2\pi i)]h(\bar z)\notag\\
&-\sum_{i=1}^N\Big[\int_0^{2\pi}(-i\delta_i e^{-i\theta})d\theta G(\delta_i e^{i\theta}+x_i)h(\delta_i e^{-i\theta}+\bar x_i)\Big], \tag{S21}\label{stoke's theorem 2}
\end{align}
where the biperiodic property of $h(\bar z)$ is used in the first two terms. In the rest of this Appendix, we will apply the above method to calculate three types of integrals, which play an important role in calculating first-order corrections.
\subsection{Integral $I_a(k,l)$}\label{B.1}
The first type of integral that we encounter is
\begin{equation}
I_a(k,l)=\int_{S_a}d^2z_aP_k(z_a,\tau_a)\bar{P}_l(\bar{z}_a,\bar{\tau}_a),\tag{S22}
\end{equation}
for $k,l\geq 0$. $z_a$ is the local coordinate of the torus $T^2_a$, and we will omit its subscript $a$ later. $S_a$ is the remainder of the torus $T^2_a$ after removing the a disk of radius $\sqrt{|\epsilon|}$ centered at $z_a=0$, and the integrand $P_k(z_a,\tau_a)\bar{P}_l(\bar{z}_a,\bar{\tau}_a)$ is regular in the domain of integration. $I_a(k,l)$ has a simple property:
\begin{equation}
I_a(l,k)=\overline{I_a(k,l)}. \tag{S23}\label{simple 1}
\end{equation}
\par
For $k\geq 3$ and $l\geq 2$, we can construct $F^z(z,\bar z)=\frac{-1}{k-1}P_{k-1}(z,\tau_a)\bar{P}_l(\bar{z},\bar{\tau}_a)$ (and $F^{\bar z}=0$) using the recursion eq.(\ref{recursion}) of $P_k(z,\tau_a)$. In this case $P_{k-1}(z,\tau_a)$ is still biperiodic on the torus, and thus the first two terms of eq.(\ref{stoke's theorem 2}) vanish. The last term contains integral over the annulus $A$ (of radius $\sqrt{|\epsilon|}$ centered at $z=0$), which can be calculated using the Laurent expansion eq.(\ref{Laurent expansion}):
\begin{align}
I_a(k,l)\Big|_{k\geq 3,l\geq 2}=&\frac{1}{2(k-1)}\int_0^{2\pi}d\theta\Big[\sqrt{|\epsilon|}^{2-k-l}e^{-i(k-l)\theta}\notag\\
&+(-1)^l\sum_{n\geq l}\bar{E}_n(\bar{\tau}_a)\left(\begin{array}{ccc}n-1\\l-1\end{array}\right)\sqrt{|\epsilon|}^{n+2-k-l}e^{-i(k-l+n)\theta}\notag\\
&+(-1)^{k-1}\sum_{m\geq k-1}E_m(\tau_a)\left(\begin{array}{ccc}m-1\\k-2\end{array}\right)\sqrt{|\epsilon|}^{m+2-k-l}e^{-i(k-l-m)\theta}\notag\\
&+(-1)^{k+l-1}\sum_{m\geq k-1}\sum_{n\geq l}E_m(\tau_a)\bar{E}_n(\bar{\tau}_a)\left(\begin{array}{ccc}m-1\\k-2\end{array}\right)\left(\begin{array}{ccc}n-1\\l-1\end{array}\right)\notag\\
&\times\sqrt{|\epsilon|}^{m+n+2-k-l}e^{-i(n-m+k-l)\theta}\Big] \notag\\
=&\frac{\pi }{k-1}\Big[\frac{1}{|\epsilon|^{k-1}}\delta_{k,l}+\sum_{n\geq l}C_{n-3,a}^{k-3,l-3}|\epsilon|^{n-l+1}\Big],\tag{S24} \label{I(k,l)}
\end{align}
Here we use the fact that $(k-l+n)\geq 3$ and $(k-l-m)\leq -1$, and the integral $\int_{0}^{2\pi}d\theta e^{-iN\theta}=2\pi\delta_{N,0}$ for integer $N$. Thus the second and third terms of eq.(\ref{I(k,l)}) vanish. The coefficient $C_{k,a}^{n,l}$ is defined as
\begin{equation}
C_{n,a}^{k,l}=(-1)^{k+l}\left(\begin{array}{ccc}n+k-l+2\\k+1\end{array}\right)\left(\begin{array}{ccc}n+2\\l+2\end{array}\right)E_{n+k-l+3}(\tau_a)\bar E_{n+3}(\bar\tau_a),\tag{S25}
\end{equation}
and it is easy to prove that
\begin{equation}
C_{n,a}^{k,l}=\frac{k+2}{l+2}\bar C_{n+k-l,a}^{l,k}.\tag{S26} \label{simple 2}
\end{equation}
From eq.(\ref{simple 1}) and eq.(\ref{simple 2}) we immediately obtain the case for $k\geq 2$ and $l\geq 3$:
\begin{align}
I_a(k,l)\Big|_{k\geq 2,l\geq 3}=&\frac{\pi }{l-1}\Big[\frac{1}{|\epsilon|^{l-1}}\delta_{l,k}+\sum_{n\geq k}\bar C_{n-3,a}^{l-3,k-3}|\epsilon|^{n-k+1}\Big]\notag\\
=&\frac{\pi}{k-1}\frac{1}{|\epsilon|^{k-1}}\delta_{k,l}+\frac{\pi}{l-1}\sum_{n'\geq l}\bar C_{n'-3+k-l,a}^{l-3,k-3}|\epsilon|^{n'-l+1}\notag\\
=&\frac{\pi}{k-1}\Big[\frac{1}{|\epsilon|^{k-1}}\delta_{k,l}+\sum_{n\geq l}C_{n-3,a}^{k-3,l-3}|\epsilon|^{n-l+1}\Big],\tag{S27}
\end{align}
which is consistent with eq.(\ref{I(k,l)}). When we consider the case $k=2$ and $l=2$, the integrand $F^z(z,\bar z)=-P_{1}(z,\tau_a)\bar{P}_2(\bar{z},\bar{\tau}_a)$ is quasi-periodic and its integral on the torus boundary can be written as 
\begin{align}
\oint_{\partial T^2}F^zd\bar z=&\int_{\bar z_0}^{\bar z_0-2\pi i}d\bar z[P_{1}(z+2\pi i\tau_a,\tau_a)-P_{1}(z,\tau_a)]\bar{P}_2(\bar{z},\bar{\tau}_a)\notag\\
&-\int_{\bar z_0}^{\bar z_0-2\pi i\bar\tau}d\bar z[P_{1}(z+2\pi i,\tau_a)-P_{1}(z,\tau_a)]\bar{P}_2(\bar{z},\bar{\tau}_a)\notag\\
=&2\eta_a'\int_{z_0}^{z_0+2\pi i}d\bar{z}\bar{P}_2(\bar{z},\bar{\tau}_a)-2\eta_a\int_{z_0}^{z_0+2\pi i\tau_a}d\bar{z}\bar{P}_2(\bar{z},\bar{\tau}_a)\notag\\
=&8 \text{Im}[\eta_a\bar{\eta_a'}],\tag{S28}
\end{align}
where $\eta_a=P_1(\pi i,\tau_a)=0$ and $\eta_a'=P_1(\pi i\tau_a,\tau_a)=-\frac{1}{2}$, thus this term has no contribution. Then we compute the integral over $A$ and finally we obtain
\begin{align}
I_a(2,2)=\frac{\pi}{|\epsilon|}-\frac{\pi}{2}\sum_{n\geq 1}n|E_{n+1}(\tau)|^2|\epsilon|^{n},\tag{S29}
\end{align}
which is also consistent with eq.(\ref{I(k,l)}).\par
For $k\geq 3$ and $l=0$, we construct $F^z=\frac{-1}{k-1}P_{k-1}(z,\tau_a)$ and obtain
\begin{align}
I_a(k,0)\Big|_{k\geq 3}=&\frac{1}{2(k-1)}\int_0^{2\pi}d\theta\Big[\sqrt{|\epsilon|}^{2-k}e^{-ik\theta}\notag\\
&+(-1)^{k-1}\sum_{m\geq k-1}E_m(\tau_a)\left(\begin{array}{ccc}m-1\\k-2\end{array}\right)\sqrt{|\epsilon|}^{m+2-k}e^{i(m-k)\theta}\Big]\notag\\
=&{-\pi}|\epsilon|E_k(\tau_a),\tag{S30}
\end{align}
and for the case $k=0$ and $l\geq 3$ we have
\begin{equation}
I_a(0,l)\Big|_{l\geq 3}={-\pi}|\epsilon|\bar E_l(\bar\tau_a).\tag{S31}
\end{equation}
For $k=2$ and $l=0$ we have
\begin{align}
I_a(2,0)=&\frac{i}{2}\int_{\bar z_0}^{\bar z_0-2\pi i}d\bar z[P_{1}(z+2\pi i\tau_a,\tau_a)-P_{1}(z,\tau_a)]\notag\\
&-\frac{i}{2}\int_{\bar z_0}^{\bar z_0-2\pi i\bar\tau}d\bar z[P_{1}(z+2\pi i,\tau_a)-P_{1}(z,\tau_a)]\notag\\
&+\frac{1}{2}\int_0^{2\pi}d\theta\Big[e^{-i2\theta}+\sum_{m\geq 1}E_m(\tau_a)\sqrt{|\epsilon|}^{m}e^{i(m-2)\theta}\Big]\notag\\
=&-\pi-\pi|\epsilon|E_2(\tau_a).\tag{S32}
\end{align}
For $k=0$ and $l=2$ we have
\begin{equation}
I_a(0,2)=-\pi-\pi|\epsilon|\bar E_2(\bar\tau_a).\tag{S33}
\end{equation}
\subsection{Integral $[J_a(x,\bar x)](k,l)$}\label{B.2}
 The second type of integral is
\begin{equation}
[J_a(x,\bar x)](k,l)=\lim_{\delta\rightarrow 0}\int_{S_a\backslash D_{\delta}}d^2z_aP_{k}(z_a,\tau_a)\bar{\tilde{P}}_l(\bar z_a,\bar x,\bar\tau_a),\tag{S34}
\end{equation}
where $D_{\delta}$ is a small disk of radius $\delta$ centered on the singularity $(x,\bar x)$. $\tilde{P}_l$ is defined as
\begin{equation}
\tilde{P}_l(z,x,\tau)=
\begin{cases}
P_1(z-x,\tau)-P_1(z,\tau)&(l=1), \\
P_l(z-x,\tau)&(l\geq2).\tag{S35}
\end{cases}
\end{equation}
This integral depends on the coordinate of the singularity $(x,\bar x)$, and from the recursion eq.(\ref{recursion}) and Laurent expansion eq.(\ref{Laurent expansion}) we have 
\begin{align}
\partial_x[J_a(x,\bar x)](k,l)=&\lim_{\delta\rightarrow 0}\int_{S_a\backslash D_{\delta}}d^2zP_{k}(z,\tau_a)\partial_{x}\Big(\frac{1}{(l-1)!}\partial_{\bar x}^{l-1}\frac{1}{\bar z-\bar x}\Big)\notag\\
=&-\frac{\pi}{(l-1)!}\partial_{\bar x}^{l-1}P_{k}(x,\tau_a)\notag\\
=&-\pi P_{k}(x,\tau_a)\delta_{l,1},\tag{S36}\\
\partial_{\bar x}[J_a(x,\bar x)](k,l)=&\lim_{\delta\rightarrow 0}\int_{S_a\backslash D_{\delta}}d^2zP_{k}(z,\tau_a)\partial_{\bar x}\bar P_l(\bar z-\bar x,\bar\tau_a)\notag\\
=&l[J_a(x,\bar x)](k,l+1),\tag{S37} \label{properties J}
\end{align}
where in the last step we restrict the insertion point $x\neq 0$.\par
For $k\geq 3$ and $l=1$, we construct $F^z=\frac{-1}{k-1}P_{k-1}(z,\tau_a)\big[\bar{P}_1(\bar{z}-\bar{x},\bar{\tau}_a)-\bar{P}_1(\bar{z},\bar{\tau}_a)\big]$. Since $x\neq 0$, $\bar{P}_1(\bar{z}-\bar{x},\bar{\tau}_a)$ is regular around the origin and has Taylor expansion
\begin{equation}
\bar{P}_1(\bar{z}-\bar{x},\bar{\tau}_a)=-\sum_{l\geq1}\bar{P}_l(\bar{x},\bar{\tau}_a)\bar{z}^{l-1}.\tag{S38}
\end{equation}
Using eq.(\ref{stoke's theorem 2}) we have
\begin{align}
[J_a(x,\bar x)](k,1)\Big|_{k\geq 3}=&-\frac{i}{2(k-1)}\oint_{A}d\bar{z}P_{k-1}(z,\tau_a)\Big[\sum_{l\geq1}\bar{P}_l(\bar{x},\bar{\tau}_a)\bar{z}^{l-1}+\bar{P}_1(\bar{z},\bar{\tau}_a)\Big]\notag\\
&+\frac{i}{2(k-1)}P_{k-1}(x,\tau_a)\oint_{\partial D_{\delta}}d\bar{z}\bar{P}_1(\bar{z}-\bar x,\bar{\tau}_a)\Big|_{\delta\rightarrow0}\notag\\
=&\frac{\pi}{k-1}\Big[P'_{k-1}(x,\tau_a)-\sum_{l\geq1}\frac{k-1}{l}|\epsilon|^lA_{k-3,a}^{l,0}\bar P'_l(\bar x,\bar\tau_a)\Big], \tag{S39}\label{J (k,1)}
\end{align}
where the coefficient $A_{k,a}^{l,n}$ is defined in eq.(\ref{coefficient A}), and $P'_k(x,\tau)=P_k(x,\tau)-E_k(\tau)$. For $k=2$ and $l=1$ we have
\begin{align}
[J_a(x,\bar x)](2,1)=&-\frac{i}{2}\oint_{\partial T^2}d\bar{z}P_{1}(z,\tau_a)\Big[\bar{P}_1(\bar{z}-\bar{x},\bar{\tau}_a)-\bar{P}_1(\bar{z},\bar{\tau}_a)\Big]\notag\\
&-\frac{i}{2}\oint_{A}d\bar{z}P_{1}(z,\tau_a)\Big[\sum_{l\geq1}\bar{P}_l(\bar{x},\bar{\tau}_a)\bar{z}^{l-1}+\bar{P}_1(\bar{z},\bar{\tau}_a)\Big]\notag\\
&+\frac{i}{2}P_{1}(x,\tau_a)\oint_{\partial D_{\delta}}d\bar{z}\bar{P}_1(\bar{z}-\bar x,\bar{\tau}_a)\Big|_{\delta\rightarrow0}\notag\\
=&\pi P_{1}(x,\tau_a)+\pi\sum_{l\geq1}|\epsilon|^lE_{l+1}(\tau_a)\bar P'_l(\bar x,\bar\tau_a),\tag{S40}
\end{align}
where the integral over $\partial T^2$ is calculated using the quasi-periodic property of $Q(z,\tau_a)$ in eq.(\ref{properties Q}). For $k=0$ and $l=1$, we construct $F^z=z\big[\bar{P}_1(\bar{z}-\bar{x},\bar{\tau}_a)-\bar{P}_1(\bar{z},\bar{\tau}_a)\big]$ and have
\begin{align}
[J_a(x,\bar x)](0,1)=&\frac{i}{2}\oint_{\partial T^2}d\bar{z}z\Big[\bar{P}_1(\bar{z}-\bar{x},\bar{\tau}_a)-\bar{P}_1(\bar{z},\bar{\tau}_a)\Big]\notag\\
&+\frac{i}{2}\oint_{A}d\bar{z}z\Big[\sum_{l\geq1}\bar{P}_l(\bar{x},\bar{\tau}_a)\bar{z}^{l-1}+\bar{P}_1(\bar{z},\bar{\tau}_a)\Big]\notag\\
&-\frac{i}{2}x\oint_{\partial D_{\delta}}d\bar{z}\bar{P}_1(\bar{z}-\bar x,\bar{\tau}_a)\Big|_{\delta\rightarrow0}\notag\\
=&-2\pi\text{Re}[x]+\pi|\epsilon|\bar P_1(\bar x,\bar \tau_a).\tag{S41}
\end{align}\par
For $k\geq 3$ and $l=2$, we construct $F^z=\frac{-1}{k-1}P_{k-1}(z,\tau_a)\bar{P}_2(\bar{z}-\bar{x},\bar{\tau}_a)$ and have
\begin{align}
[J_a(x,\bar x)](k,2)\Big|_{k\geq 3}=&\frac{i}{2(k-1)}\oint_{A}d\bar{z}\sum_{l\geq1}l\bar{P}_{l+1}(\bar{x},\bar{\tau}_a)P_{k-1}(z,\tau_a)\bar{z}^{l-1}\notag\\
&+\frac{i}{2(k-1)}P_{k-1}(x,\tau_a)\oint_{\partial D_{\delta}}d\bar{z}\bar{P}_2(\bar{z}-\bar x,\bar{\tau}_a)\Big|_{\delta\rightarrow0}\notag\\
=&\pi\sum_{l\geq1}|\epsilon|^lA_{k-3,a}^{l,0}\bar P_{l+1}(\bar x,\bar\tau_a).\tag{S42}
\end{align}
For $k=2$ and $l=2$ we have
\begin{align}
[J_a(x,\bar x)](2,2)=&-\frac{i}{2}\oint_{\partial T^2}d\bar{z}P_{1}(z,\tau_a)\bar{P}_2(\bar{z}-\bar{x},\bar{\tau}_a)\notag\\
&+\frac{i}{2(k-1)}\oint_{A}d\bar{z}\sum_{l\geq1}l\bar{P}_{l+1}(\bar{x},\bar{\tau}_a)P_{1}(z,\tau_a)\bar{z}^{l-1}\notag\\
&+\frac{i}{2(k-1)}P_{1}(x,\tau_a)\oint_{\partial D_{\delta}}d\bar{z}\bar{P}_2(\bar{z}-\bar x,\bar{\tau}_a)\Big|_{\delta\rightarrow0}\notag\\
=&-\pi\sum_{l\geq1}|\epsilon|^lE_{l+1}(\tau_a)l\bar{P}_{l+1}(\bar{x},\bar{\tau}_a).\tag{S43}
\end{align}
For $k=0$ and $l=2$, we construct $F^z=z\bar{P}_2(\bar{z}-\bar{x},\bar{\tau}_a)$ and have
\begin{align}
[J_a(x,\bar x)](0,2)=&\frac{i}{2}\oint_{\partial T^2}d\bar{z}z\bar{P}_2(\bar{z}-\bar{x},\bar{\tau}_a)\notag\\
&-\frac{i}{2}\oint_{A}d\bar{z}\sum_{l\geq1}l\bar{P}_{l+1}(\bar{x},\bar{\tau}_a)z\bar{z}^{l-1}\notag\\
&-\frac{i}{2}x\oint_{\partial D_{\delta}}d\bar{z}\bar{P}_2(\bar{z}-\bar x,\bar{\tau}_a)\Big|_{\delta\rightarrow0}\notag\\
=&-\pi-\pi|\epsilon|\bar P_{2}(\bar x,\bar \tau_a).\tag{S44}
\end{align}
These results can also be obtained by using eq.(\ref{properties J}).\par
For $k\geq 3$ and $l=4$, we construct $F^z=\frac{-1}{k-1}P_{k-1}(z,\tau_a)\bar{P}_4(\bar{z}-\bar{x},\bar{\tau}_a)$ and obtain
\begin{equation}
[J_a(x,\bar x)](k,4)=\pi\sum_{l\geq1}|\epsilon|^lA_{k-3,a}^{l,0}\frac{l^2+3l+2}{6}\bar{P}_{l+3}(\bar x,\bar\tau_a).\tag{S45}
\end{equation}
For $k=2$ and $l=4$ we have
\begin{equation}
[J_a(x,\bar x)](2,4)=-\pi\sum_{l\geq1}|\epsilon|^lE_{l+1}(\tau_a)\frac{l^3+3l^2+2l}{6}\bar{P}_{l+3}(\bar x,\bar\tau_a).\tag{S46}
\end{equation}
For $k=0$ and $l=4$, we construct $F^z=z\bar{P}_4(\bar{z}-\bar{x},\bar{\tau}_a)$ and obtain
\begin{equation}
[J_a(x,\bar x)](0,4)=-\pi|\epsilon|\bar P_{4}(\bar x,\bar \tau_a).	\tag{S47}
\end{equation}
\subsection{Integral $[K_a(x_1,\bar x_1;x_2,\bar x_2)](k,l)$}\label{B.3}
The last type of integral is 
\begin{equation}
[K_a(x_1,\bar x_1;x_2,\bar x_2)](k,l)=\lim_{\delta_1,\delta_2\rightarrow 0}\int_{S_a\backslash(D_{\delta_1}\cup D_{\delta_2})}d^2z_a\tilde P_k(z_a,x_1,\tau_a)\bar{\tilde{P}}_l(\bar z_a,\bar x_2,\bar\tau_a)\tag{S48}
\end{equation}
for $k,l\geq 1$. $D_{\delta_1}$ and $D_{\delta_2}$ are small disks centered on the singularities. This integral appears in the first order deformation of the two-point function (for the case where $x_1$ and $x_2$ are inserted in the same torus.) In particular, if we fix $x_1$ and $x_2$ at the same point $x$, the corresponding integral is
\begin{equation}
[K_a(x,\bar x)](k,l)=\lim_{\delta\rightarrow 0}\int_{S_a\backslash D_\delta}d^2z_a\tilde P_k(z_a,x,\tau_a)\bar{\tilde{P}}_l(\bar z_a,\bar x,\bar\tau_a),\tag{S49}
\end{equation}
which appears in the first-order deformation of one-point function. Using the definition of $\tilde P_l(z,x,\tau)$, the recursion eq.(\ref{recursion}) and the Laurent expansion eq.(\ref{Laurent expansion}), we obtain some properties of $K_a(x,\bar x)$:
\begin{align}
\partial_x[K_a(x,\bar x)](k,l)=&\lim_{\delta\rightarrow 0}\int_{S_a\backslash D_\delta}d^2z\Big[\partial_x P_k(z-x,\tau_a)\bar{\tilde{P}}_l(\bar z,\bar x,\bar\tau_a)+\tilde P_k(z_a,x,\tau_a)\partial_x \bar P_l(\bar z-\bar x,\bar\tau_a)\Big]\notag\\
=&k[K_a(x,\bar x)](k+1,l)+\frac{-\pi}{(l-1)!}\int_{S_a}d^2z\partial_{\bar z}^{l-1}\tilde P_k(z,x,\tau_a)\delta^{(2)}(z-x)\notag\\
=&k[K_a(x,\bar x)](k+1,l)+\pi P_1(x,\tau_a)\delta_{k,1}\delta_{l,1}\notag\\
&+\frac{(-1)^{k-1}\pi^2}{(k-1)!(l-1)!}\int_{S_a}d^2z\delta^{(2)}(z-x)\partial_z^{k-1}\partial_{\bar z}^{l-2}\delta^{(2)}(z-x),\tag{S50}\\
\partial_{\bar x}[K_a(x,\bar x)](k,l)
=&l[K_a(x,\bar x)](k,l+1)+\pi\bar P_1(\bar x,\bar\tau_a)\delta_{k,1}\delta_{l,1}\notag\\
&+\frac{(-1)^{l-1}\pi^2}{(k-1)!(l-1)!}\int_{S_a}d^2z\delta^{(2)}(z-x)\partial_z^{k-2}\partial_{\bar z}^{l-1}\delta^{(2)}(z-x). \tag{S51}\label{properties K 1}
\end{align}
The last term in each of these equations is purely divergent, and we discard it to regularize the integral. We have similar properties for $K_a(x_1,\bar x_1;x_2,\bar x_2)$:
\begin{align}
\partial_{x_1}[K_a(x_1,\bar x_1;x_2,\bar x_2)](k,l)=&k[K_a(x_1,\bar x_1;x_2,\bar x_2)](k+1,l),\tag{S52}\\
\partial_{\bar x_1}[K_a(x_1,\bar x_1;x_2,\bar x_2)](k,l)=&-\pi\bar{\tilde{P}}_l(\bar x_1,\bar x_2,\bar\tau_a)\delta_{k,1},\tag{S53}\\
\partial_{x_2}[K_a(x_1,\bar x_1;x_2,\bar x_2)](k,l)=&-\pi\tilde P_k(x_2,x_1,\tau_a)\delta_{l,1},\tag{S54}\\
\partial_{\bar x_2}[K_a(x_1,\bar x_1;x_2,\bar x_2)](k,l)=&l[K_a(x_1,\bar x_1;x_2,\bar x_2)](k,l+1),\tag{S55}
\end{align}
where we restrict the insertion points $x_1,x_2\neq 0$ and $x_1\neq x_2$.\par
First we calculate the integral $[K_a(x,\bar x)](k,l)$, which appears in the deformed one-point function of primary field. For $k=2$ and $l=2$ we have
\begin{align}
[K_a(x,\bar x)](2,2)=&-\frac{i}{2}\oint_{\partial\hat T^2}d\bar zP_1(z-x,\tau_a)\bar P_2(\bar z-\bar x,\bar\tau_a)\notag\\
&-\frac{i}{2}\oint_{A}d\bar z\sum_{k,l\geq 1}lP_k(x,\tau_a)\bar P_{l+1}(\bar x,\bar\tau_a)z^{k-1}\bar z^{l-1}\notag\\
&+\frac{i}{2}\oint_{\partial D_{\delta}}d\bar zP_1(z-x,\tau_a)\bar P_2(\bar z-\bar x,\bar\tau_a)\Big|_{\delta\rightarrow0}\notag\\
=&-\pi\sum_{k\geq 1}|\epsilon|^kk|P_{k+1}(x,\tau_a)|^2+\pi\frac{1}{\delta}\Big|_{\delta\rightarrow0},\tag{S56}
\end{align}
and we can simply discard the last term. For $k=2$ and $l=1$ we have
\begin{align}
[K_a(x,\bar x)](2,1)=&-\frac{i}{2}\oint_{\partial\hat T^2}d\bar zP_1(z-x,\tau_a)\Big[\bar P_1(\bar z-\bar x,\bar\tau_a)-\bar P_1(\bar z,\bar \tau_a)\Big]\notag\\
&+\frac{i}{2}\oint_{A}d\bar z\sum_{k\geq 1}P_k(x,\tau_a)z^{k-1}\Big[\sum_{l\geq 1}\bar P_{l+1}(\bar x,\bar\tau_a)\bar z^{l-1}+\bar P_1(\bar z,\bar\tau_a)\Big]\notag\\
&+\frac{i}{2}\oint_{\partial D_{\delta}}d\bar zP_1(z-x,\tau_a)\Big[\bar P_1(\bar z-\bar x,\bar\tau_a)-\bar P_1(\bar x,\bar \tau_a)\Big]\Big|_{\delta\rightarrow0}\notag\\
=&\pi P_1(x,\tau_a)+\pi\sum_{k\geq 1}|\epsilon|^kP_{k+1}(x,\tau_a)\bar P'_k(\bar x,\bar\tau_a),\tag{S57}
\end{align}
and we immediately obtain the case for $k=1$ and $l=2$:
\begin{align}
[K_a(x,\bar x)](1,2)=&\overline{[K_a(x,\bar x)](2,1)}\notag\\
=&\pi\bar P_1(\bar x,\bar\tau_a)+\pi\sum_{k\geq 1}|\epsilon|^kP'_{k}(x,\tau_a)\bar P_{k+1}(\bar x,\bar\tau_a).\tag{S58}
\end{align}
We know from eq.(\ref{properties K 1}) that $[K_a(x,\bar x)](1,1)$ satisfies
\begin{align}
\partial_x[K_a(x,\bar x)](1,1)=&[K_a(x,\bar x)](2,1)+\pi P_1(x,\tau_a)-\pi\int_{S_a}d^2z\delta^{(2)}(z-x)\frac{1}{z-x}\notag\\
=&2\pi P_1(x,\tau_a)+\pi\sum_{k\geq 1}|\epsilon|^kP_{k+1}(x,\tau_a)\bar P'_k(\bar x,\bar\tau_a)+A,\tag{S59}\\
\partial_{\bar x}[K_a(x,\bar x)](1,1)=&[K_a(x,\bar x)](1,2)-2\pi i\bar P_1(\bar x,\bar\tau_a)-\pi\int_{S_a}d^2z\delta^{(2)}(z-x)\frac{1}{\bar z-\bar x}\notag\\
=&2\pi\bar P_1(\bar x,\bar\tau_a)+\pi\sum_{k\geq 1}|\epsilon|^kP'_{k}(x,\tau_a)\bar P_{k+1}(\bar x,\bar\tau_a)+A,\tag{S60}
\end{align}
where $A$ is purely divergent term. We discard $A$ and integrate them to obtain the regularized $[K_a(x,\bar x)](1,1)$:
\begin{align}
[K_a(x,\bar x)](1,1)=2\pi\log|Q(x,\tau_a)|^2-\pi\sum_{k\geq 1}|\epsilon|^k\frac{1}{k}|P'_{k}(x,\tau_a)|^2.\tag{S61}
\end{align}\par
Then we calculate the integral $[K_a(x_1,\bar x_1;x_2,\bar x_2)](k,l)$, which appears in the deformed two-point function. For $k=2$ and $l=2$ we have
\begin{align}
[K_a(x_1,\bar x_1;x_2,\bar x_2)](2,2)=&-\frac{i}{2}\oint_{\partial\hat T^2}d\bar zP_1(z-x_1,\tau_a)\bar P_2(\bar z-\bar x_2,\bar\tau_a)\notag\\
&-\frac{i}{2}\oint_{A}d\bar z\sum_{k,l\geq 1}lP_k(x_1,\tau_a)\bar P_{l+1}(\bar x_2,\bar\tau_a)z^{k-1}\bar z^{l-1}\notag\\
&+\frac{i}{2}\bar P_2(\bar x_1-\bar x_2,\bar\tau_a)\oint_{\partial D_{\delta_1}}d\bar zP_1(z-x_1,\tau_a)\Big|_{\delta_1\rightarrow0}\notag\\
&+\frac{i}{2}P_1(x_2-x_1,\tau_a)\oint_{\partial D_{\delta_2}}d\bar z\bar P_2(\bar z-\bar x_2,\bar\tau_a)\Big|_{\delta_2\rightarrow0}\notag\\
=&-\pi\sum_{k\geq 1}|\epsilon|^kkP_{k+1}(x_1,\tau_a)\bar P_{k+1}(\bar x_2,\bar\tau_a).\tag{S62}
\end{align}
For $k=2$ and $l=1$ we have
\begin{align}
[K_a(x_1,\bar x_1;x_2,\bar x_2)](2,1)=&\pi P_1(x_1,\tau_a)+\pi P_1(x_2-x_1,\tau_a)\notag\\
&+\pi\sum_{k\geq 1}|\epsilon|^kP_{k+1}(x_1,\tau_a)\bar P'_k(\bar x_2,\bar\tau_a).\tag{S63}
\end{align}
For $k=1$ and $l=2$ we have
\begin{align}
[K_a(x_1,\bar x_1;x_2,\bar x_2)](1,2)=&\pi\bar P_1(\bar x_2,\bar\tau_a)+\pi\bar P_1(\bar x_1-\bar x_2,\bar\tau_a)\notag\\
&+\pi i\sum_{k\geq 1}|\epsilon|^kP'_k(x_1,\tau_a)\bar P_{k+1}(\bar x_2,\bar\tau_a).\tag{S64}
\end{align}
For $[K_a(x_1,\bar x_1;x_2,\bar x_2)](1,1)$ it satisfies
\begin{align}
\partial_{x_1}[K_a(x_1,\bar x_1;x_2,\bar x_2)](1,1)=&\pi P_1(x_1,\tau_a)+\pi P_1(x_2-x_1,\tau_a)\notag\\
&+\pi\sum_{k\geq 1}|\epsilon|^kP_{k+1}(x_1,\tau_a)\bar P'_k(\bar x_2,\bar\tau_a),\tag{S65}\\
\partial_{\bar x_1}[K_a(x_1,\bar x_1;x_2,\bar x_2)](1,1)=&-\pi\Big[\bar P_1(\bar x_1-\bar x_2,\bar\tau_a)-\bar P_1(\bar x_1,\bar\tau_a)\Big],\tag{S66}\\
\partial_{x_2}[K_a(x_1,\bar x_1;x_2,\bar x_2)](1,1)=&-\pi\Big[P_1(x_2-x_1,\tau_a)-P_1(x_2,\tau_a)\Big],\tag{S67}\\
\partial_{\bar x_2}[K_a(x_1,\bar x_1;x_2,\bar x_2)](1,1)=&\pi\bar P_1(\bar x_2,\bar\tau_a)+\pi\bar P_1(\bar x_1-\bar x_2,\bar\tau_a)\notag\\
&+\pi\sum_{k\geq 1}|\epsilon|^kP'_k(x_1,\tau_a)\bar P_{k+1}(\bar x_2,\bar\tau_a).\tag{S68}
\end{align}
We integrate these equations to obtain $[K_a(x_1,\bar x_1;x_2,\bar x_2)](1,1)$:
\begin{align}
[K_a(x_1,\bar x_1;x_2,\bar x_2)](1,1)=&\pi\log\frac{|Q(x_1,\tau_a)Q(x_2,\tau_a)|^2}{|Q(x_1-x_2,\tau_a)|^2}\notag\\
&-\pi\sum_{k\geq 1}|\epsilon|^k\frac{1}{k}P'_{k}(x_1,\tau_a)\bar P'_k(\bar x_2,\bar\tau_a).\tag{S69}
\end{align}
For $k=4$ and $l=1$ we have
\begin{align}
[K_a(x_1,\bar x_1;x_2,\bar x_2)](4,1)=&\frac{\pi}{3}P_3(x_1,\tau_a)+\frac{\pi}{3}P_3(x_2-x_1,\tau_a)\notag\\
&+\frac{\pi}{3}\sum_{k\geq 1}|\epsilon|^k\frac{k^2+3k+2}{2}P_{k+3}(x_1,\tau_a)\bar P'_k(\bar x_2,\bar\tau_a).\tag{S70}
\end{align}
For $k=1$ and $l=4$ we have 
\begin{align}
[K_a(x_1,\bar x_1;x_2,\bar x_2)](1,4)=&\frac{\pi}{3}\bar P_3(\bar x_2,\bar\tau_a)+\frac{\pi}{3}\bar P_3(\bar x_1-\bar x_2,\bar\tau_a)\notag\\
&+\frac{\pi}{3}\sum_{k\geq 1}|\epsilon|^k\frac{k^2+3k+2}{2}P'_{k}(x_1,\tau_a)\bar P_{k+3}(\bar x_2,\bar\tau_a).\tag{S71}
\end{align}
For $k=4$ and $l=2$ we have
\begin{align}
[K_a(x_1,\bar x_1;x_2,\bar x_2)](4,2)=&-\frac{\pi}{3}\sum_{k\geq 1}|\epsilon|^k\frac{k^3+3k^2+2k}{2}P_{k+3}(x_1,\tau_a)\bar P_{k+1}(\bar x_2,\bar\tau_a).	\tag{S72}
\end{align}
For $k=2$ and $l=4$ we have
\begin{align}
[K_a(x_1,\bar x_1;x_2,\bar x_2)](2,4)=&-\frac{\pi}{3}\sum_{k\geq 1}|\epsilon|^k\frac{k^3+3k^2+2k}{2}P_{k+1}(x_1,\tau_a)\bar P_{k+3}(\bar x_2,\bar\tau_a).	\tag{S73}
\end{align}
For $k=4$ and $l=4$ we have 
\begin{align}
[K_a(x_1,\bar x_1;x_2,\bar x_2)](4,4)=&-\frac{\pi}{3}\sum_{k\geq 1}|\epsilon|^k\frac{k(k+1)^2(k+2)^2}{12}P_{k+3}(x_1,\tau_a)\bar P_{k+3}(\bar x_2,\bar\tau_a).\tag{S74}
\end{align}
\section{Complete integral results}\label{C}
In this appendix, we demonstrate the complete integral results that appear in the first-order deformation of the partition function and the correlation function. These integrals can be classified into three types, e.g. $\mathcal{F}\bar{\mathcal{F}}$-type, $\mathcal{F}\bar{\mathcal{P}}$-type (or $\mathcal{P}\bar{\mathcal{F}}$-type), and $\mathcal{P}\bar{\mathcal{P}}$-type. The $\mathcal{F}\bar{\mathcal{F}}$-type appears in zero-point contribution $\int d^2zD_z\bar D_{\bar z}\langle{X}\rangle$.
The $\mathcal{F}\bar{\mathcal{P}}$-type (or $\mathcal{P}\bar{\mathcal{F}}$-type) appears in one-point contribution $\int d^2zD_z\bar{\mathcal{P}}_{\bar z,\bar x_i}\langle{X}\rangle$.
The $\mathcal{P}\bar{\mathcal{P}}$-type appears in two-point contribution $\int d^2z\mathcal{P}_{z,x_i}\bar{\mathcal{P}}_{\bar z,\bar x_j}\langle{X}\rangle$.
All of these integrals can be represented by $I_a$, $J_a(x,\bar x)$ and $K_a(x_1,\bar x_1;x_2,\bar x_2)$ in Appendix \ref{B}, as well as column vectors $\alpha_a$, $\beta_a$, $\theta_a$, $\xi_a(x)$ and $\zeta_a(x)$ defined in Subsection \ref{2.2}.
As an example, consider the following $\mathcal{P}\bar{\mathcal{P}}$-type integral:
\begin{align}
&\int_{\Sigma^{(2)}}d^2z\big[{}^2{\mathcal{P}}_1(z,x_1;\tau_a,\tau_{\bar a},\epsilon){}^2{\bar{\mathcal{P}}}_2(\bar z,\bar x_2;\bar\tau_a,\bar\tau_{\bar a},\bar\epsilon)\big]\notag\\
=&\int_{S_a}d^2z\bigg\lbrace\Big[\tilde P_1(z,x_1,\tau_a)+\mathop{\sum}\limits_{k\geq 1} P_{k+3}(z,\tau_a)[\xi^{(0)}_a(x_1)](k)\Big]\notag\\
&\times\Big[\bar{\tilde P}_2(\bar{z},\bar{x}_2,\bar{\tau}_a)+\mathop{\sum}\limits_{l\geq 1}\bar{P}_{l+3}(\bar{z},\bar{\tau}_a)[\bar{\xi}^{(1)}_a(\bar{x}_2)](l)\Big]\bigg\rbrace\notag\\
&+\int_{S_{\bar a}}d^2z\bigg\lbrace\Big[-\epsilon P_3(z,\tau_{\bar{a}})+\mathop{\sum}\limits_{k\geq 1}P_{k+3}(z,\tau_{\bar{a}})[\zeta^{(0)}_{\bar{a}}(x_1)](k)\Big]\Big[\mathop{\sum}\limits_{l\geq 1}\bar{P}_{l+3}(\bar{z},\bar{\tau}_{\bar{a}})[\bar{\zeta}^{(1)}_{\bar{a}}(\bar{x}_2)](l)\Big]\bigg\rbrace\notag\\
=&[K_a(x_1,\bar x_1;x_2,\bar x_2)](1,2)+\sum_{k,l\geq 1}I_a(k+3,l+3)[\xi^{(0)}_a(x_1)](k)[\bar{\xi}^{(1)}_a(\bar{x}_2)](l)\notag\\
&+\sum_{k,l\geq 1}I_{\bar a}(k+3,l+3)[\zeta^{(0)}_{\bar{a}}(x_1)](k)[\bar{\zeta}^{(1)}_{\bar{a}}(\bar{x}_2)](l)+\sum_{k\geq 1}[J_a(x_2,\bar x_2)](k+3,2)[\xi^{(0)}_a(x_1)](k)\notag\\
&+\sum_{k\geq 1}\overline{[J_a(x_1,\bar x_1)](k+3,1)}[\bar{\xi}^{(1)}_a(\bar{x}_2)](k)-\epsilon\sum_{l\geq 1}I_{\bar a}(3,l+3)[\bar{\zeta}^{(1)}_{\bar{a}}(\bar{x}_2)](l).\tag{S75}
\end{align}
\subsection{The full form of three types of integrals}\label{C.1}
The following are the $\mathcal{F}\bar{\mathcal{F}}$-type integrals that appear in the deformed partition function:
\begin{align}
&\int_{\Sigma^{(2)}}d^2z(\sideset{^2}{_a}{\mathop{\mathcal{F}}}\sideset{^2}{_a}{\mathop{\bar{\mathcal{F}}}})\notag\\
=&4\pi^2\text{Im}[\tau_a]-\pi|\epsilon|-2\pi|\epsilon|\sum_{k\geq 1}\text{Re}[E_{k+3}(\tau_a)\alpha_a(k)]\notag\\
&+\sum_{k,l\geq 1}\frac{\pi}{k+2}\Big[\frac{1}{|\epsilon|^{k+2}}\delta_{k,l}-\sum_{n\geq l}C_{n,a}^{k,l}|\epsilon|^{n-l+1}\Big]\alpha_a(k)\bar{\alpha}_a(l)\notag\\
&+\sum_{k,l\geq 1}\frac{\pi}{k+2}\Big[\frac{1}{|\epsilon|^{k+2}}\delta_{k,l}-\sum_{n\geq l}C_{n,{\bar a}}^{k,l}|\epsilon|^{n-l+1}\Big]\beta_{\bar a}(k)\bar{\beta}_{\bar a}(l),\tag{S76}    \label{FF a a}\\
&\int_{\Sigma^{(2)}}d^2z(\sideset{^2}{_a}{\mathop{\mathcal{F}}}\sideset{^2}{_{\bar a}}{\mathop{\bar{\mathcal{F}}}})\notag\\
=&-\pi|\epsilon|\sum_{k\geq 1}\Big[\bar{E}_{k+3}(\bar{\tau}_a)\bar{\beta}_a(k)+E_{k+3}(\tau_{\bar a})\beta_{\bar a}(k)\Big]\notag\\
&+\sum_{k,l\geq 1}\frac{\pi}{k+2}\Big[\frac{1}{|\epsilon|^{k+2}}\delta_{k,l}-\sum_{n\geq l}C_{n,a}^{k,l}|\epsilon|^{n-l+1}\Big]\alpha_a(k)\bar{\beta}_a(l)\notag\\
&+\sum_{k,l\geq 1}\frac{\pi}{k+2}\Big[\frac{1}{|\epsilon|^{k+2}}\delta_{k,l}-\sum_{n\geq l}C_{n,{\bar a}}^{k,l}|\epsilon|^{n-l+1}\Big]\beta_{\bar a}(k)\bar{\alpha}_{\bar a}(l),\tag{S77}    \label{FF a abar}\\
&\int_{\Sigma^{(2)}}d^2z\bar{\epsilon}^{\frac{1}{2}}(\sideset{^2}{_a}{\mathop{\mathcal{F}}}\sideset{^2}{^\Pi}{\mathop{\bar{\mathcal{F}}}})\notag\\
=&\pi\bar\epsilon-\pi|\epsilon|\sum_{l\geq 1}\bar{E}_{l+1}(\bar{\tau}_a)\bar{\theta}_a(l)\notag\\
&-\sum_{n,k\geq 1}\frac{\pi}{k+2}|\epsilon|^{n-1}\bar\epsilon\Big[C_{n-2,a}^{k,-1}\alpha_a(k)+C_{n-2,\bar{a}}^{k,-1}\beta_{\bar a}(k)\Big]\notag\\
&+\sum_{k,l\geq 1}\frac{\pi}{k+2}\Big[\frac{1}{|\epsilon|^{k+2}}\delta_{k,l}-\sum_{n\geq l}C_{n,a}^{k,l}|\epsilon|^{n-l+1}\Big]\alpha_a(k)\bar{\theta}_a(l+2)\notag\\
&+\sum_{k,l\geq 1}\frac{\pi}{k+2}\Big[\frac{1}{|\epsilon|^{k+2}}\delta_{k,l}-\sum_{n\geq l}C_{n,{\bar a}}^{k,l}|\epsilon|^{n-l+1}\Big]\beta_{\bar a}(k)\bar{\theta}_{\bar a}(l+2),\tag{S78}  \label{FF a pi}\\
&\int_{\Sigma^{(2)}}d^2z\epsilon^{\frac{1}{2}}\bar{\epsilon}^{\frac{1}{2}}({}^2\mathcal{F}^{\Pi}{}^2\bar{\mathcal{F}}^{\Pi})\notag\\
=&2\pi|\epsilon|-\pi|\epsilon|^3\sum_{a=1,2}\sum_{n\geq 1}|\epsilon|^{n-1}n|E_{n+1}(\tau_a)|^2\notag\\
&-\sum_{a=1,2}\sum_{n,k\geq 1}\frac{2\pi}{k+2}|\epsilon|^n\text{Re}[C_{n-2,a}^{k,-1}\bar\epsilon{\theta}_a(k+2)]\notag\\
&+\sum_{a=1,2}\sum_{k,l\geq 1}\frac{\pi}{k+2}\Big[\frac{1}{|\epsilon|^{k+2}}\delta_{k,l}-\sum_{n\geq l}C_{n,a}^{k,l}|\epsilon|^{n-l+1}\Big]\theta_a(k+2)\bar{\theta}_a(l+2).\tag{S79}    \label{FF pi pi}
\end{align}
The following are the $\mathcal{F}\bar{\mathcal{P}}$-type integrals that appear in the deformed one-point functions of primary field and of stress tensor:
\begin{align}
&\int_{\Sigma^{(2)}\backslash{D_{\delta}}}d^2z(\sideset{^2}{_a}{\mathop{\mathcal{F}}}{}^2\bar{\mathcal{P}}_1)\notag\\
=&-2\pi\text{Re}[x]+\pi|\epsilon|\bar P_1(\bar x,\bar \tau_a)\notag\\
&+\bar{\epsilon}\sum_{k,n\geq 1}\frac{\pi}{k+2}C_{k,\bar a}^{n-1,0}|\epsilon|^{n}\beta_{\bar{a}}(k)+2\pi i|\epsilon|\sum_{l\geq 1}\bar E_{l+3}(\bar\tau_a)[\bar{\xi}^{(0)}_a(\bar{x})](l)\notag\\
&+\sum_{k\geq 1}\frac{\pi}{k+2}\Big[P'_{k+2}(x,\tau_a)-\sum_{l\geq1}\frac{k+2}{l}|\epsilon|^lA_{k,a}^{l,0}\bar P'_l(\bar x,\bar\tau_a)\Big]\alpha_a(k)\notag\\
&+\sum_{k,l\geq 1}\frac{\pi}{k+2}\Big[\frac{1}{|\epsilon|^{k+2}}\delta_{k,l}-\sum_{n\geq l}C_{k,a}^{n,l}|\epsilon|^{n-l+1}\Big]\alpha_a(k)[\bar{\xi}^{(0)}_a(\bar{x})](l)\notag\\
&+\sum_{k,l\geq 1}\frac{\pi}{k+2}\Big[\frac{1}{|\epsilon|^{k+2}}\delta_{k,l}-\sum_{n\geq l}C_{k,\bar a}^{n,l}|\epsilon|^{n-l+1}\Big]\beta_{\bar{a}}(k)[\bar{\zeta}^{(0)}_{\bar{a}}(\bar{x})](l),    \tag{S80} \label{FP a 1}\\
&\int_{\Sigma^{(2)}\backslash{D_{\delta}}}d^2z(\sideset{^2}{_{\bar a}}{\mathop{\mathcal{F}}}{}^2\bar{\mathcal{P}}_1)\notag\\
=&\bar{\epsilon}\sum_{k,n\geq 1}\frac{\pi}{k+2}C_{k,\bar a}^{n-1,0}|\epsilon|^{n}\alpha_{\bar{a}}(k)+2\pi i|\epsilon|\sum_{l\geq 1}\bar E_{l+3}(\bar\tau_{\bar a})[\bar{\zeta}^{(0)}_{\bar a}(\bar{x})](l)\notag\\
&+\sum_{k\geq 1}\frac{\pi}{k+2}\Big[P'_{k+2}(x,\tau_a)-\sum_{l\geq1}\frac{k+2}{l}|\epsilon|^lA_{k,a}^{l,0}\bar P'_l(\bar x,\bar\tau_a)\Big]\beta_a(k)\notag\\
&+\sum_{k,l\geq 1}\frac{\pi}{k+2}\Big[\frac{1}{|\epsilon|^{k+2}}\delta_{k,l}-\sum_{n\geq l}C_{k,a}^{n,l}|\epsilon|^{n-l+1}\Big]\beta_a(k)[\bar{\xi}^{(0)}_a(\bar{x})](l)\notag\\
&+\sum_{k,l\geq 1}\frac{\pi}{k+2}\Big[\frac{1}{|\epsilon|^{k+2}}\delta_{k,l}-\sum_{n\geq l}C_{k,\bar a}^{n,l}|\epsilon|^{n-l+1}\Big]\alpha_{\bar{a}}(k)[\bar{\zeta}^{(0)}_{\bar{a}}(\bar{x})](l),   \tag{S81} \label{FP abar 1}\\
&\int_{\Sigma^{(2)}\backslash{D_{\delta}}}d^2z(\sideset{^2}{^\Pi}{\mathop{\mathcal{F}}}\epsilon^{\frac{1}{2}}{}^2\bar{\mathcal{P}}_1)\notag\\
=&\pi\epsilon P_{1}(x,\tau_a)-2\pi i\epsilon\sum_{l\geq1}\bar{P}'_l(\bar{x},\bar{\tau}_a)E_{l+1}(\tau_a)|\epsilon|^l+\pi i\bar\epsilon\sum_{n\geq k\geq1}\bar C_{0,a}^{n,k}|\epsilon|^{n-k+1}\theta_{\bar a}(k+2)\notag\\
&+\sum_{k\geq 1}\frac{\pi}{k+2}\Big[P'_{k+2}(x,\tau_a)-\sum_{l\geq1}\frac{k+2}{l}|\epsilon|^lA_{k,a}^{l,0}\bar P'_l(\bar x,\bar\tau_a)\Big]\theta_a(k+2)\notag\\
&+\sum_{k,l\geq 1}\frac{\pi}{l+2}\Big[\frac{1}{|\epsilon|^{l+2}}\delta_{k,l}+\sum_{n\geq k}\bar C_{l,a}^{n,k}|\epsilon|^{n-k+1}\Big]\theta_a(k+2)[\bar{\xi}^{(0)}_a(\bar{x})](l)\notag\\
&+\sum_{k,l\geq 1}\frac{\pi}{l+2}\Big[\frac{1}{|\epsilon|^{l+2}}\delta_{k,l}+\sum_{n\geq k}\bar C_{l,\bar a}^{n,k}|\epsilon|^{n-k+1}\Big]\theta_{\bar{a}}(k+2)[\bar{\zeta}^{(0)}_{\bar{a}}(\bar{x})](l)\notag\\
&+\sum_{n,l\geq 1}\frac{\pi}{l+2}\epsilon|\epsilon|^{n-1}\Big[\bar C_{l,a}^{n-2,-1}[\bar{\xi}^{(0)}_a(\bar{x})](l)+\bar C_{l,\bar a}^{n-2,-1}[\bar{\zeta}^{(0)}_{\bar{a}}(\bar{x})](l)\Big],\tag{S82}   \label{FP pi 1}\\
&\int_{\Sigma^{(2)}\backslash{D_{\delta}}}d^2z(\sideset{^2}{_a}{\mathop{\mathcal{F}}}{}^2\bar{\mathcal{P}}_2)\notag\\
=&-\pi-\pi|\epsilon|\sum_{l}\bar E_{l+3}(\bar\tau_a)[\bar{\xi}^{(1)}_a(\bar{x})](l)\notag\\
&-\pi\bar P_{2}(\bar x,\bar \tau_a)|\epsilon|+\pi\sum_{k,l\geq 1}|\epsilon|^lA_{k,a}^{l,0}\bar P_{l+1}(\bar x,\bar\tau_a)\alpha_a(k)\notag\\
&+\sum_{k,l\geq 1}\frac{\pi}{k+2}\Big[\frac{1}{|\epsilon|^{k+2}}\delta_{k,l}-\sum_{n\geq l}C_{k,a}^{n,l}|\epsilon|^{n-l+1}\Big]\alpha_a(k)[\bar{\xi}^{(1)}_a(\bar{x})](l)\notag\\
&+\sum_{k,l\geq 1}\frac{\pi}{k+2}\Big[\frac{1}{|\epsilon|^{k+2}}\delta_{k,l}-\sum_{n\geq l}C_{k,\bar a}^{n,l}|\epsilon|^{n-l+1}\Big]\beta_{\bar{a}}(k)[\bar{\zeta}^{(1)}_{\bar{a}}(\bar{x})](l), \tag{S83}  \label{FP a 2}\\
&\int_{\Sigma^{(2)}\backslash{D_{\delta}}}d^2z(\sideset{^2}{_{\bar a}}{\mathop{\mathcal{F}}}{}^2\bar{\mathcal{P}}_2)\notag\\
=&-\pi|\epsilon|\sum_{l}\bar E_{l+3}(\bar\tau_a)[\bar{\zeta}^{(1)}_{\bar a}(\bar{x})](l)+\pi\sum_{k,l\geq 1}|\epsilon|^lA_{k,a}^{l,0}\bar P_{l+1}(\bar x,\bar\tau_a)\beta_a(k)\notag\\
&+\sum_{k,l\geq 1}\frac{\pi}{k+2}\Big[\frac{1}{|\epsilon|^{k+2}}\delta_{k,l}-\sum_{n\geq l}C_{k,a}^{n,l}|\epsilon|^{n-l+1}\Big]\beta_a(k)[\bar{\xi}^{(1)}_a(\bar{x})](l)\notag\\
&+\sum_{k,l\geq 1}\frac{\pi}{k+2}\Big[\frac{1}{|\epsilon|^{k+2}}\delta_{k,l}-\sum_{n\geq l}C_{k,\bar a}^{n,l}|\epsilon|^{n-l+1}\Big]\alpha_{\bar{a}}(k)[\bar{\zeta}^{(1)}_{\bar{a}}(\bar{x})](l),   \tag{S84} \label{FP abar 2}\\
&\int_{\Sigma^{(2)}\backslash{D_{\delta}}}d^2z(\sideset{^2}{^\Pi}{\mathop{\mathcal{F}}}\epsilon^{\frac{1}{2}}{}^2\bar{\mathcal{P}}_2)\notag\\
=&-\pi\epsilon\sum_{l\geq1}l\bar{P}_{l+1}(\bar{x},\bar{\tau}_a)E_{l+1}(\tau_a)|\epsilon|^l+\pi\sum_{k,l\geq 1}|\epsilon|^lA_{k,a}^{l,0}\bar P_{l+1}(\bar x,\bar\tau_a)\theta_a(k+2)\notag\\
&+\sum_{k,l\geq 1}\frac{\pi}{l+2}\Big[\frac{1}{|\epsilon|^{l+2}}\delta_{k,l}+\sum_{n\geq k}\bar C_{l,a}^{n,k}|\epsilon|^{n-k+1}\Big]\theta_a(k+2)[\bar{\xi}^{(1)}_a(\bar{x})](l)\notag\\
&+\sum_{k,l\geq 1}\frac{\pi}{l+2}\Big[\frac{1}{|\epsilon|^{l+2}}\delta_{k,l}+\sum_{n\geq k}\bar C_{l,\bar a}^{n,k}|\epsilon|^{n-k+1}\Big]\theta_{\bar{a}}(k+2)[\bar{\zeta}^{(1)}_{\bar{a}}(\bar{x})](l)\notag\\
&+\sum_{n,l\geq 1}\frac{\pi}{l+2}\epsilon|\epsilon|^{n-1}\Big[\bar C_{l,a}^{n-2,-1}[\bar{\xi}^{(1)}_a(\bar{x})](l)+\bar C_{l,\bar a}^{n-2,-1}[\bar{\zeta}^{(1)}_{\bar{a}}(\bar{x})](l)\Big],  \tag{S85} \label{FP pi 2}\\
&\int_{\Sigma^{(2)}\backslash{D_{\delta}}}d^2z({}^2{\mathcal{P}}_4\sideset{^2}{_a}{\mathop{\bar{\mathcal{F}}}})\notag\\
=&-\pi P_{4}(x,\tau_a)|\epsilon|-\pi\sum_{k\geq 1}|\epsilon|E_{k+3}(\tau_a)[\xi^{(3)}_a(x)](k)\notag\\
&+\pi\sum_{k,l\geq 1}\frac{l^2+3l+2}{6}|\epsilon|^l\bar A_{k,a}^{l,0}{P}_{l+3}(x,\tau_a)\bar\alpha_a(k)\notag\\
&+\sum_{k,l\geq 1}\frac{\pi}{k+2}\Big[\frac{1}{|\epsilon|^{k+2}}\delta_{k,l}-\sum_{n\geq l}C_{k,a}^{n,l}|\epsilon|^{n-l+1}\Big][\xi^{(3)}_a(x)](k)\bar\alpha_a(l)\notag\\
&+\sum_{k,l\geq 1}\frac{\pi}{k+2}\Big[\frac{1}{|\epsilon|^{k+2}}\delta_{k,l}-\sum_{n\geq l}C_{k,\bar a}^{n,l}|\epsilon|^{n-l+1}\Big][\zeta^{(3)}_{\bar{a}}(x)](k)\bar\beta_{\bar{a}}(l),\tag{S86}   \label{FP a 4}\\
&\int_{\Sigma^{(2)}\backslash{D_{\delta}}}d^2z({}^2{\mathcal{P}}_4\sideset{^2}{_{\bar a}}{\mathop{\bar{\mathcal{F}}}})\notag\\
=&-\pi\sum_{k\geq 1}|\epsilon|E_{k+3}(\tau_{\bar a})[\zeta^{(3)}_{\bar a}(x)](k)+\pi\sum_{k,l\geq 1}\frac{l^2+3l+2}{6}|\epsilon|^l\bar A_{k,a}^{l,0}{P}_{l+3}(x,\tau_a)\bar\beta_a(k)\notag\\
&+\sum_{k,l\geq 1}\frac{\pi}{k+2}\Big[\frac{1}{|\epsilon|^{k+2}}\delta_{k,l}-\sum_{n\geq l}C_{k,a}^{n,l}|\epsilon|^{n-l+1}\Big][\xi^{(3)}_a(x)](k)\bar\beta_a(l)\notag\\
&+\sum_{k,l\geq 1}\frac{\pi}{k+2}\Big[\frac{1}{|\epsilon|^{k+2}}\delta_{k,l}-\sum_{n\geq l}C_{k,\bar a}^{n,l}|\epsilon|^{n-l+1}\Big][\zeta^{(3)}_{\bar{a}}(x)](k)\bar\alpha_{\bar{a}}(l),\tag{S87}   \label{FP abar 4}\\
&\int_{\Sigma^{(2)}\backslash{D_{\delta}}}d^2z({}^2{\mathcal{P}}_4{}^2\bar{\mathcal{F}}^{\Pi}\bar\epsilon^{\frac{1}{2}})\notag\\
=&-\pi\sum_{l\geq1}\frac{l^2+3l+2}{6}|\epsilon|^l{P}_{l+3}(x,\tau_a)\Big[\bar\epsilon l\bar E_{l+1}(\bar\tau_a)-\bar A_{k,a}^{l,0}\bar\theta_a(k+2)\Big]\notag\\
&+\sum_{k,l\geq 1}\frac{\pi}{k+2}\Big[\frac{1}{|\epsilon|^{k+2}}\delta_{k,l}+\sum_{n\geq k}C_{k,a}^{n,l}|\epsilon|^{n-l+1}\Big][\xi^{(3)}_a(x)](k)\bar\theta_a(l+2)\notag\\
&+\sum_{k,l\geq 1}\frac{\pi}{k+2}\Big[\frac{1}{|\epsilon|^{k+2}}\delta_{k,l}+\sum_{n\geq k}C_{k,\bar a}^{n,l}|\epsilon|^{n-l+1}\Big][\zeta^{(3)}_{\bar{a}}(x)](k)\bar\theta_{\bar{a}}(l+2).\tag{S88}   \label{FP pi 4}
\end{align}
The following are the $\mathcal{P}\bar{\mathcal{P}}$-type integrals that appear in the deformed one-point functions of primary field:
\begin{align}
&\int_{\Sigma^{(2)}\backslash{D_{\delta}}}d^2z({}^2{\mathcal{P}}_1{}^2\bar {\mathcal{P}}_1)\notag\\
=&2\pi\log|Q(x,\tau_a)|^2-\pi\sum_{k\geq 1}\frac{|\epsilon|^k}{k}|P'_{k}(x,\tau_a)|^2\notag\\
&+\sum_{k,n\geq 1}\frac{\pi}{k+2}|\epsilon|^{n}2\text{Re}\Big[C_{k,\bar a}^{n-1,0}\bar\epsilon[\zeta^{(0)}_{\bar{a}}(x)](k)\Big]\notag\\
&+\sum_{k\geq 1}\frac{\pi}{k+2}2\text{Re}\bigg[\Big[P'_{k+2}(x,\tau_a)-\sum_{l\geq1}\frac{k+2}{l}|\epsilon|^lA_{k,a}^{l,0}\bar P'_l(\bar x,\bar\tau_a)\Big][\xi^{(0)}_a(x)](k)\bigg]\notag\\
&+\sum_{n\geq 1}C_{0,\bar a}^{n,0}|\epsilon|^{n+3}+\sum_{k,l\geq 1}\frac{\pi}{k+2}\Big[\frac{1}{|\epsilon|^{k+2}}\delta_{k,l}-\sum_{n\geq l}C_{k,a}^{n,l}|\epsilon|^{n-l+1}\Big][\xi^{(0)}_a(x)](k)[\bar{\xi}^{(0)}_a(\bar{x})](l)\notag\\
&+\sum_{k,l\geq 1}\frac{\pi}{k+2}\Big[\frac{1}{|\epsilon|^{k+2}}\delta_{k,l}-\sum_{n\geq l}C_{k,\bar a}^{n,l}|\epsilon|^{n-l+1}\Big][\zeta^{(0)}_{\bar{a}}(x)](k)[\bar{\zeta}^{(0)}_{\bar{a}}(\bar{x})](l),\tag{S89}   \label{PP 1 1}\\
&\int_{\Sigma^{(2)}\backslash{D_{\delta}}}d^2z({}^2{\mathcal{P}}_1{}^2\bar {\mathcal{P}}_2)\notag\\
=&\pi\bar P_1(\bar x,\bar\tau_a)+\sum_{k,n\geq 1}\frac{-2\pi i}{k+2}\bar C_{k,\bar a}^{n-1,0}\epsilon|\epsilon|^{n}[\bar \zeta^{(1)}_{\bar{a}}(\bar x)](k)\notag\\
&+\pi\sum_{l\geq 1}|\epsilon|^l\bar P_{l+1}(\bar x,\bar\tau_a)\Big[P'_l(x,\tau_a)+\sum_{k\geq 1}A_{k,a}^{l,0}[\xi^{(0)}_a(x)](k)\Big]\notag\\
&+\sum_{k\geq 1}\frac{\pi}{k+2}\Big[\bar P'_{k+2}(\bar x,\bar\tau_a)-\sum_{l\geq1}\frac{k+2}{l}|\epsilon|^l\bar A_{k,a}^{l,0}P'_l(x,\tau_a)\Big][\bar\xi^{(1)}_a(\bar x)](k)\notag\\
&+\sum_{k,l\geq 1}\frac{\pi}{k+2}\Big[\frac{1}{|\epsilon|^{k+2}}\delta_{k,l}-\sum_{n\geq l}C_{k,a}^{n,l}|\epsilon|^{n-l+1}\Big][\xi^{(0)}_a(x)](k)[\bar{\xi}^{(1)}_a(\bar{x})](l)\notag\\
&+\sum_{k,l\geq 1}\frac{\pi}{k+2}\Big[\frac{1}{|\epsilon|^{k+2}}\delta_{k,l}-\sum_{n\geq l}C_{k,\bar a}^{n,l}|\epsilon|^{n-l+1}\Big][\zeta^{(0)}_{\bar{a}}(x)](k)[\bar{\zeta}^{(1)}_{\bar{a}}(\bar{x})](l),\tag{S90}   \label{PP 1 2}\\
&\int_{\Sigma^{(2)}\backslash{D_{\delta}}}d^2z({}^2{\mathcal{P}}_2{}^2\bar{\mathcal{P}}_2)\notag\\
=&-\pi\sum_kk|P_{k+1}(x,\tau_a)|^2|\epsilon|^k+\pi\sum_{k,l\geq 1}|\epsilon|^l2\text{Re}\Big[\bar P_{l+1}(\bar x,\bar\tau_a)A_{k,a}^{l,0}[\xi^{(1)}_a(x)](k)\Big]\notag\\
&+\sum_{k,l\geq 1}\frac{\pi}{k+2}\Big[\frac{1}{|\epsilon|^{k+2}}\delta_{k,l}-\sum_{n\geq l}C_{k,a}^{n,l}|\epsilon|^{n-l+1}\Big][\xi^{(1)}_a(x)](k)[\bar{\xi}^{(1)}_a(\bar{x})](l)\notag\\
&+\sum_{k,l\geq 1}\frac{\pi}{k+2}\Big[\frac{1}{|\epsilon|^{k+2}}\delta_{k,l}-\sum_{n\geq l}C_{k,\bar a}^{n,l}|\epsilon|^{n-l+1}\Big][\zeta^{(1)}_{\bar{a}}(x)](k)[\bar{\zeta}^{(1)}_{\bar{a}}(\bar{x})](l).\tag{S91}   \label{PP 2 2}
\end{align}
The following are the $\mathcal{P}\bar{\mathcal{P}}$-type integrals that appear in the deformed two-point function, and the two insertion points $x_1$ and $x_2$ are on the same torus $S_a$:
\begin{align}
&\int_{\Sigma^{(2)}\backslash{D_{\delta}}}d^2z({}^2{\mathcal{P}}_1^{(x_1)}{}^2\bar {\mathcal{P}}_1^{(\bar x_2)})\notag\\
=&\pi\log\frac{|Q(x_1,\tau_a)Q(x_2,\tau_a)|^2}{|Q(x_1-x_2,\tau_a)|^2}+\frac{\pi}{2}-\pi\sum_k\frac{|\epsilon|^k}{k}P'_{k}(x_1,\tau_a)\bar P'_k(\bar x_2,\bar\tau_a)\notag\\
&+\sum_{n\geq 1}C_{0,\bar a}^{n,0}|\epsilon|^{n+3}+\sum_{k,n\geq 1}\frac{\pi}{k+2}|\epsilon|^{n}\Big[C_{k,\bar a}^{n-1,0}\bar\epsilon[\zeta^{(0)}_{\bar{a}}(x_1)](k)+\bar C_{k,\bar a}^{n-1,0}\epsilon[\bar\zeta^{(0)}_{\bar{a}}(\bar x_2)](k)\Big]\notag\\
&+\sum_{k\geq 1}\frac{\pi}{k+2}\Big[\bar P'_{k+2}(\bar x_1,\bar\tau_a)-\sum_{l\geq1}\frac{k+2}{l}|\epsilon|^l\bar A_{k,a}^{l,0}P'_l(x_1,\tau_a)\Big][\bar\xi^{(0)}_a(\bar x_2)](k)\notag\\
&+\sum_{k\geq 1}\frac{\pi}{k+2}\Big[P'_{k+2}(x_2,\tau_a)-\sum_{l\geq1}\frac{k+2}{l}|\epsilon|^lA_{k,a}^{l,0}\bar P'_l(\bar x_2,\bar\tau_a)\Big][\xi^{(0)}_a(x_1)](k)\notag\\
&+\sum_{k,l\geq 1}\frac{\pi}{k+2}\Big[\frac{1}{|\epsilon|^{k+2}}\delta_{k,l}-\sum_{n\geq l}C_{k,a}^{n,l}|\epsilon|^{n-l+1}\Big][\xi^{(0)}_a(x_1)](k)[\bar{\xi}^{(0)}_a(\bar{x}_2)](l)\notag\\
&+\sum_{k,l\geq 1}\frac{\pi}{k+2}\Big[\frac{1}{|\epsilon|^{k+2}}\delta_{k,l}-\sum_{n\geq l}C_{k,\bar a}^{n,l}|\epsilon|^{n-l+1}\Big][\zeta^{(0)}_{\bar{a}}(x_1)](k)[\bar{\zeta}^{(0)}_{\bar{a}}(\bar{x}_2)](l),\tag{S92}   \label{P1P2 1 1}\\
&\int_{\Sigma^{(2)}\backslash{D_{\delta}}}d^2z({}^2{\mathcal{P}}_1^{(x_1)}{}^2\bar {\mathcal{P}}_2^{(\bar x_2)})\notag\\
=&\pi\Big[\bar P_1(\bar x_2,\bar\tau_a)+\bar P_1(\bar x_1-\bar x_2,\bar\tau_a)\Big]+\sum_{k,n\geq 1}\frac{\pi}{k+2}\bar C_{k,\bar a}^{n-1,0}\epsilon|\epsilon|^{n}[\bar \zeta^{(1)}_{\bar{a}}(\bar x_2)](k)\notag\\
&\pi\sum_{l\geq 1}|\epsilon|^l\bar P_{l+1}(\bar x_2,\bar\tau_a)\Big[P'_l(x_1,\tau_a)+\sum_{k\geq 1}A_{k,a}^{l,0}[\xi^{(0)}_a(x_1)](k)\Big]\notag\\
&+\sum_{k\geq 1}\frac{\pi}{k+2}\Big[\bar P'_{k+2}(\bar x_1,\bar\tau_a)-\sum_{l\geq1}\frac{k+2}{l}|\epsilon|^l\bar A_{k,a}^{l,0}P'_l(x_1,\tau_a)\Big][\bar\xi^{(1)}_a(\bar x_2)](k)\notag\\
&+\sum_{k,l\geq 1}\frac{\pi}{k+2}\Big[\frac{1}{|\epsilon|^{k+2}}\delta_{k,l}-\sum_{n\geq l}C_{k,a}^{n,l}|\epsilon|^{n-l+1}\Big][\xi^{(0)}_a(x_1)](k)[\bar{\xi}^{(1)}_a(\bar{x}_2)](l)\notag\\
&+\sum_{k,l\geq 1}\frac{\pi}{k+2}\Big[\frac{1}{|\epsilon|^{k+2}}\delta_{k,l}-\sum_{n\geq l}C_{k,\bar a}^{n,l}|\epsilon|^{n-l+1}\Big][\zeta^{(0)}_{\bar{a}}(x_1)](k)[\bar{\zeta}^{(1)}_{\bar{a}}(\bar{x}_2)](l),\tag{S93}   \label{P1P2 1 2}\\
&\int_{\Sigma^{(2)}\backslash{D_{\delta}}}d^2z({}^2{\mathcal{P}}_2^{(x_1)}{}^2\bar {\mathcal{P}}_2^{(\bar x_2)})\notag\\
=&-\pi\sum_{k\geq 1}|\epsilon|^kkP_{k+1}(x_1,\tau_a)\bar P_{k+1}(\bar x_2,\bar\tau_a)\notag\\
&\pi\sum_{k,l\geq 1}|\epsilon|^l\Big[P_{l+1}(x_1,\tau_a)\bar A_{k,a}^{l,0}[\bar\xi^{(1)}_a(\bar x_2)](k)+\bar P_{l+1}(\bar x_2,\bar\tau_a)A_{k,a}^{l,0}[\xi^{(1)}_a(x_1)](k)\Big]\notag\\
&+\sum_{k,l\geq 1}\frac{\pi}{k+2}\Big[\frac{1}{|\epsilon|^{k+2}}\delta_{k,l}-\sum_{n\geq l}C_{k,a}^{n,l}|\epsilon|^{n-l+1}\Big][\xi^{(1)}_a(x_1)](k)[\bar{\xi}^{(1)}_a(\bar{x}_2)](l)\notag\\
&+\sum_{k,l\geq 1}\frac{\pi}{k+2}\Big[\frac{1}{|\epsilon|^{k+2}}\delta_{k,l}-\sum_{n\geq l}C_{k,\bar a}^{n,l}|\epsilon|^{n-l+1}\Big][\zeta^{(1)}_{\bar{a}}(x_1)](k)[\bar{\zeta}^{(1)}_{\bar{a}}(\bar{x}_2)](l),\tag{S94}   \label{P1P2 2 2}\\
&\int_{\Sigma^{(2)}\backslash{D_{\delta}}}d^2z({}^2{\mathcal{P}}_1^{(x_1)}{}^2\bar{\mathcal{P}}_4^{(\bar x_2)})\notag\\
=&\frac{\pi}{3}\Big[\bar P_3(\bar x_2,\bar\tau_a)+\bar P_3(\bar x_1-\bar x_2,\bar\tau_a)\Big]+\sum_{k,n\geq 1}\frac{\pi}{k+2}\bar C_{k,\bar a}^{n-1,0}\epsilon|\epsilon|^{n}[\bar \zeta^{(3)}_{\bar{a}}(\bar x_2)](k)\notag\\
&+\pi\sum_{l\geq 1}\frac{l^2+3l+2}{6}|\epsilon|^l\bar{P}_{l+3}(\bar x_2,\bar\tau_a)\Big[P'_l(x_1,\tau_a)+\sum_{k\geq 1}A_{k,a}^{l,0}[\xi^{(0)}_a(x_1)](k)\Big]\notag\\
&+\sum_{k\geq 1}\frac{\pi}{k+2}\Big[\bar P'_{k+2}(\bar x_1,\bar\tau_a)-\sum_{l\geq1}\frac{k+2}{l}|\epsilon|^l\bar A_{k,a}^{l,0}P'_l(x_1,\tau_a)\Big][\bar\xi^{(3)}_a(\bar x_2)](k)\notag\\
&+\sum_{k,l\geq 1}\frac{\pi}{k+2}\Big[\frac{1}{|\epsilon|^{k+2}}\delta_{k,l}-\sum_{n\geq l}C_{k,a}^{n,l}|\epsilon|^{n-l+1}\Big][\xi^{(0)}_a(x_1)](k)[\bar{\xi}^{(3)}_a(\bar{x}_2)](l)\notag\\
&+\sum_{k,l\geq 1}\frac{\pi}{k+2}\Big[\frac{1}{|\epsilon|^{k+2}}\delta_{k,l}-\sum_{n\geq l}C_{k,\bar a}^{n,l}|\epsilon|^{n-l+1}\Big][\zeta^{(0)}_{\bar{a}}(x_1)](k)[\bar{\zeta}^{(3)}_{\bar{a}}(\bar{x}_2)](l),\tag{S95}   \label{P1P2 1 4}\\
&\int_{\Sigma^{(2)}\backslash{D_{\delta}}}d^2z({}^2{\mathcal{P}}_2^{(x_1)}{}^2\bar{\mathcal{P}}_4^{(\bar x_2)})\notag\\
=&\pi\sum_{k,l\geq 1}|\epsilon|^l\bar A_{k,a}^{l,0}P_{l+1}(x_1,\tau_a)[\bar\xi^{(3)}_a(\bar x_2)](k)\notag\\
&-\pi\sum_{l\geq 1}\frac{l^2+3l+2}{6}|\epsilon|^l\bar{P}_{l+3}(\bar x_2,\bar\tau_a)\Big[lP_{l+1}(x_1,\tau_a)-\sum_{k\geq 1}A_{k,a}^{l,0}[\xi^{(1)}_a(x_1)](k)\Big]\notag\\
&+\sum_{k,l\geq 1}\frac{\pi}{k+2}\Big[\frac{1}{|\epsilon|^{k+2}}\delta_{k,l}-\sum_{n\geq l}C_{k,a}^{n,l}|\epsilon|^{n-l+1}\Big][\xi^{(1)}_a(x_1)](k)[\bar{\xi}^{(3)}_a(\bar{x}_2)](l)\notag\\
&+\sum_{k,l\geq 1}\frac{\pi}{k+2}\Big[\frac{1}{|\epsilon|^{k+2}}\delta_{k,l}-\sum_{n\geq l}C_{k,\bar a}^{n,l}|\epsilon|^{n-l+1}\Big][\zeta^{(1)}_{\bar{a}}(x_1)](k)[\bar{\zeta}^{(3)}_{\bar{a}}(\bar{x}_2)](l),\tag{S96}   \label{P1P2 2 4}\\
&\int_{\Sigma^{(2)}\backslash{D_{\delta}}}d^2z({}^2{\mathcal{P}}_4^{(x_1)}{}^2\bar{\mathcal{P}}_4^{(\bar x_2)})\notag\\
=&\frac{-\pi}{3}\sum_{l\geq 1}\frac{l(l+1)^2(l+2)^2}{12}|\epsilon|^lP_{l+3}(x_1,\tau_a)\bar P_{l+3}(\bar x_2,\bar\tau_a)\notag\\
&+\pi\sum_{k,l\geq 1}\frac{l^2+3l+2}{6}|\epsilon|^l\bar{P}_{l+3}(\bar x_2,\bar\tau_a)A_{k,a}^{l,0}[\xi^{(3)}_a(x_1)](k)\notag\\
&+\pi\sum_{k,l\geq 1}\frac{l^2+3l+2}{6}|\epsilon|^l{P}_{l+3}(x_1,\tau_a)\bar A_{k,a}^{l,0}[\bar\xi^{(3)}_a(\bar x_2)](k)\notag\\
&+\sum_{k,l\geq 1}\frac{\pi}{k+2}\Big[\frac{1}{|\epsilon|^{k+2}}\delta_{k,l}-\sum_{n\geq l}C_{k,a}^{n,l}|\epsilon|^{n-l+1}\Big][\xi^{(3)}_a(x_1)](k)[\bar{\xi}^{(3)}_a(\bar{x}_2)](l)\notag\\
&+\sum_{k,l\geq 1}\frac{\pi}{k+2}\Big[\frac{1}{|\epsilon|^{k+2}}\delta_{k,l}-\sum_{n\geq l}C_{k,\bar a}^{n,l}|\epsilon|^{n-l+1}\Big][\zeta^{(3)}_{\bar{a}}(x_1)](k)[\bar{\zeta}^{(3)}_{\bar{a}}(\bar{x}_2)](l).\tag{S97}   \label{P1P2 4 4}
\end{align}
The following are the $\mathcal{P}\bar{\mathcal{P}}$-type integrals that appear in the deformed two-point function, and the two insertion points $x$ and $y$ are on different tori:
\begin{align}
&\int_{\Sigma^{(2)}\backslash{D_{\delta}}}d^2z({}^2{\mathcal{P}}_1^{(x)}{}^2\bar {\mathcal{P}}_1^{(\bar y)})\notag\\
=&-\frac{\pi}{2}\bar\epsilon\Big[\bar P'_{2}(\bar x,\bar\tau_a)-\sum_{l\geq1}|\epsilon|^l(l+1)P'_l(x,\tau_a)\bar E_{l+2}(\bar\tau_a)\Big]\notag\\
&-\frac{\pi}{2}\epsilon\Big[P'_{2}(y,\tau_{\bar a})-\sum_{l\geq1}|\epsilon|^l(l+1)\bar P'_l(\bar y,\bar\tau_{\bar a})E_{l+2}(\tau_{\bar a})\Big]\notag\\
&+\sum_{k,n\geq 1}\frac{\pi}{k+2}\Big[C_{k,a}^{n-1,0}\bar\epsilon|\epsilon|^{n}[\xi^{(0)}_a(x)](k)+\bar C_{k,\bar a}^{n-1,0}\epsilon|\epsilon|^{n}[\bar{\xi}^{(0)}_{\bar a}(\bar{y})](k)\Big]\notag\\
&+\sum_{k\geq 1}\frac{\pi}{k+2}\Big[\bar P'_{k+2}(\bar x,\bar\tau_a)-\sum_{l\geq1}\frac{k+2}{l}|\epsilon|^l\bar A_{k,a}^{l,0}P'_l(x,\tau_a)\Big][\bar{\zeta}^{(0)}_{a}(\bar{y})](k)\notag\\
&+\sum_{k\geq 1}\frac{\pi}{k+2}\Big[P'_{k+2}(y,\tau_{\bar a})-\sum_{l\geq1}\frac{k+2}{l}|\epsilon|^lA_{k,{\bar a}}^{l,0}\bar P'_l(\bar y,\bar\tau_{\bar a})\Big][\zeta^{(0)}_{\bar{a}}(x)](k)\notag\\
&+\sum_{k,l\geq 1}\frac{\pi}{k+2}\Big[\frac{1}{|\epsilon|^{k+2}}\delta_{k,l}-\sum_{n\geq l}C_{k,a}^{n,l}|\epsilon|^{n-l+1}\Big][\xi^{(0)}_a(x)](k)[\bar{\zeta}^{(0)}_{a}(\bar{y})](l)\notag\\
&+\sum_{k,l\geq 1}\frac{\pi}{k+2}\Big[\frac{1}{|\epsilon|^{k+2}}\delta_{k,l}-\sum_{n\geq l}C_{k,\bar a}^{n,l}|\epsilon|^{n-l+1}\Big][\zeta^{(0)}_{\bar{a}}(x)](k)[\bar{\xi}^{(0)}_{\bar a}(\bar{y})](l), \tag{S98}  \label{PxPy 1 1}\\
&\int_{\Sigma^{(2)}\backslash{D_{\delta}}}d^2z({}^2{\mathcal{P}}_1^{(x)}{}^2\bar {\mathcal{P}}_2^{(\bar y)})\notag\\
=&-\frac{\pi}{2}\epsilon\sum_{l}|\epsilon|^ll(l+1)\bar P_{l+1}(\bar y,\bar\tau_{\bar a})E_{l+2}(\tau_{\bar a})+\sum_{k,n\geq 1}\frac{\pi}{k+2}\bar C_{k,\bar a}^{n-1,0}\epsilon|\epsilon|^{n}[\bar\xi^{(1)}_{\bar{a}}(\bar y)](k)\notag\\
&+\pi\sum_{k,l\geq 1}|\epsilon|^l\bar P_{l+1}(\bar y,\bar\tau_{\bar a})A_{k,{\bar a}}^{l,0}[\zeta^{(0)}_{\bar{a}}(x)](k)\notag\\
&+\sum_{k\geq 1}\frac{\pi}{k+2}\Big[\bar P'_{k+2}(\bar x,\bar\tau_a)-\sum_{l\geq1}\frac{k+2}{l}|\epsilon|^l\bar A_{k,a}^{l,0}P'_l(x,\tau_a)\Big][\bar{\zeta}^{(1)}_{a}(\bar{y})](k)\notag\\
&+\sum_{k,l\geq 1}\frac{\pi}{k+2}\Big[\frac{1}{|\epsilon|^{k+2}}\delta_{k,l}-\sum_{n\geq l}C_{k,a}^{n,l}|\epsilon|^{n-l+1}\Big][\xi^{(0)}_a(x)](k)[\bar{\zeta}^{(1)}_{a}(\bar{y})](l)\notag\\
&+\sum_{k,l\geq 1}\frac{\pi}{k+2}\Big[\frac{1}{|\epsilon|^{k+2}}\delta_{k,l}-\sum_{n\geq l}C_{k,\bar a}^{n,l}|\epsilon|^{n-l+1}\Big][\zeta^{(0)}_{\bar{a}}(x)](k)[\bar{\xi}^{(1)}_{\bar a}(\bar{y})](l), \tag{S99}  \label{PxPy 1 2}\\
&\int_{\Sigma^{(2)}\backslash{D_{\delta}}}d^2z({}^2{\mathcal{P}}_2^{(x)}{}^2\bar {\mathcal{P}}_2^{(\bar y)})\notag\\
=&\pi\sum_{k,l\geq 1}|\epsilon|^l\Big[P_{l+1}(x,\tau_a)\bar A_{k,a}^{l,0}[\bar{\zeta}^{(1)}_{a}(\bar{y})](k)+\bar P_{l+1}(\bar y,\bar\tau_{\bar a})A_{k,{\bar a}}^{l,0}[\zeta^{(1)}_{\bar{a}}(x)](k)\Big]\notag\\
&+\sum_{k,l\geq 1}\frac{\pi}{k+2}\Big[\frac{1}{|\epsilon|^{k+2}}\delta_{k,l}-\sum_{n\geq l}C_{k,a}^{n,l}|\epsilon|^{n-l+1}\Big][\xi^{(1)}_a(x)](k)[\bar{\zeta}^{(1)}_{a}(\bar{y})](l)\notag\\
&+\sum_{k,l\geq 1}\frac{\pi}{k+2}\Big[\frac{1}{|\epsilon|^{k+2}}\delta_{k,l}-\sum_{n\geq l}C_{k,\bar a}^{n,l}|\epsilon|^{n-l+1}\Big][\zeta^{(1)}_{\bar{a}}(x)](k)[\bar{\xi}^{(1)}_{\bar a}(\bar{y})](l), \tag{S100}  \label{PxPy 2 2}\\
&\int_{\Sigma^{(2)}\backslash{D_{\delta}}}d^2z({}^2{\mathcal{P}}_1^{(x)}{}^2\bar{\mathcal{P}}_4^{(\bar y)})\notag\\
=&-\frac{\pi}{2}\epsilon\sum_{l\geq1}\frac{l(l+1)^2(l+2)}{6}|\epsilon|^l\bar{P}_{l+3}(\bar y,\bar\tau_{\bar a})E_{l+2}(\tau_{\bar a})+\sum_{k,n\geq 1}\frac{\pi}{k+2}\bar C_{k,\bar a}^{n-1,0}\epsilon|\epsilon|^{n}[\bar{\xi}^{(3)}_{\bar a}(\bar{y})](k)\notag\\
&+\pi\sum_{k,l\geq 1}\frac{l^2+3l+2}{6}|\epsilon|^l\bar{P}_{l+3}(\bar y,\bar\tau_{\bar a})A_{k,{\bar a}}^{l,0}[\zeta^{(0)}_{\bar{a}}(x)](k)\notag\\
&+\sum_{k\geq 1}\frac{\pi}{k+2}\Big[\bar P'_{k+2}(\bar x,\bar\tau_a)-\sum_{l\geq1}\frac{k+2}{l}|\epsilon|^l\bar A_{k,a}^{l,0}P'_l(x,\tau_a)\Big][\bar{\zeta}^{(3)}_{a}(\bar{y})](k)\notag\\
&+\sum_{k,l\geq 1}\frac{\pi}{k+2}\Big[\frac{1}{|\epsilon|^{k+2}}\delta_{k,l}-\sum_{n\geq l}C_{k,a}^{n,l}|\epsilon|^{n-l+1}\Big][\xi^{(0)}_a(x)](k)[\bar{\zeta}^{(3)}_{a}(\bar{y})](l)\notag\\
&+\sum_{k,l\geq 1}\frac{\pi}{k+2}\Big[\frac{1}{|\epsilon|^{k+2}}\delta_{k,l}-\sum_{n\geq l}C_{k,\bar a}^{n,l}|\epsilon|^{n-l+1}\Big][\zeta^{(0)}_{\bar{a}}(x)](k)[\bar{\xi}^{(3)}_{\bar a}(\bar{y})](l), \tag{S101}  \label{PxPy 1 4}\\
&\int_{\Sigma^{(2)}\backslash{D_{\delta}}}d^2z({}^2{\mathcal{P}}_2^{(x)}{}^2\bar{\mathcal{P}}_4^{(\bar y)})\notag\\
=&\pi\sum_{k,l\geq 1}|\epsilon|^l\bar A_{k,a}^{l,0}P_{l+1}(x,\tau_a)[\bar{\zeta}^{(3)}_{a}(\bar{y})](k)\notag\\
&+\pi\sum_{l\geq 1}\frac{l^2+3l+2}{6}|\epsilon|^l\bar{P}_{l+3}(\bar y,\bar\tau_{\bar a})A_{k,{\bar a}}^{l,0}[\zeta^{(1)}_{\bar{a}}(x)](k)\notag\\
&+\sum_{k,l\geq 1}\frac{\pi}{k+2}\Big[\frac{1}{|\epsilon|^{k+2}}\delta_{k,l}-\sum_{n\geq l}C_{k,a}^{n,l}|\epsilon|^{n-l+1}\Big][\xi^{(1)}_a(x)](k)[\bar{\zeta}^{(3)}_{a}(\bar{y})](l)\notag\\
&+\sum_{k,l\geq 1}\frac{\pi}{k+2}\Big[\frac{1}{|\epsilon|^{k+2}}\delta_{k,l}-\sum_{n\geq l}C_{k,\bar a}^{n,l}|\epsilon|^{n-l+1}\Big][\zeta^{(1)}_{\bar{a}}(x)](k)[\bar{\xi}^{(3)}_{\bar a}(\bar{y})](l), \tag{S102}  \label{PxPy 2 4}\\
&\int_{\Sigma^{(2)}\backslash{D_{\delta}}}d^2z({}^2{\mathcal{P}}_4^{(x)}{}^2\bar{\mathcal{P}}_4^{(\bar y)})\notag\\
=&\pi\sum_{k,l\geq 1}\frac{l^2+3l+2}{6}|\epsilon|^l\Big[\bar{P}_{l+3}(\bar y,\bar\tau_{\bar a})A_{k,{\bar a}}^{l,0}[\zeta^{(3)}_{\bar a}(x)](k)+{P}_{l+3}(x,\tau_a)\bar A_{k,a}^{l,0}[\bar{\zeta}^{(3)}_a(\bar{y})](l)\Big]\notag\\
&+\sum_{k,l\geq 1}\frac{\pi}{k+2}\Big[\frac{1}{|\epsilon|^{k+2}}\delta_{k,l}-\sum_{n\geq l}C_{k,\bar a}^{n,l}|\epsilon|^{n-l+1}\Big][\xi^{(3)}_a(x)](k)[\bar{\zeta}^{(3)}_a(\bar{y})](l)\notag\\
&+\sum_{k,l\geq 1}\frac{\pi}{k+2}\Big[\frac{1}{|\epsilon|^{k+2}}\delta_{k,l}-\sum_{n\geq l}C_{k,\bar a}^{n,l}|\epsilon|^{n-l+1}\Big][\zeta^{(3)}_{\bar{a}}(x)](k)[\bar{\xi}^{(3)}_{\bar{a}}(\bar{y})](l). \tag{S103}  \label{PxPy 4 4}
\end{align}
\subsection{Details of deformed partition function}\label{C.2}
In this Appendix we discuss the detailed calculation of the first-order deformed partition function. Applying eqs.(\ref{WI0})(\ref{Dz})(\ref{deltalambdaZ}) we have
\begin{align}
\delta_{\lambda}Z=&\int_{\Sigma^{(2)}}d^2zD_z\bar D_{\bar z}Z\notag\\
=&\frac{1}{4\pi^2}\int_{\Sigma^{(2)}}d^2z\bigg\lbrace(\sideset{^2}{_1}{\mathop{\mathcal{F}}}\sideset{^2}{_1}{\mathop{\bar{\mathcal{F}}}})\partial_{\tau_1}\partial_{\bar{\tau}_1}+(\sideset{^2}{_1}{\mathop{\mathcal{F}}}\sideset{^2}{_2}{\mathop{\bar{\mathcal{F}}}})\partial_{\tau_1}\partial_{\bar{\tau}_2}-2\pi i\bar{\epsilon}^{\frac{1}{2}}(\sideset{^2}{_1}{\mathop{\mathcal{F}}}{}^2\bar{\mathcal{F}}^{\Pi})\partial_{\tau_1}\partial_{\bar{\epsilon}}\notag\\
&+(\sideset{^2}{_2}{\mathop{\mathcal{F}}}\sideset{^2}{_1}{\mathop{\bar{\mathcal{F}}}})\partial_{\tau_2}\partial_{\bar{\tau}_1}+(\sideset{^2}{_2}{\mathop{\mathcal{F}}}\sideset{^2}{_2}{\mathop{\bar{\mathcal{F}}}})\partial_{\tau_2}\partial_{\bar{\tau}_2}-2\pi i\bar{\epsilon}^{\frac{1}{2}}(\sideset{^2}{_2}{\mathop{\mathcal{F}}}{}^2\bar{\mathcal{F}}^{\Pi})\partial_{\tau_2}\partial_{\bar{\epsilon}}\notag\\
&+2\pi i\epsilon^{\frac{1}{2}}({}^2\mathcal{F}^{\Pi}\sideset{^2}{_1}{\mathop{\bar{\mathcal{F}}}})\partial_\epsilon\partial_{\bar{\tau}_1}+2\pi i\epsilon^{\frac{1}{2}}({}^2\mathcal{F}^{\Pi}\sideset{^2}{_2}{\mathop{\bar{\mathcal{F}}}})\partial_\epsilon\partial_{\bar{\tau}_2}+4\pi^2\epsilon^{\frac{1}{2}}\bar{\epsilon}^{\frac{1}{2}}({}^2\mathcal{F}^{\Pi}{}^2\bar{\mathcal{F}}^{\Pi})\partial_\epsilon\partial_{\bar{\epsilon}}\bigg\rbrace Z.\tag{S104} \label{DDZ}
\end{align}
The $\mathcal{F}\bar{\mathcal{F}}$-type integrals are calculated in eqs.~(\ref{FF a a})(\ref{FF a abar})(\ref{FF a pi})(\ref{FF pi pi}). For simplicity, we approximate the result up to $|\epsilon|^3$:
\begin{align}
\delta_{\lambda}Z=&\sum_{a=1,2}\Big[\text{Im}[\tau_a]-\frac{|\epsilon|}{6\pi}\Big]\partial_{\tau_a}\partial_{\bar{\tau}_a}Z+\Big[2\pi|\epsilon|-\frac{2\pi}{3}|\epsilon|^3\sum_{a=1,2}|E_2(\tau_a)|^2\Big]\partial_\epsilon\partial_{\bar{\epsilon}}Z\notag\\
&+\sum_{a=1,2}2\text{Re}\bigg[\Big[-\frac{i}{2}\bar\epsilon+\frac{i}{3}\bar\epsilon|\epsilon|\bar E_2(\bar\tau_a)+\frac{i}{2}\epsilon|\epsilon|^2\bar E_2(\bar\tau_{\bar a})E_4(\tau_{\bar a})\Big]\partial_{\tau_a}\partial_{\bar{\epsilon}}Z\bigg]\notag\\
&+2\text{Re}\bigg[\Big[-\frac{|\epsilon|}{6\pi}\Big(\epsilon^2E_4(\tau_2)+\bar\epsilon^2\bar E_4(\bar\tau_1)\Big)\Big]\partial_{\tau_1}\partial_{\bar{\tau}_2}Z\bigg]+O(|\epsilon|^4).\tag{S105} \label{1st PF}
\end{align}
\subsection{Details of deformed one-point function}\label{C.3}
In this Appendix we discuss the detailed calculation of the first-order deformed one-point functions. For primary one-point function, the first-order correction depends on integral eq.(\ref{1pt correction}). The first term in eq.(\ref{1pt correction}) has been calculated in eq.(\ref{1st PF}) already. For the second term in eq.(\ref{1pt correction}), we apply eqs.(\ref{Dz})(\ref{Pzx}) and obtain
\begin{align}
&\frac{1}{Z}\int_{\Sigma^{(2)}\backslash{D_{\delta}}}d^2zD_z\bar{\mathcal{P}}_{\bar z,\bar x}\big(Z\langle{V}\rangle\big)\notag\\
=&\frac{1}{Z}\int_{\Sigma^{(2)}\backslash{D_{\delta}}}d^2z\bigg\lbrace\sum_{b=a,\bar a}(\sideset{^2}{_b}{\mathop{\mathcal{F}}}{}^2\bar{\mathcal{P}}_1)\frac{1}{2\pi i}\partial_{\tau_b}\partial_{\bar{x}}+({}^2\mathcal{F}^{\Pi}\epsilon^{\frac{1}{2}}{}^2\bar{\mathcal{P}}_1)\partial_{\epsilon}\partial_{\bar{x}}\notag\\
&+\sum_{b=a,\bar a}(\sideset{^2}{_b}{\mathop{\mathcal{F}}}{}^2\bar{\mathcal{P}}_2)\frac{1}{2\pi i}\partial_{\tau_b}wt[\bar{u}]+({}^2\mathcal{F}^{\Pi}\epsilon^{\frac{1}{2}}{}^2\bar{\mathcal{P}}_2)\partial_{\epsilon}wt[\bar{u}]\bigg\rbrace\big(Z\langle{V}\rangle\big). \tag{S106}\label{1p secondterm}
\end{align}
The $\mathcal{F}\mathcal{\bar{P}}$-type integrals  are calculated in eqs.(\ref{FP a 1})(\ref{FP abar 1})(\ref{FP pi 1})(\ref{FP a 2})(\ref{FP abar 2})(\ref{FP pi 2}), we approximate eq.(\ref{1p secondterm}) up to $|\epsilon|^3$:
\begin{align}
&\frac{1}{Z}\int_{\Sigma^{(2)}\backslash{D_{\delta}}}d^2zD_z\bar{\mathcal{P}}_{\bar z,\bar x}\big(Z\langle{V}\rangle\big)\notag\\
=&\frac{1}{Z}\bigg\lbrace i\Big[\text{Re}[x]-\frac{|\epsilon|}{3}\bar P_1(\bar x,\bar \tau_a)\Big]\partial_{\tau_a}\partial_{\bar x}+i\Big[\frac{1}{2}+\frac{|\epsilon|}{3}\bar P_2(\bar x,\bar \tau_a)\Big]wt[\bar u]\partial_{\tau_a}\notag\\
&-\frac{i}{6}\Big[\epsilon^2P_3(x,\tau_a)+2|\epsilon|\Big(\epsilon^2E_4(\tau_a)+\bar\epsilon^2\bar E_4(\bar\tau_{\bar a})\Big)\bar P_1(\bar x,\bar\tau_a)\Big]\partial_{\tau_{\bar a}}\partial_{\bar x}\notag\\
&+\frac{i|\epsilon|}{3}\Big[\epsilon^2E_4(\tau_a)+\bar\epsilon^2\bar E_4(\bar\tau_{\bar a})\Big]\bar P_2(\bar x,\bar\tau_a)wt[\bar u]\partial_{\tau_{\bar a}}\notag\\
&+\pi\Big[\epsilon P_1(x,\tau_a)+\Big(\frac{2\epsilon|\epsilon|}{3} E_2(\tau_a)-\frac{\bar\epsilon}{\pi}|\epsilon E_4(\tau_{\bar a})|^2\Big)\bar P_1(\bar x,\bar\tau_a)+\frac{\epsilon^3}{3}E_2(\tau_{\bar a})P_3(x,\tau_a)\Big]\partial_{\epsilon}\partial_{\bar x}\notag\\
&-\Big[\frac{2\pi}{3}\epsilon|\epsilon|E_2(\tau_a)-\bar\epsilon|\epsilon E_4(\tau_{\bar a})|^2\Big]\bar P_2(\bar x,\bar\tau_a)wt[\bar u]\partial_{\epsilon}\bigg\rbrace\big(Z\langle{V}\rangle\big)+O(|\epsilon|^4),\tag{S107} \label{1st 1p DP}
\end{align}
One can obtain the third term contribution in eq.(\ref{1pt correction}) by taking complex conjugate of eq.(\ref{1st 1p DP}). For the fourth term in eq.(\ref{1pt correction}),  we can divide the fourth term into four parts using eq.(\ref{Pzx}):
\begin{align}
&\int_{\Sigma^{(2)}\backslash{D_{\delta}}}d^2z\mathcal{P}_{z,x}\bar{\mathcal{P}}_{\bar z,\bar x}\langle{V}\rangle\notag\\
=&\int_{\Sigma^{(2)}\backslash{D_{\delta}}}d^2z\bigg\lbrace({}^2\mathcal{P}_1{}^2\bar{\mathcal{P}}_1)\partial_x\partial_{\bar x}+({}^2\mathcal{P}_1{}^2\bar{\mathcal{P}}_2)\partial_x wt[\bar u]\notag\\
&+({}^2\mathcal{P}_2{}^2\bar{\mathcal{P}}_1)wt[u]\partial_{\bar x}+({}^2\mathcal{P}_2{}^2\bar{\mathcal{P}}_2)wt[u]wt[\bar u]\bigg\rbrace\langle{V}\rangle.\tag{S108} \label{1st fourthterm}
\end{align}
Using $\mathcal{P}\bar{\mathcal{P}}$-type integrals eqs.(\ref{PP 1 1})(\ref{PP 2 2})(\ref{PP 1 2}) , up to $|\epsilon|^3$, then it becomes
\begin{align}
&\int_{\Sigma^{(2)}\backslash{D_{\delta}}}d^2z\mathcal{P}_{z,x}\bar{\mathcal{P}}_{\bar z,\bar x}\langle{V}\rangle\notag\\
=&\Big[2\pi\log|Q(x,\tau_a)|^2+\frac{\pi}{2}\Big]\partial_x\partial_{\bar x}\langle{V}\rangle+2\text{Re}\Big[\pi\bar P_{1}(\bar x,\bar\tau_a)\partial_x wt[\bar u]\langle{V}\rangle\Big]\notag\\
&-\sum_{k=1}^3\frac{2\pi}{k(k+2)}|\epsilon|^kP_x^{(k)}\bar P_{\bar x}^{(k)}\langle{V}\rangle+O(|\epsilon|^4).\tag{S109} \label{1p PP}
\end{align} 
where the operator $P^{(k)}_x$ is introduced to simplify the result:
\begin{equation}
P^{(k)}_x=P'_k(x,\tau_a)\partial_x-kP_{k+1}(x,\tau_a)wt[u],  \tag{S110}
\end{equation}
for $x\in S_a$. $\log Q(x,\tau_a)$ in eq.(\ref{1p PP}) is the primitive function of $P_1(x,\tau_a)$ defined by eq.(\ref{Q}).\par
For one-point function of stress tensor, the first-order correction is shown as eq.(\ref{1st int T}). The first two terms in eq.(\ref{1st int T}) have been calculated in eq.(\ref{1st PF}) and eq.(\ref{1st 1p DP}) respectively. The third term in eq.(\ref{1st int T}) can be further written by eq.(\ref{Dz}) as
\begin{align}
&\frac{c}{2Z}\int_{\Sigma^{(2)}\backslash{D_{\delta}}}d^2z{}^2{\mathcal{P}}_4\bar D_{\bar z}Z\notag\\
=&\frac{c}{2Z}\int_{\Sigma^{(2)}\backslash{D_{\delta}}}d^2z\bigg\lbrace\sum_{b=a,\bar a}\big({}^2\mathcal{P}_4\sideset{^2}{_b}{\mathop{\bar{\mathcal{F}}}}\big)\frac{-1}{2\pi i}\partial_{\bar{\tau}_b}+\big({}^2{\mathcal{P}}_4{}^2\bar{\mathcal{F}}^{\Pi}\bar{\epsilon}^{\frac{1}{2}}\big)\partial_{\bar\epsilon}\bigg\rbrace Z,\tag{S111}
\end{align}
which is obtained using eqs.(\ref{FP a 4})(\ref{FP abar 4})(\ref{FP pi 4}), up to $|\epsilon|^3$:
\begin{align}
&\frac{c}{2Z}\int_{\Sigma^{(2)}\backslash{D_{\delta}}}d^2z{}^2{\mathcal{P}}_4\bar D_{\bar z}Z\notag\\
=&-\frac{c}{6Z}\bigg\lbrace i|\epsilon|P_{4}(x,\tau_a)\partial_{\bar{\tau}_a}+i|\epsilon|\Big[\epsilon^2E_4(\tau_{\bar a})+\bar\epsilon^2\bar E_4(\bar\tau_a)\Big]P_4(x,\tau_a)\partial_{\bar{\tau}_{\bar a}}\notag\\
&+2\pi\bar\epsilon|\epsilon|\bar E_2(\bar\tau_a)P_4(x,\tau_a)\partial_{\bar\epsilon}\bigg\rbrace Z+O(|\epsilon|^4).\tag{S112} \label{1st T PD}
\end{align}
\subsection{Details of deformed two-point function}\label{C.4}
In this Appendix we discuss the detailed calculation of the first-order deformed two-point functions. For primary two-point function, the first-order correction depends on eq.(\ref{2p WI}). After equality the first lines have been calculated in eqs.(\ref{1st PF})(\ref{1st 1p DP}). In the second line, the integrals of $\mathcal{P}_{z,x_1}\bar{\mathcal{P}}_{\bar{z},\bar{x}_1}$ and $\mathcal{P}_{z,x_2}\bar{\mathcal{P}}_{\bar{z},\bar{x}_2}$ have been calculated in eq.(\ref{1p PP}).
To calculate remaining integrals of the second line, there are two different profiles. The one profile is that two insertion points live on the same torus $x_1,x_2\in S_a$. Using eqs.(\ref{P1P2 1 1})(\ref{P1P2 1 2})(\ref{P1P2 2 2}), we approximate the result up to $|\epsilon|^3$:
\begin{align}
&\int_{\Sigma^{(2)}\backslash{D_{\delta}}}d^2z\mathcal{P}_{z,x_1}\bar{\mathcal{P}}_{\bar z,\bar x_2}\langle{V_1V_2}\rangle
\notag\\=&\Big[\pi\log\frac{|Q(x_1,\tau_a)Q(x_2,\tau_a)|^2}{|Q(x_1-x_2,\tau_a)|^2}+\frac{\pi}{2}\Big]\partial_{x_1}\partial_{\bar x_2}\langle{V_1V_2}\rangle\notag\\
&+\Big[\pi\bar P_1(\bar x_1-\bar x_2,\bar\tau_a)+\pi\bar P_1(\bar x_2,\bar\tau_a)\Big]\partial_{x_1} wt[\bar u_2]\langle{V_1V_2}\rangle\notag\\
&+\Big[\pi P_1(x_2-x_1,\tau_a)+\pi P_1(x_1,\tau_a)\Big]wt[u_1]\partial_{\bar x_2}\langle{V_1V_2}\rangle\notag\\
&-\sum_{k=1}^3\frac{2\pi}{k(k+2)}|\epsilon|^kP_{x_1}^{(k)}\bar P_{\bar x_2}^{(k)}\langle{V_1V_2}\rangle+O(|\epsilon|^4).\tag{S113} \label{2p PP 1}
\end{align}
The logarithmic divergence $\log{|Q(x_1-x_2,\tau_a)|^2}$ in eq.(\ref{2p PP 1}) in the deformed two point functions has been observed in \cite{He:2019vzf}.
The other profile is that two insertion points live on different tori $x_1=x\in S_a$ and $x_2=y\in S_{\bar a}$, and their corresponding primary states are $u$ and $v$, respectively. Using eqs.(\ref{PxPy 1 1})(\ref{PxPy 1 2})(\ref{PxPy 2 2}) we obtain the integral of $\mathcal{P}_{z,x}\bar{\mathcal{P}}_{\bar z,\bar y}$ up to $|\epsilon|^3$:
\begin{align}
&\int_{\Sigma^{(2)}\backslash{D_{\delta}}}d^2z\mathcal{P}_{z,x}\bar{\mathcal{P}}_{\bar z,\bar y}\langle{V_xV_y}\rangle\notag\\
=&\sum_{k=1}^3\frac{-\pi}{k+1}\Big[\epsilon^kP'_{k+1}(y,\tau_{\bar a})P_x^{(k-1)}\partial_{\bar y}+\bar\epsilon^k\bar P'_{k+1}(\bar x,\bar\tau_a)\partial_{x}\bar P_{\bar y}^{(k-1)}\Big]\langle{V_xV_y}\rangle\notag\\
&+\frac{3\pi}{2}|\epsilon|^2\Big[\bar\epsilon\bar E_4(\bar\tau_a)P_x^{(2)}\partial_{\bar y}+\epsilon E_4(\tau_{\bar a})\partial_x\bar P_{\bar y}^{(2)}\Big]\langle{V_xV_y}\rangle\notag\\
&-\frac{2\pi}{3}|\epsilon|\Big[\bar\epsilon^2\bar E_{4}(\bar\tau_a)+\epsilon^2E_4(\tau_{\bar a})\Big]P_x^{(1)}\bar P_{\bar y}^{(1)}\langle{V_xV_y}\rangle+O(|\epsilon|^4).\tag{S114} \label{2p PP 2}
\end{align}\par
For two-point function of stress tensor, there are two types to consider: $\langle{T_1T_2}\rangle$ and $\langle{T_1\bar T_2}\rangle$. The first-order correction of $\langle{T_1T_2}\rangle$ is shown as eq.(\ref{1st T1T2}), and all the integrals have been computed in eqs.(\ref{1st PF})(\ref{1st 1p DP})(\ref{1st T PD}) above.  The first-order correction of $\langle{T_1\bar T_2}\rangle$ is shown as eq.(\ref{1st T1T2bar}),
and all the integrals in eq.(\ref{1st T1T2bar}) have been computed in eqs.(\ref{1st PF})(\ref{1st 1p DP})(\ref{2p PP 1})(\ref{2p PP 2})(\ref{1st T PD}) except for the last three ones.
In the case of two insertion points live on the same torus $x_1,x_2\in S_a$, The integral of ${}^2\bar{\mathcal{P}}_4(\bar z,\bar x_2){\mathcal{P}}_{z,x_1}$ is obtain\footnote{The integral of ${}^2{\mathcal{P}}_4(z,x_1)\bar{\mathcal{P}}_{\bar z,\bar x_2}$ is complex conjugate of eq.(\ref{P4P 12}).} using eqs.(\ref{P1P2 1 4})(\ref{P1P2 2 4}), up to $|\epsilon|^3$:
\begin{align}
&\frac{c}{2}\int_{\Sigma^{(2)}\backslash{D_{\delta}}}d^2z\ {}^2\bar{\mathcal{P}}_4(\bar z,\bar x_2){\mathcal{P}}_{z,x_1}\langle{T_1}\rangle\notag\\
=&\frac{\pi c}{6}\Big[\bar P_3(\bar x_1-\bar x_2,\bar\tau_a)+\bar P_3(\bar x_2,\bar\tau_a)\Big]\partial_{x_1}\langle{T_1}\rangle\notag\\
&+\frac{\pi c}{6}\sum_{k=1}^3(k+1)|\epsilon|^k\bar P_{k+3}(\bar x_2,\bar\tau_a)P_{x_1}^{(k)}\langle{T_1}\rangle+O(|\epsilon|^4).\tag{S115} \label{P4P 12}
\end{align}
The integral of ${}^2\mathcal{P}_4(z,x_1){}^2\bar{\mathcal{P}}_4(\bar z,\bar x_2)$ is computed in eq.(\ref{P1P2 4 4}):
\begin{align}
&\frac{c^2}{4}\int_{\Sigma^{(2)}\backslash{D_{\delta}}}d^2z\ {}^2\mathcal{P}_4(z,x_1){}^2\bar{\mathcal{P}}_4(\bar z,\bar x_2)\notag\\
=&-\sum_{k=1}^3\frac{k(k+1)^2(k+2)\pi c^2}{72}|\epsilon|^kP_{k+3}(x_1,\tau_a)\bar P_{k+3}(\bar x_2,\bar\tau_a)+O(|\epsilon|^4).\tag{S116}\label{P4P4 12}
\end{align}
In the case of two insertion points live on different tori $x_1=x\in S_a$ and $x_2=y\in S_{\bar a}$, the integral of ${}^2\bar{\mathcal{P}}_4(\bar z,\bar y){\mathcal{P}}_{z,x}$ is obtained using eqs.(\ref{PxPy 1 4})(\ref{PxPy 2 4}):
\begin{align}
&\frac{c}{2}\int_{\Sigma^{(2)}\backslash{D_{\delta}}}d^2z\ {}^2\bar{\mathcal{P}}_4(\bar z,\bar y){\mathcal{P}}_{z,x}\langle{T_x}\rangle\notag\\
=&\Big[\frac{\pi c}{6}\Big(\bar\epsilon^2\bar P_3(\bar x,\bar\tau_a)\bar P_4(\bar y,\bar\tau_{\bar a})+\frac{\pi c}{2}\bar\epsilon^3\bar P'_4(\bar x,\bar\tau_a)\bar P_5(\bar y,\bar\tau_{\bar a})-3\pi c\epsilon|\epsilon|^2E_4(\tau_{\bar a})\bar P_5(\bar y,\bar\tau_{\bar a})\Big]\partial_{x}\langle{T_x}\rangle\notag\\
&+\frac{\pi c}{3}|\epsilon|\Big[\epsilon^2E_4(\tau_{\bar a})+\bar\epsilon^2\bar E_4(\bar\tau_a)\Big]\bar P_4(\bar y,\bar\tau_{\bar a})P_x^{(1)}\langle{T_x}\rangle+O(|\epsilon|^4).\tag{S117} \label{P4P xy}
\end{align}
rThe integral of ${}^2\mathcal{P}_4(z,x){}^2\bar{\mathcal{P}}_4(\bar z,\bar y)$ is computed in eq.(\ref{PxPy 4 4}):
\begin{align}
&\frac{c^2}{4}\int_{\Sigma^{(2)}\backslash{D_{\delta}}}d^2z\ {}^2\mathcal{P}_4(z,x){}^2\bar{\mathcal{P}}_4(\bar z,\bar y)\notag\\
=&-\frac{\pi c^2}{6}|\epsilon|\Big[\epsilon^2E_4(\tau_{\bar a})+\bar\epsilon^2\bar E_4(\bar\tau_a)\Big]P_4(x,\tau_a)\bar P_4(\bar y,\bar\tau_{\bar a})+O(|\epsilon|^4).\tag{S118} \label{P4P4 xy}
\end{align}\par

\bibliographystyle{JHEP}
\bibliography{reference}

\end{document}